\documentclass{article}

\usepackage{arxiv}

\usepackage[utf8]{inputenc} 
\usepackage[T1]{fontenc}    
\usepackage{hyperref}       
\usepackage{url}            
\usepackage{booktabs}       
\usepackage{amsfonts}       
\usepackage{nicefrac}       
\usepackage{microtype}      
\usepackage{lipsum}		
\usepackage{graphicx}
\usepackage{natbib}
\usepackage{doi}
\usepackage{caption}
\usepackage{subcaption}
\usepackage{algorithm}
\usepackage{algorithmic}
\usepackage{amsmath}

\title{Web-Based Dynamic Paintings: Real-Time Interactive Artworks in Web Using a 2.5D Pipeline}

\author{ 
 \href{https://orcid.org/0000-0003-3618-4166}{\includegraphics[scale=0.06]{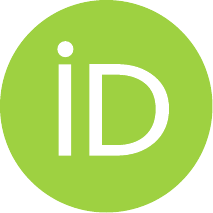}\hspace{1mm}Ergun Akleman}\thanks{Joint with Computer Science and Engineering Department.} \\
	Visual Computing \& Computational Media,\\ Texas A\&M University, College Station, TX, 77831\\
	\texttt{ergun@tamu.edu} \\
      \And
      Youyou Wang\\
Department of Computer Science and Engineering\\ 
Texas A\&M University, College Station, TX, 77831\\
	\texttt{kingyy2010@gmail.com} \\
     \And
      	 Yinan Xiong\\
Department of Visualization\\ 
Texas A\&M University, College Station, TX, 77831\\
	\texttt{yvonnexiong1114@gmail.com } \\
    \And
    	 Anusha Shanker\\
Department of Visualization\\ 
Texas A\&M University, College Station, TX, 77831\\
	\texttt{anushishanker@gmail.com} \\
   \And
	Fermi Perumal\\
Department of Visualization\\ 
Texas A\&M University, College Station, TX, 77831\\
	\texttt{ferminiveditha@gmail.com} \\
      \And
	\"Ozg\"ur G\"onen\\
Department of Architecture\\ 
Texas A\&M University, College Station, TX, 77831\\
	\texttt{ozgur.gonen@gmail.com} \\
  \And
	 Motahareh Fard\\
Department of Visualization\\ 
Texas A\&M University, College Station, TX, 77831\\
	\texttt{motah.fard@gmail.com } \\
 }
 
\renewcommand{\shorttitle}{Web-Based  Dynamic Paintings with 2.5D Pipeline}

\hypersetup{
pdftitle={A template for the arxiv style},
pdfsubject={q-bio.NC, q-bio.QM},
pdfauthor={David S.~Hippocampus, Elias D.~Striatum},
pdfkeywords={First keyword, Second keyword, More},
}

\begin{document}
\maketitle

\begin{abstract}

In this work, we present a 2.5D pipeline approach to creating dynamic paintings that can be re-rendered interactively in real-time on the Web. Using this 2.5D approach, any existing simple painting such as portraits can be turned into an interactive dynamic web-based artwork. Our interactive system provides most global illumination effects such as reflection, refraction, shadow, and subsurface scattering by processing images. In our system, the scene is defined only by a set of images. These include (1) a shape image, (2) two diffuse images, (3) a background image, (4) one foreground image, and (5) one transparency image. A shape image is either a normal map or a height. Two diffuse images are usually hand-painted. They are interpolated using illumination information. The transparency image is used to define the transparent and reflective regions that can reflect the foreground image and refract the background image, both of which are also hand-drawn. This framework, which mainly uses hand-drawn images, provides qualitatively convincing painterly global illumination effects such as reflection and refraction. We also include parameters to provide additional artistic controls. For instance, using our piecewise linear Fresnel function, it is possible to control the ratio of reflection and refraction. This system is the result of a long line of research contributions. On the other hand, the art-directed Fresnel function that provides physically plausible compositing of reflection and refraction with artistic control is completely new. Art-directed warping equations that provide qualitatively convincing refraction and reflection effects with linearized artistic control are also new. You can try our web-based system for interactive dynamic real-time paintings at http://mock3d.tamu.edu/. 
\end{abstract}

\section{Introduction and Motivation}

This document is an extended version of our ISEA 2022 paper \cite{akleman2022dynamic}. Non-photorealistic rendering (NPR) has emerged as a subfield of computer graphics during the 1990s to produce computer-generated images that invoke the appearance of being created "by hand" \cite{strothotte2002non,gooch2001non} by emulating broad artistic styles such as outlines and silhouettes \cite{hertzmann1999silhouettes}, technical illustrations \cite{gooch1998non}, pen and ink drawings  \cite{deussen2000computer,markosian1997real,du2017designing}, impressionist  \cite{litwinowicz1997processing}, implicit painting \cite{akleman1998,akleman1998a}, and cubist paintings \cite{meadows2000a,smith2004,morrison2020remote}, Chinese painting \cite{chan2002two,liu2015chinese}, charcoals \cite{majumder2002real,du2016charcoal}, and stippling \cite{lu2002non}; as well as artistic tools and mediums such as brush strokes \cite{vanderhaeghe2013stroke,yeh2002animals,hertzmann1998painterly,lin2012video}, watercolor \cite{curtis1997computer}. Convolution neural networks have been shown to be effective for style transfer \cite{gatys2016image,selim2016painting}. 

In recent years, there has also been growing interest in turning specific paintings into dynamic computer-generated images with moving lights and cameras. These paintings can have non-realistic components. For example, Murphy developed a non-photorealistic approach to match shapes and colors in the artwork of Disney background painter Eyvind Earle, who uses non-realistic shadows \cite{murphy2015developing}. "Atelier des Lumières" group developed large-scale video projections of many of Van Gogh's well-known works \cite{connaissance2019}.  Liu created a 3D version of a Jiangnan water country painting by the contemporary Chinese artist Yang Ming-Yi as a primary visual reference \cite{liu2015chinese}. Justice created dynamic time-lapse animations based on some of the works of Edgar Payne, using barycentric shading as the core of his procedure \cite{justice2018}.  Zhao recreated an Ayvazovski painting
\cite{yan2015}. Subramanian obtained painterly reflection, refraction, and caustics with a classical still life painting of wine and glass \cite{subramanian2020painterly}. Ross developed a shader that emulates the style of Georgia O’Keeffe \cite{ross2021Georgia},

\begin{figure}
\includegraphics[width=0.32\linewidth]{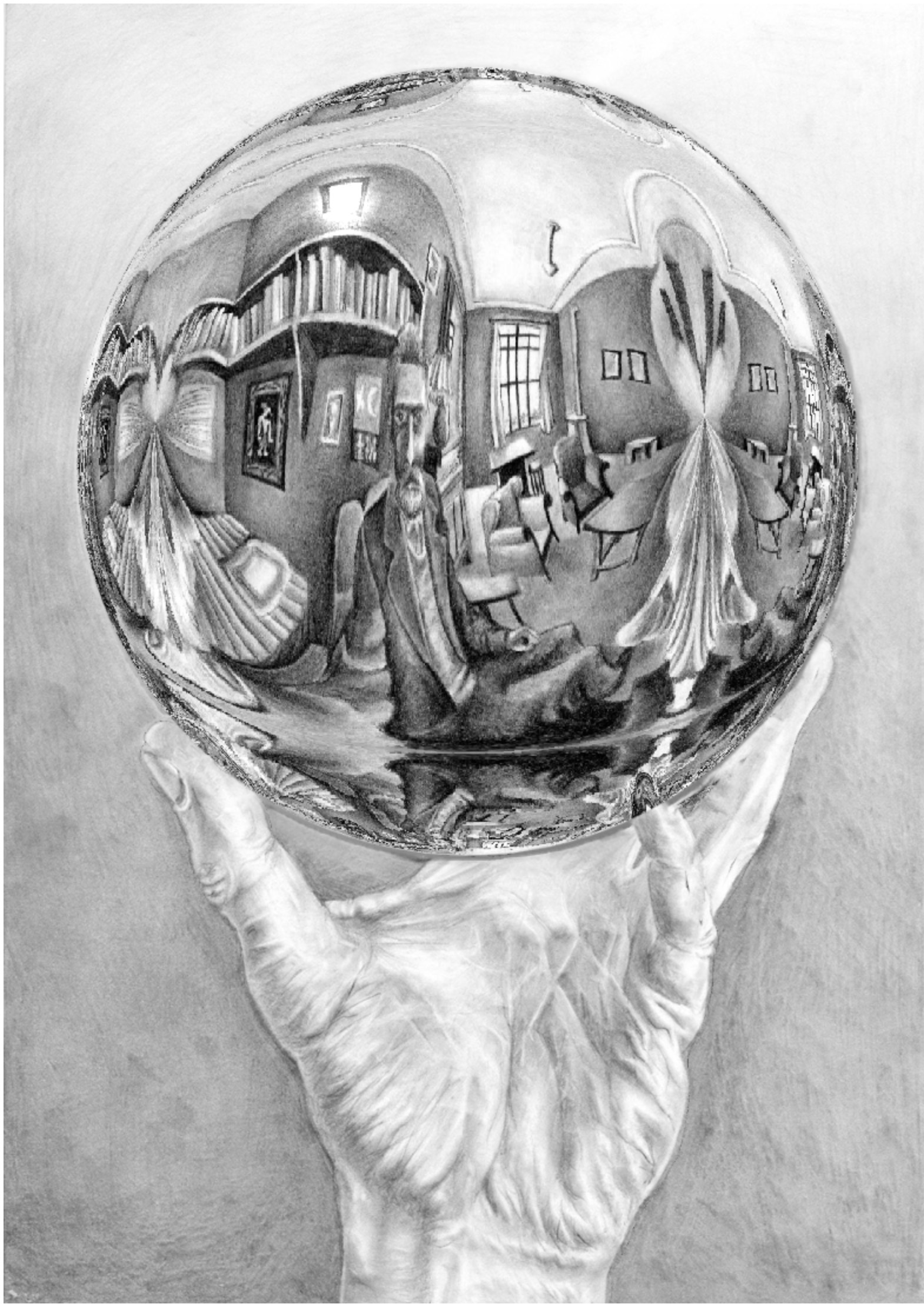}
\includegraphics[width=0.32\linewidth]{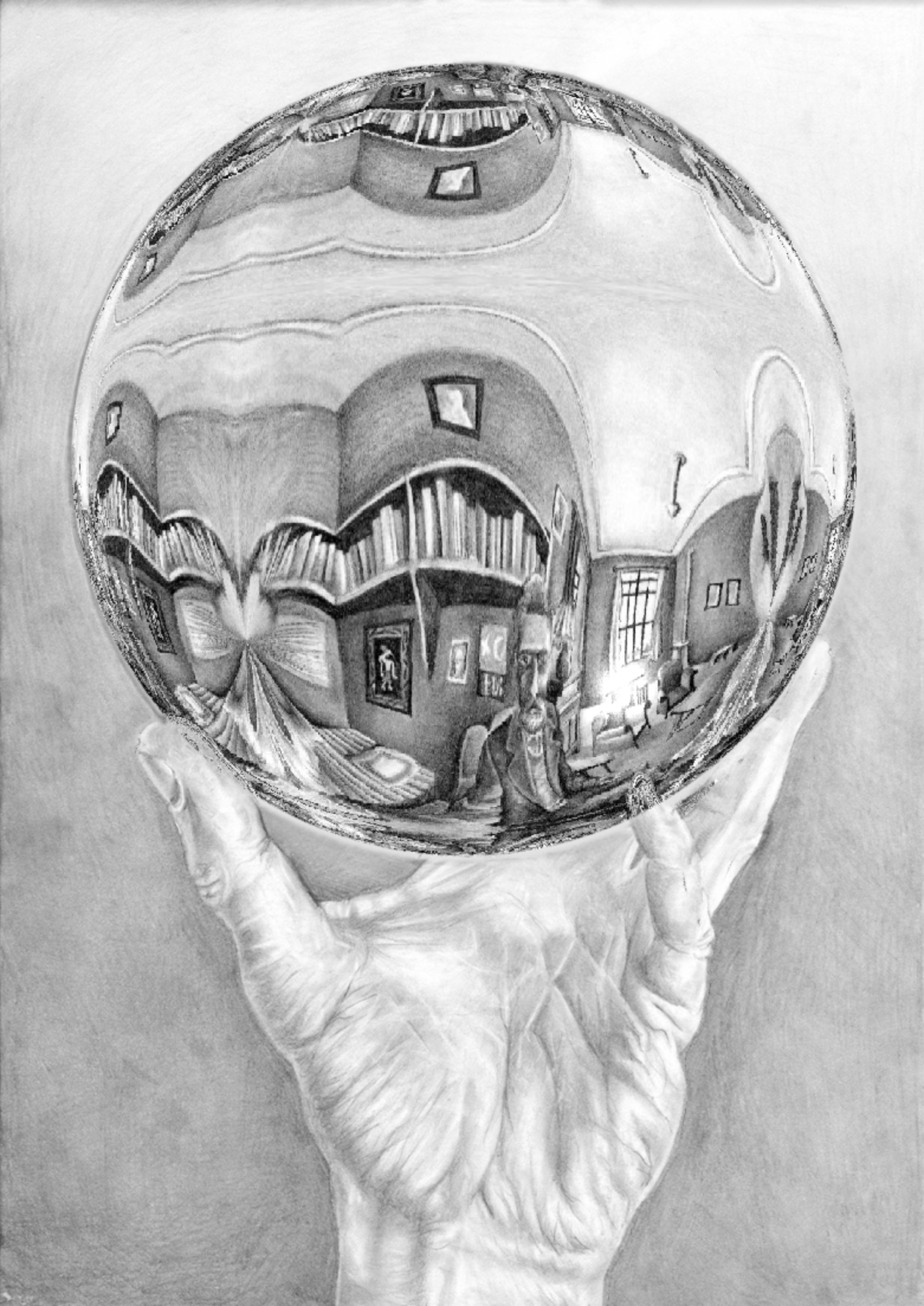}
\includegraphics[width=0.32\linewidth]{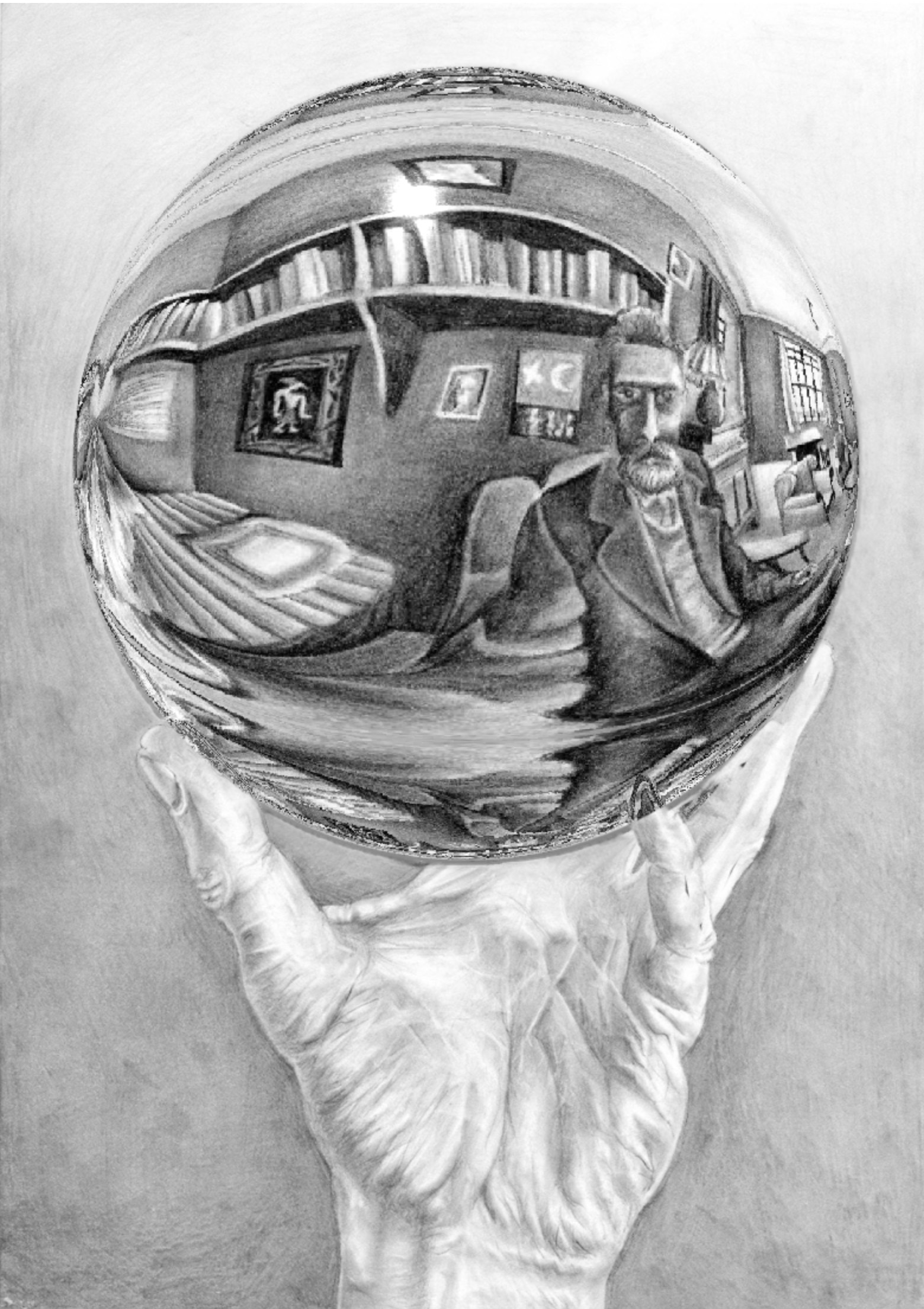}
\caption{\it Interactively created illustrations based on ``Self-Portrait in Spherical Mirror'', a lithograph by Dutch artist M. C. Escher. }
\label{figEscher}
\end{figure}

\begin{figure}[hbtp]
\centering
\includegraphics[width=1.0\textwidth]{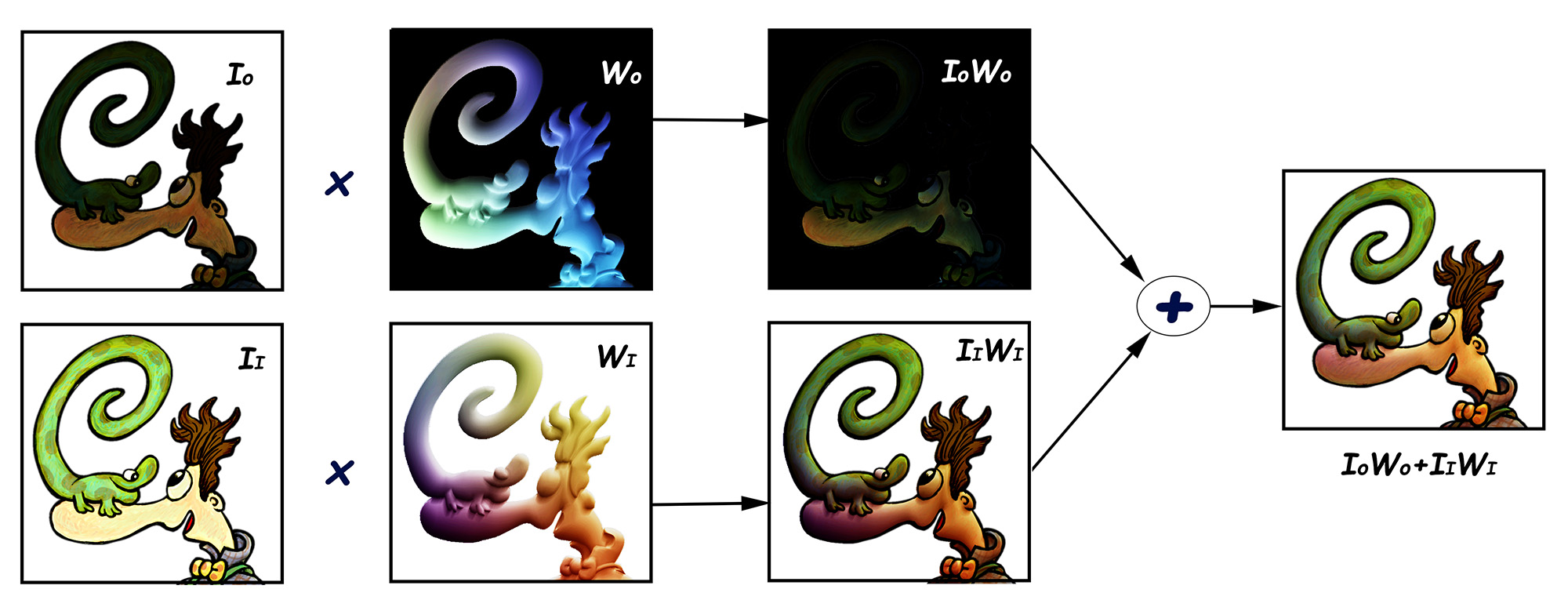}
\caption{An example demonstrating the pipeline for diffuse rendering.  $I_0$ and $I_1$  are control images, $\Omega_0$ and $\Omega_1$  are weight images that satisfy partition of unity. Here, we compute $\Omega_0$ from a shape image just using image processing. $I = I_0 \Omega_0 + I_1 \Omega_1$ is the final rendering obtained by taking a weighted average of the two control images. }
\label{figtable}
\end{figure}

\subsection{Problem Definition} 

\begin{figure*}
\includegraphics[width=0.19\linewidth]{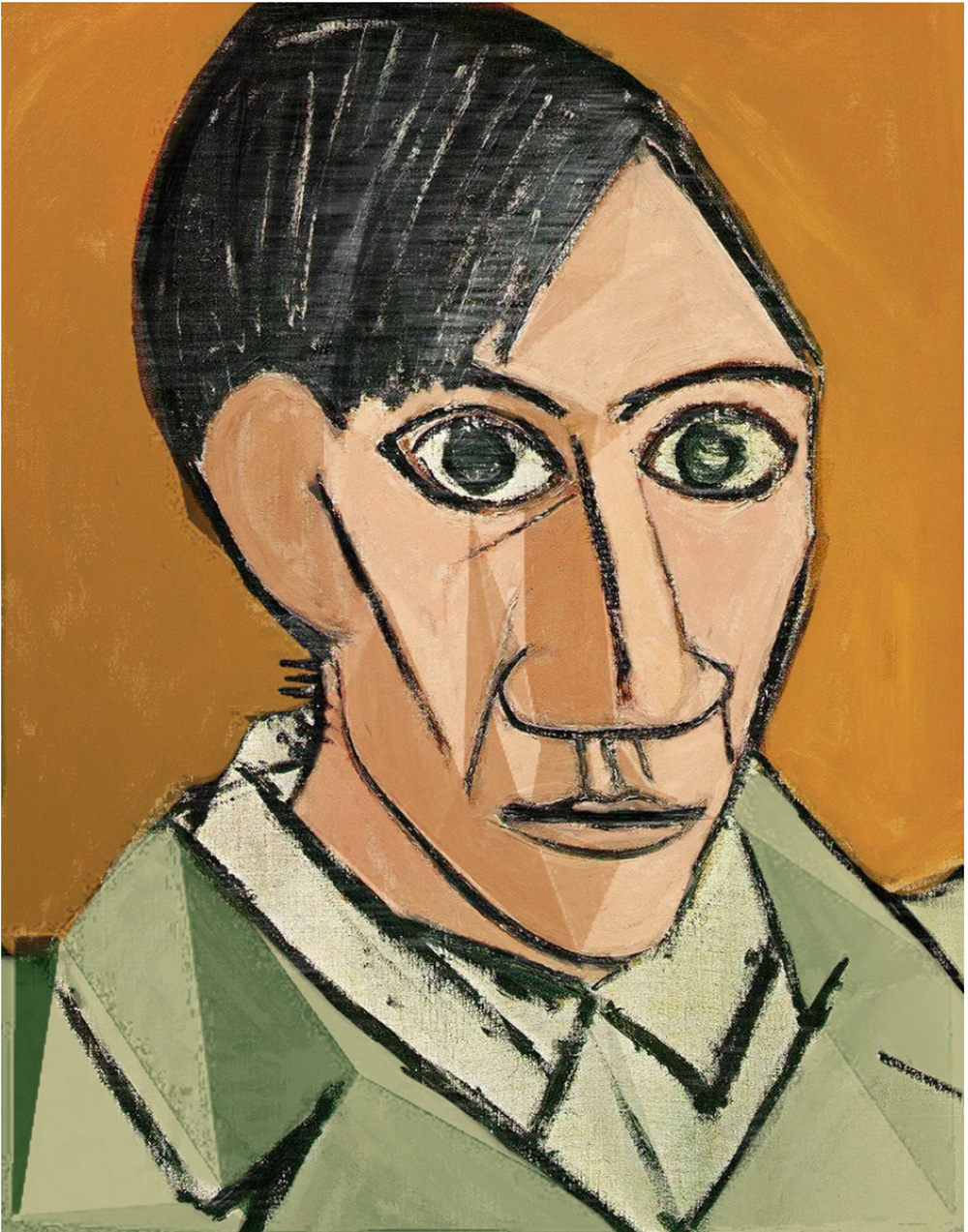}
\includegraphics[width=0.19\linewidth]{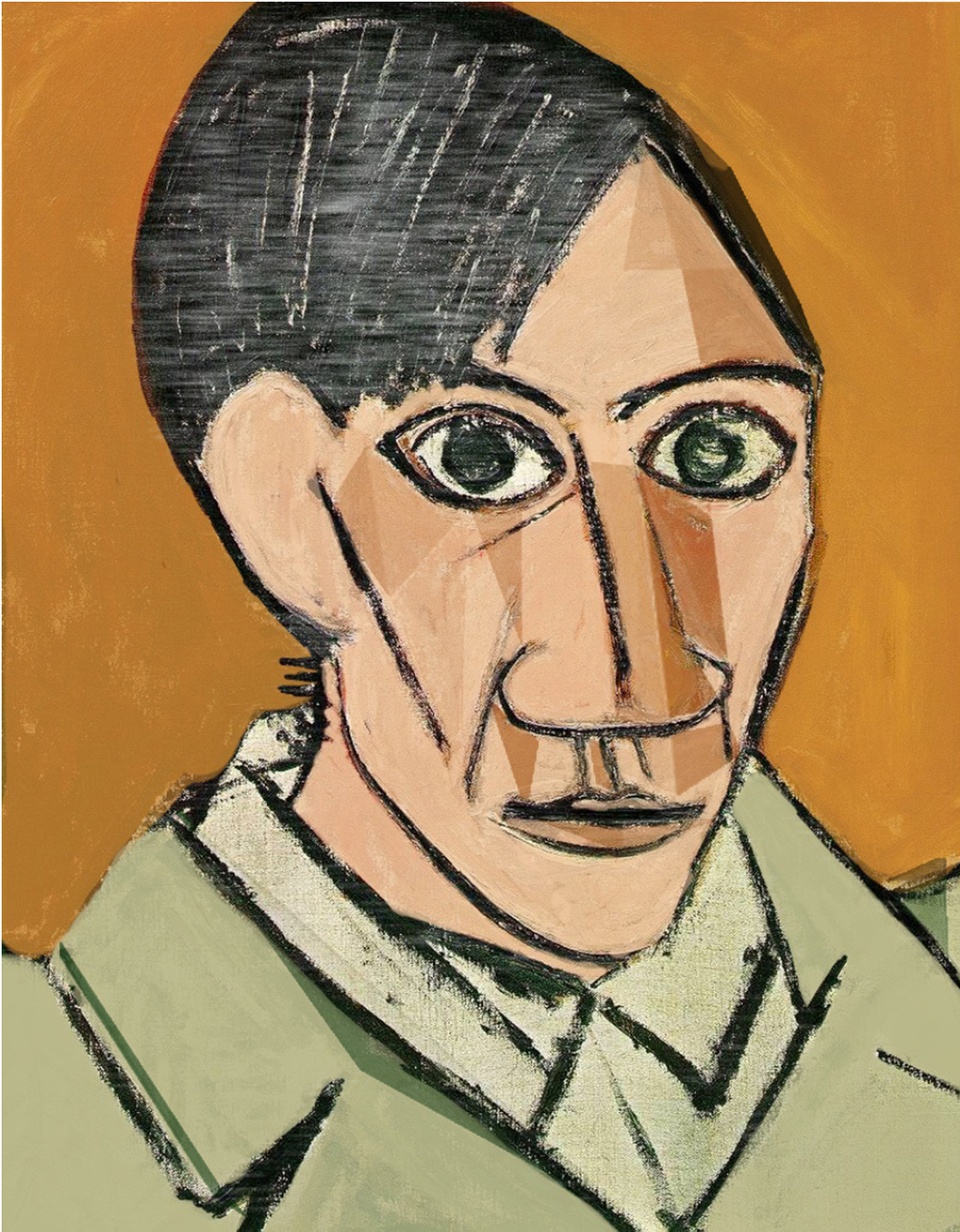}
\includegraphics[width=0.19\linewidth]{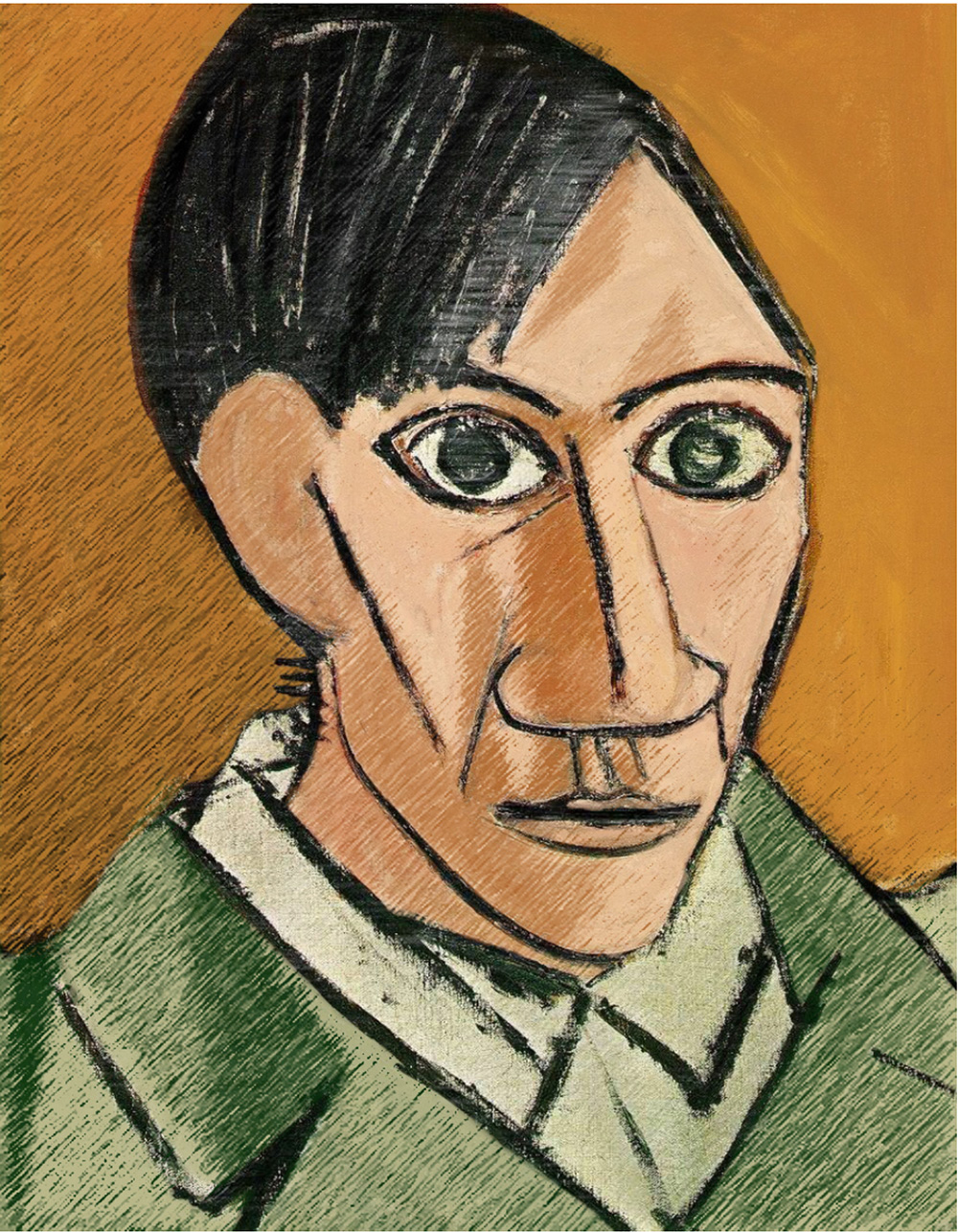}
\includegraphics[width=0.19\linewidth]{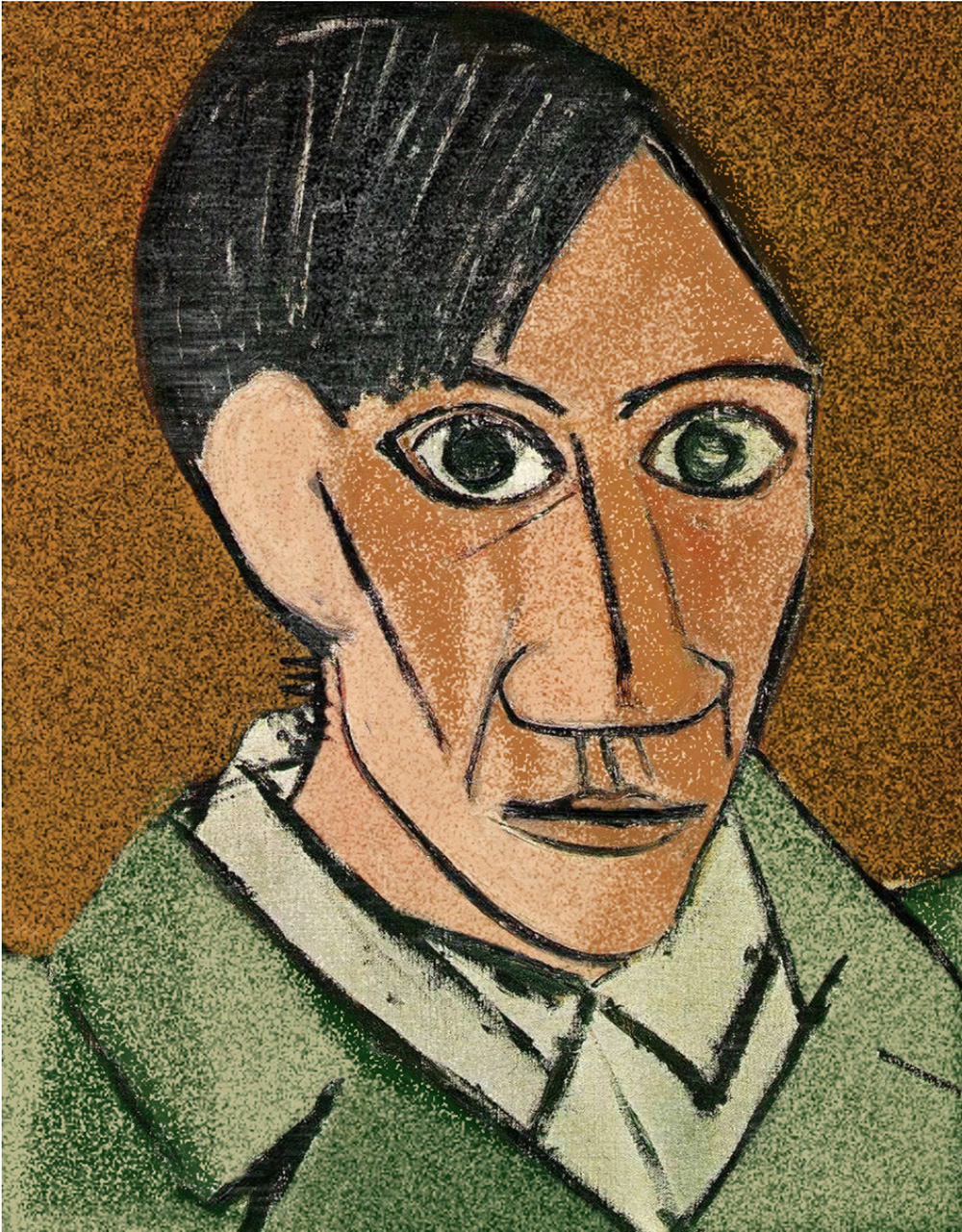}
\includegraphics[width=0.19\linewidth]{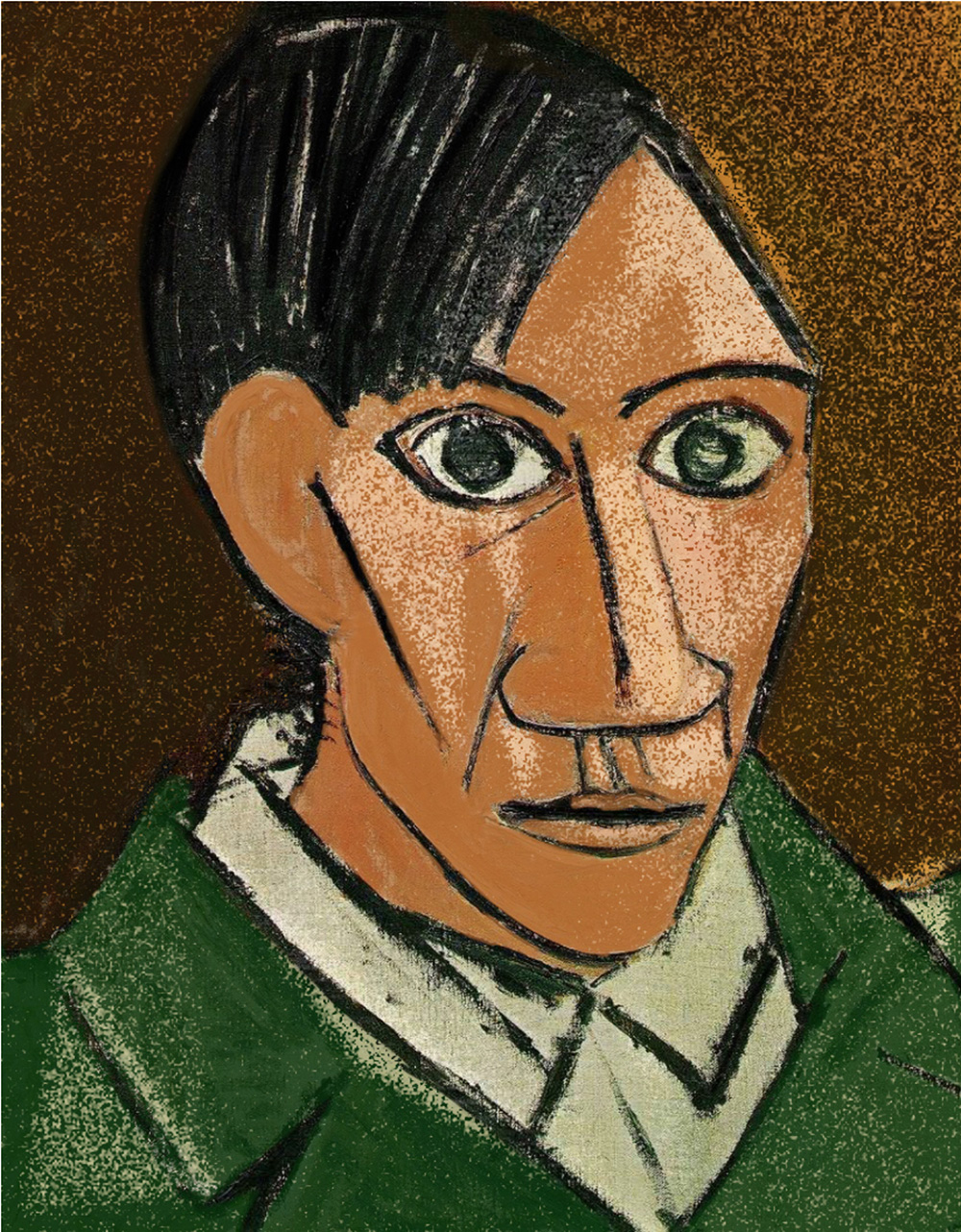}
\caption{\it Interactively created paintings based on ``Autoportrait'', an oil painting by Pablo Picasso in 1907. All images were created using a normal map.  }
\label{figPicasso1}
\vspace{0.1in}
\includegraphics[width=0.19\linewidth]{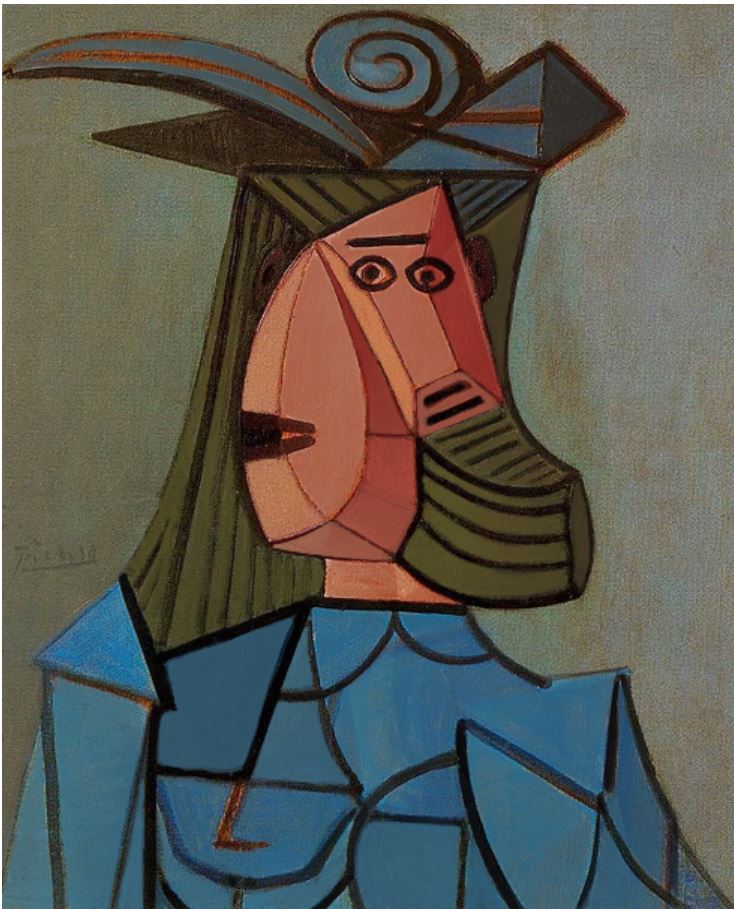}
\includegraphics[width=0.19\linewidth]{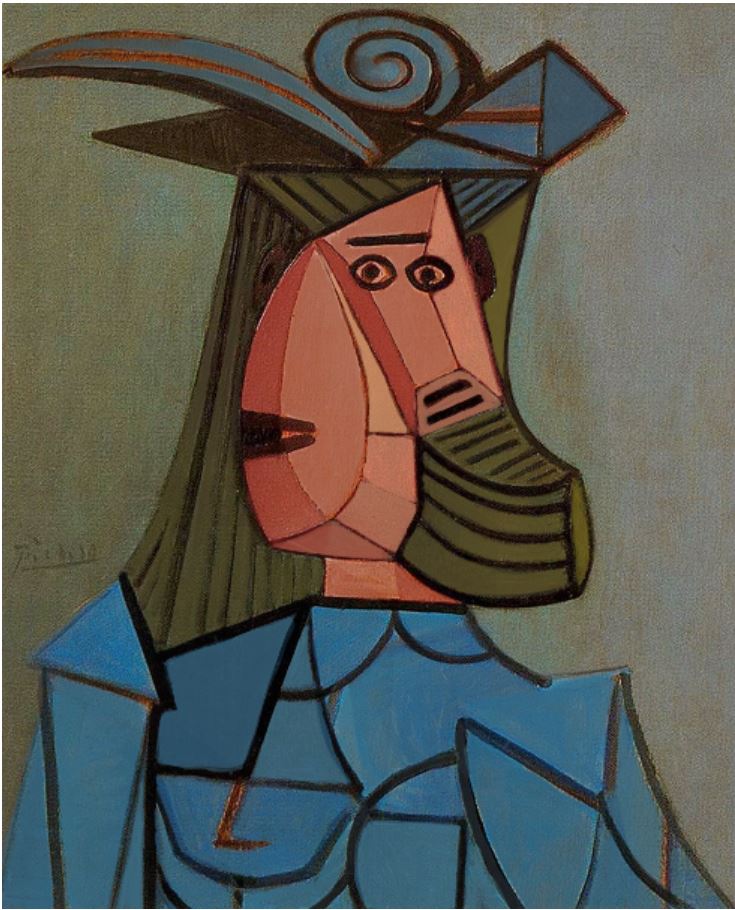}
\includegraphics[width=0.19\linewidth]{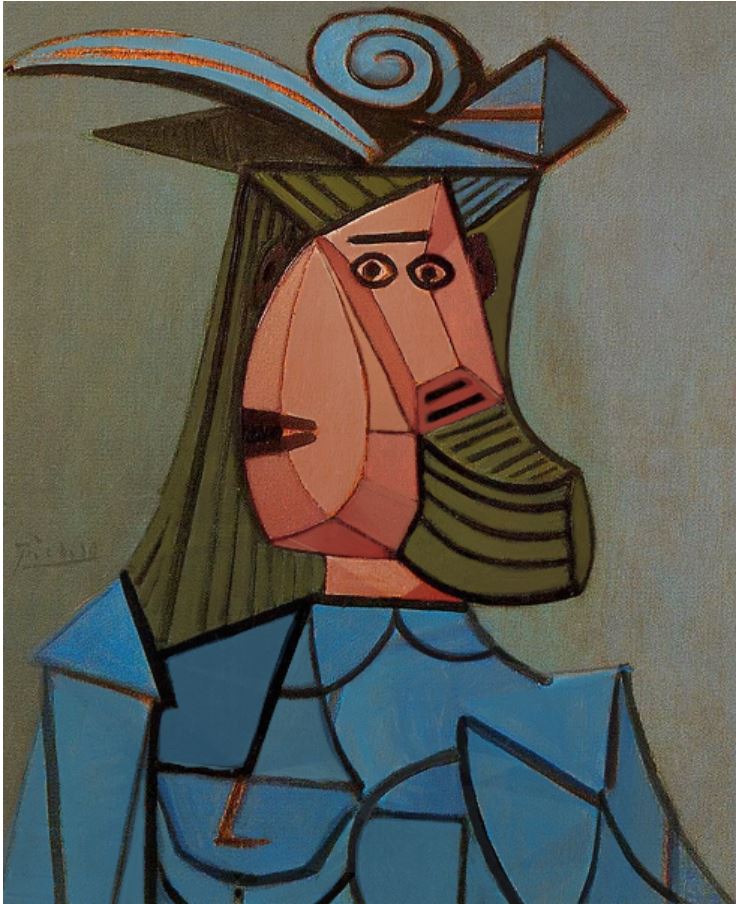}
\includegraphics[width=0.19\linewidth]{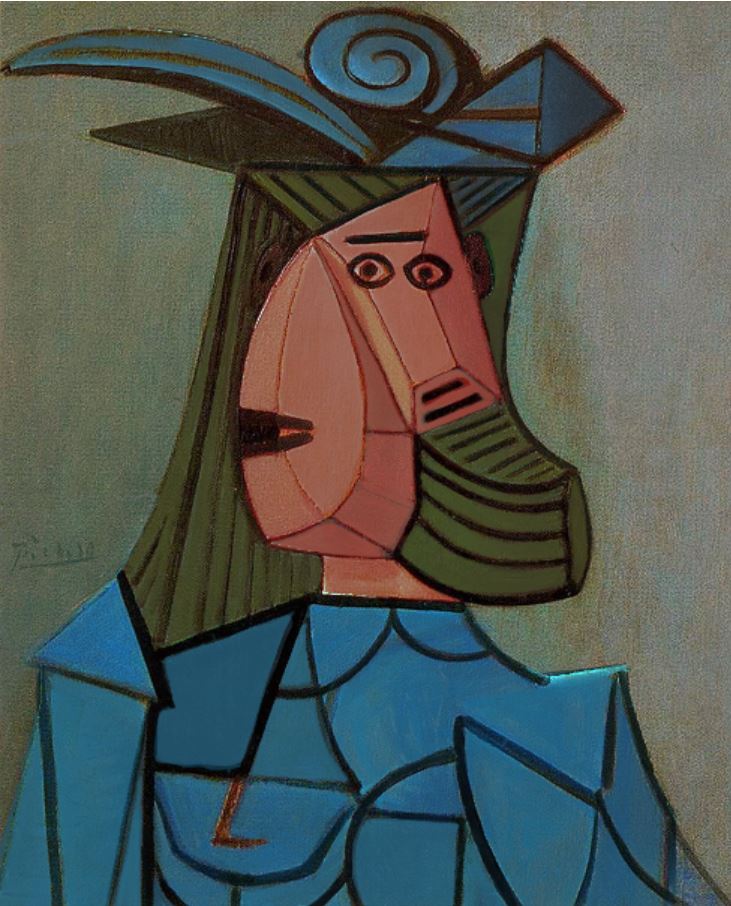}
\includegraphics[width=0.19\linewidth]{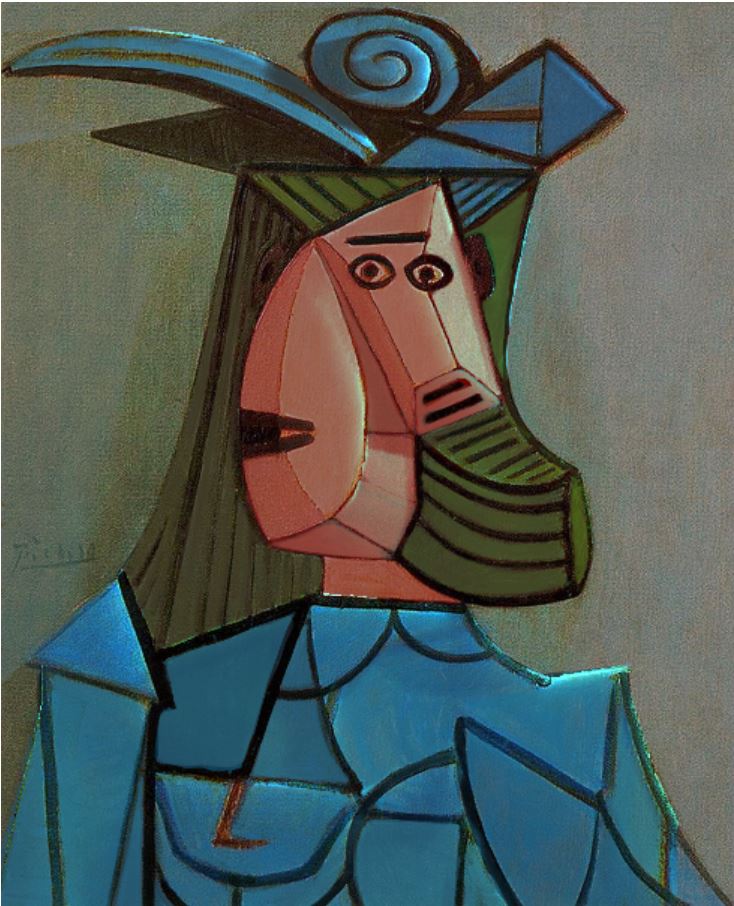}
\caption{\it Interactively created paintings based on ``Portrait of woman (Dora Maar)'', an oil painting by Pablo Picasso in 1942. The two images on the left are created using a depth map; therefore, we also obtain shadows.  }
\label{figPicasso2}
\end{figure*}

The main problem with these methods is that they are still based on the standard computer graphics pipeline. There is still a need for modeling and animation software such as Maya or Blender even if the proxy geometry is simple. Moreover, there is also a need for shader software, such as Arnold or Renderman, even for expressive depiction. In this work, we propose a completely different pipeline. All information from shapes to materials is provided by images. Shapes are defined by normal maps or depth maps. The shading parameters are also provided by a set of control images. We can obtain physically plausible local and global illumination with complete style control using these images. There exist solutions to obtain diffuse reflection, shadows, reflection, and refraction with a set of images\cite{wang2014,wang2014global,akleman2016,akleman2017}. Examples of illustrations or paintings that are interactively obtained in real time using our web-based system are shown in Figures~\ref{figEscher}, \ref{figPicasso1}, and \ref{figPicasso2}. The interface of our system is shown in Figures~\ref{figMock3D0}, and~\ref{figMock3D1}. 

\subsection{Contributions}

In this paper, our main contribution is the introduction of a 2.5D pipeline that can guarantee the non-realism needed for the creation of dynamic paintings and illustrations. We can list our contributions as elements of a basic 2.5D pipeline to create simple dynamic illustrations. 
\begin{enumerate}
\item We have developed a method for representing nonrealistic shapes as non-conservative vector fields. 
\item We demonstrated that non-conservative vector fields could be implemented as images that are viewed as normal maps. 
\item We have integrated non-realistic shadow and subsurface scattering computation into diffuse illumination to provide artistic (or in other words non-realist) versions of diffuse illumination. 
\item We have developed non-realistic versions of reflection, refraction, and Fresnel functions that can support artistic versions of transparency.
\item We demonstrated that all non-realistic shapes and effects can be  represented as images. Using images significantly simplifies the scene descriptions for dynamic illustrations and paintings.
\item We further demonstrated that non-realist rendering results can be obtained with Barycentric operations on intermediate images which are created by warping or processing input images.  
\end{enumerate}

We have also implemented a basic 2.5D pipeline to create simple dynamic illustrations. 
\begin{enumerate}
\item We have implemented a simple web-based version using Javascript and WebGL.  
\item We have also developed another standalone version using C++ and webGL. This version also provides shadows with nonconservative vector fields. An extended version also provides integrated modeling and rendering with layered images. 
\item These systems provide proof-of-concept for nonrealist modeling and rendering. Using these systems, a wide variety of artists could turn artworks in a large variety of non-realistic artistic styles (some of them their own) into dynamic documents. 
\end{enumerate}

\section{The Basic 2.5D Modeling and Rendering Pipeline} 
\label{section1}

\begin{figure}
\centering \includegraphics[width=0.7\linewidth]{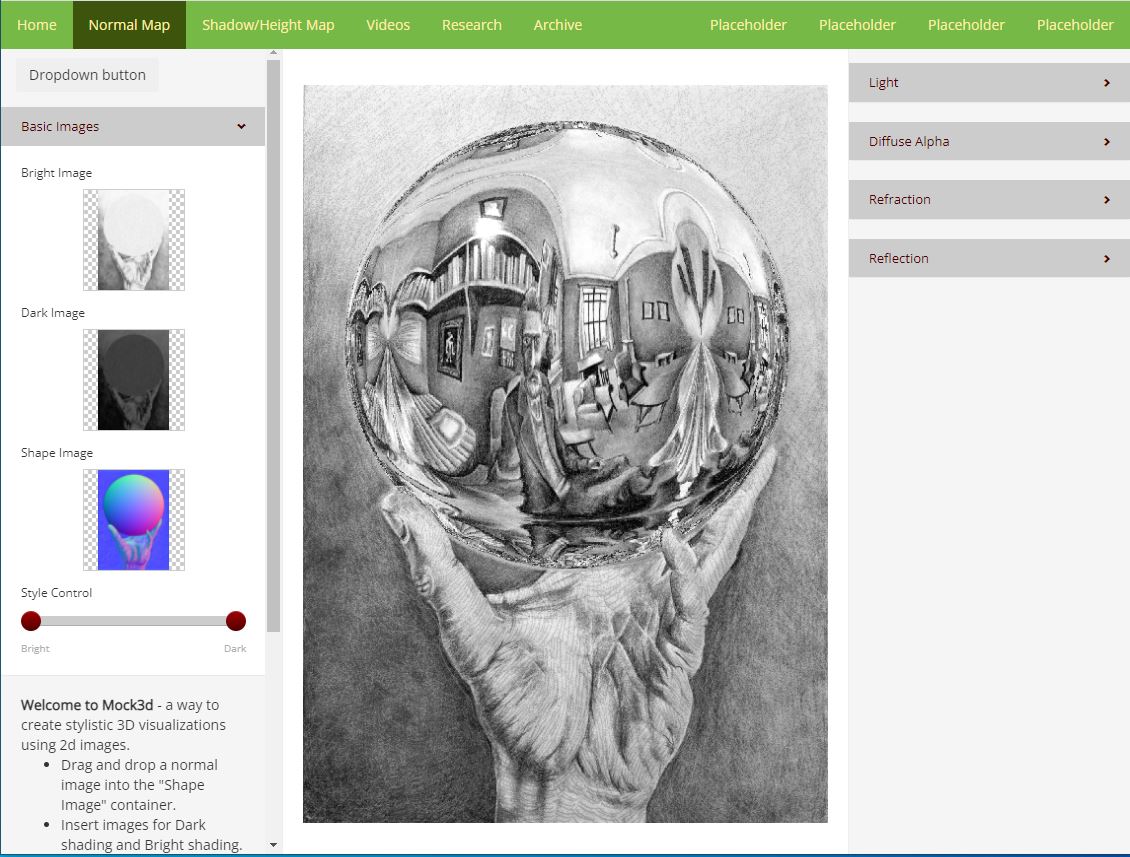}
\caption{The interface of the web-based system for normal maps. See mock3D.tamu.edu/\-normalmap/index.html.}
\label{figMock3D0} 
\vspace{0.1in}
\centering \includegraphics[width=0.7\linewidth]{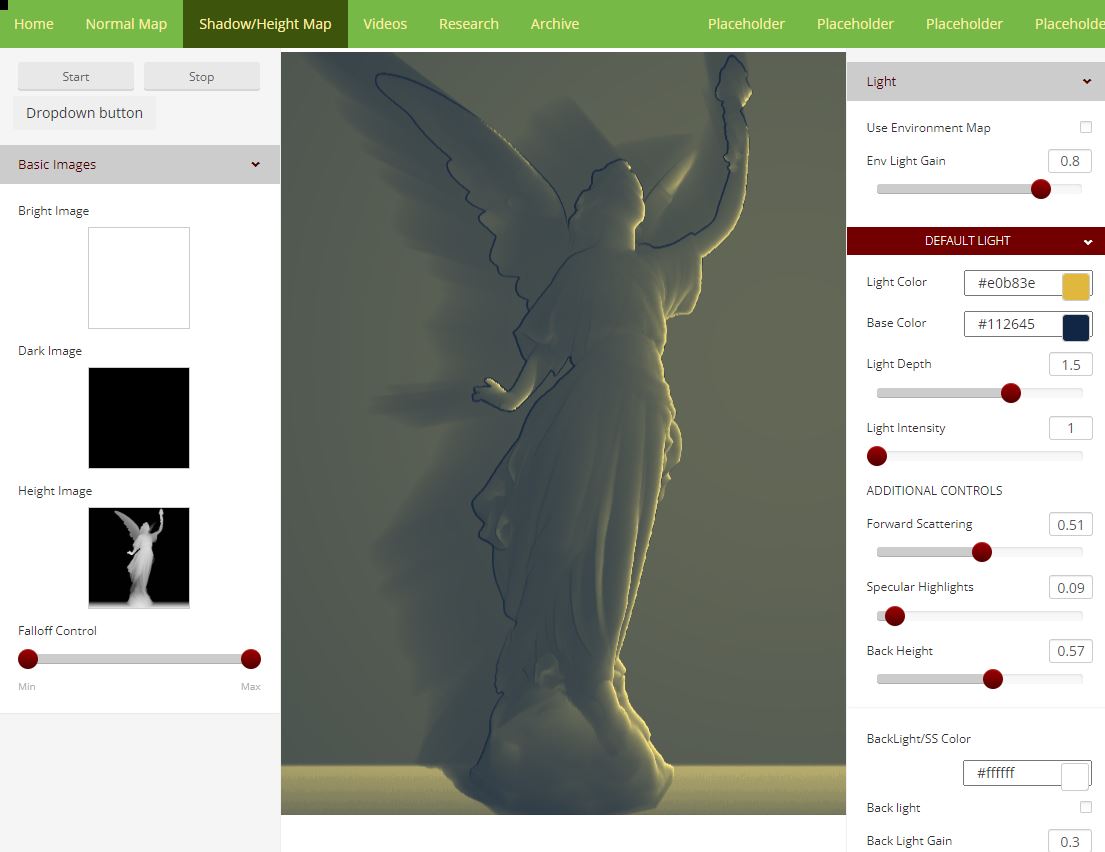}
\caption{The interface of the web-based system for height maps. See mock3D.tamu.edu/\-shadow/index.html }
\label{figMock3D1}
\end{figure}

In the paper, we present a detailed overview of the basic 2.5D pipeline. In this pipeline, all information from shapes to materials is provided by images. Shapes are defined by normal maps or depth maps. The scaling parameters are also provided by a set of control images. We can obtain physically plausible local and global illumination with complete style control. There are already solutions to obtain diffuse reflection, shadows, reflection, and refraction with a set of images \cite{wang2014,wang2014global,akleman2016,akleman2017}. An example showing how diffuse rendering is computed is shown in Figure~\ref{figtable}.  The interface of a system based on the 2.5S pipeline is shown in Figures~\ref{figMock3D0}, and~\ref{figMock3D1}.

\subsection{2.5D Proxy Objects: Mock3D Shapes with Associated Materials}
\label{section1.1}

\begin{figure}[htbp!]
  \centering
  \begin{subfigure}[t]{0.25\textwidth}
  \includegraphics[width=1.0\textwidth]{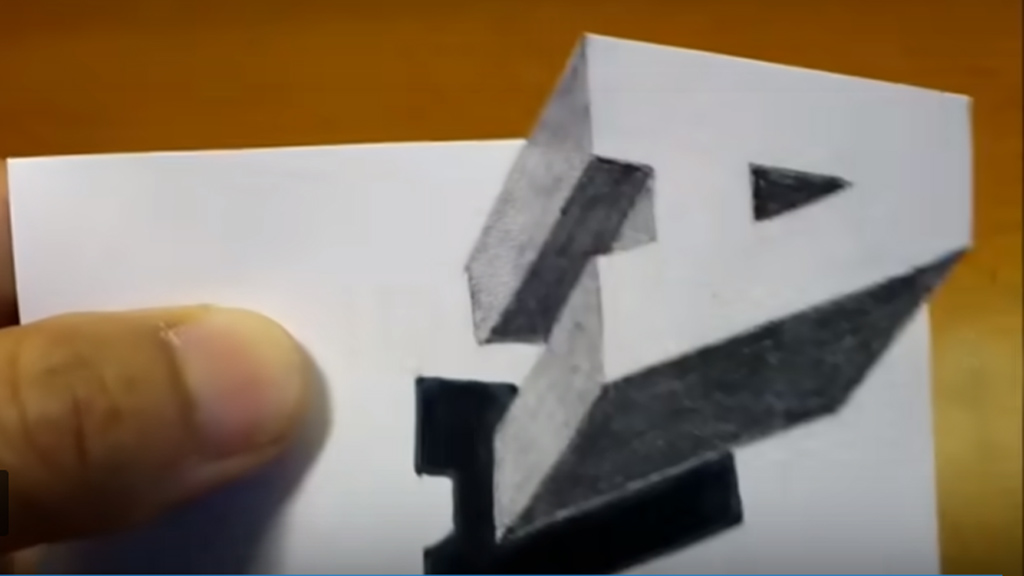}
  \caption{ An anamorphic illustration with embedded perspective, shadow, and shading.}
  \label{figbasreliefs0}
  \end{subfigure}
  \hfill
  \begin{subfigure}[t]{0.25\textwidth}
  \includegraphics[width=1.0\textwidth]{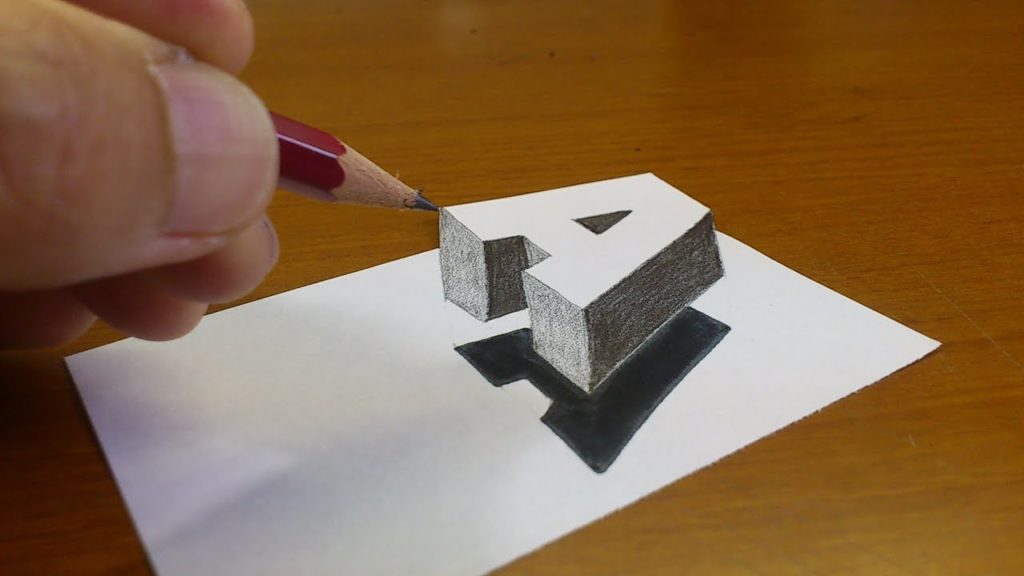}
  \caption{ 3D appearance of the anamorphic illustration from a particular viewpoint.}
  \label{figbasreliefs1}
  \end{subfigure}
  \hfill
  \begin{subfigure}[t]{0.25\textwidth}
  \includegraphics[width=1.0\textwidth]{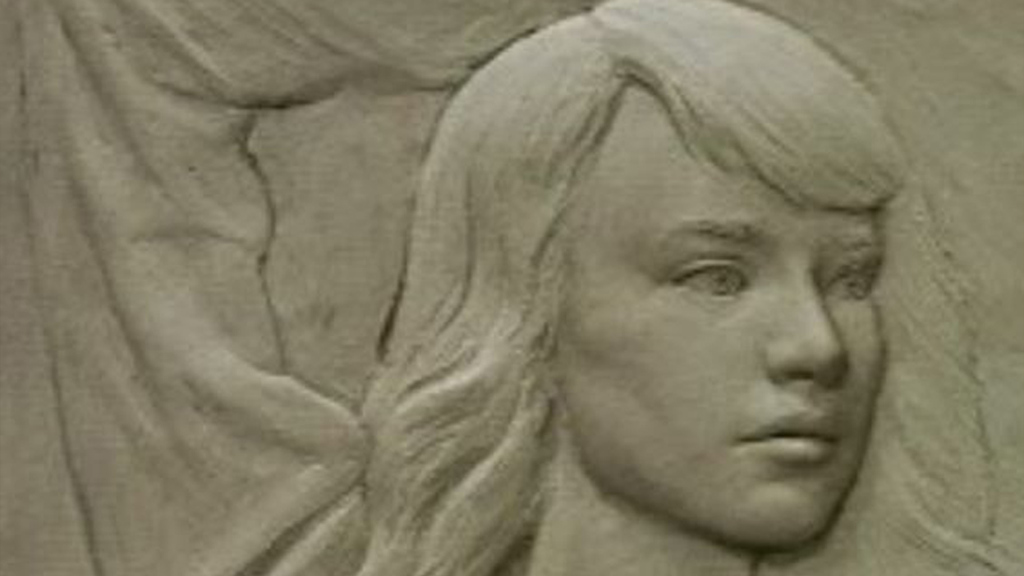}
  \caption{ A bas-relief with shadow \& shading consistent with environment.}
  \label{figbasreliefs2}
  \end{subfigure}
  \hfill
  \begin{subfigure}[t]{0.22\textwidth}
  \centering
  \includegraphics[width=0.99\textwidth]{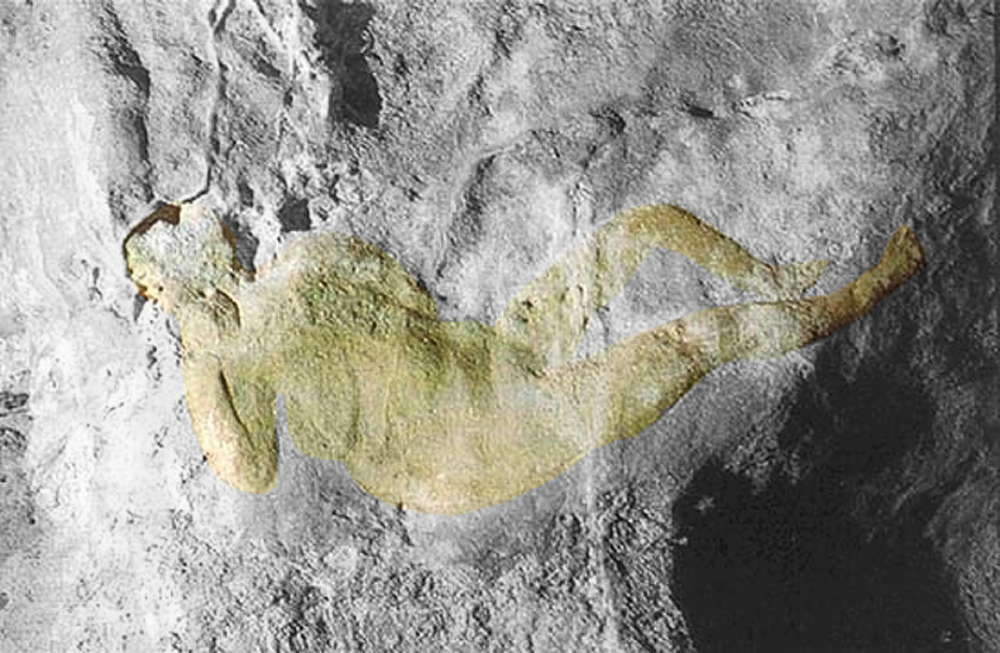}
  \caption{ An anamorphic bas-relief Madeleine cave from Paleolithic age.}
  \label{figmadeleine}
  \end{subfigure}
  \hfill
  \caption{ Examples that demonstrate Anamorphic illustrations provide perspective but do not provide consistent \& dynamic shadow and shading. Only bas-reliefs cannot provide perspective. Anamorphic bas-reliefs provide embedded perspectives with dynamic shading and shadows that are consistent with the environment. Note that the bas-relief from Madeleine Cave already has an embedded perspective. It is probably the earliest example of an anamorphic bas-relief.}
  \label{figbasreliefs}
  \end{figure}

We represent 2.5D Proxy Objects as anamorphic bas-reliefs as shown in Figure~\ref{figbasreliefs}. These are mock3D shapes with associated materials. Such objects can essentially be formulated as a layer that is represented as a parametric surface that is given as $\mathbf{P}(u,v)$ where $(u,v)$ are parametric coordinates in $[0,1]^2$. The simplest of such layers is a planar rectangle, and in most cases, it is sufficient to use a planar rectangle. We can also use tensor product B-spline surfaces or simply a wrinkled polygonal mesh. Since these are always functions from a square domain $[0,1]^2$ to the 3D space $\mathbf{P}=(x,y,z)$, it is easy to texture map them. We can simply use the parametric coordinates $(u,v)$ as texture coordinates. Without loss of generality in the rest of the paper, we assume that it is a unit square in $z=0$ as $\mathbf{P}(u,v)=(u,v,c)$ where $(u,v) \in [0,1]^2$ and $c$ is a constant, which is usually zero. 

\begin{figure}[ht]
\centering  
  \begin{subfigure}[t]{0.24\textwidth}
  \includegraphics[width=1.0\textwidth]{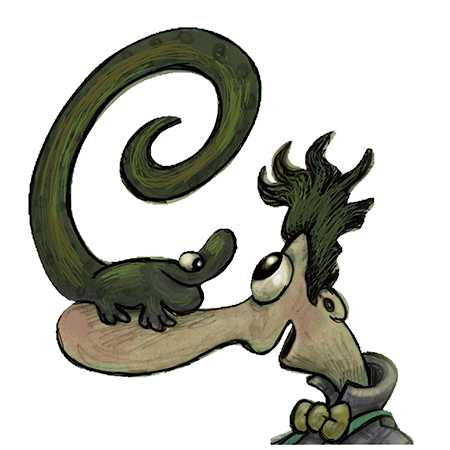}
  \caption{ Original Image.}
  \label{figBukalemunoriginal}
  \end{subfigure}
  \hfill
  \begin{subfigure}[t]{0.24\textwidth}
  \includegraphics[width=1.0\textwidth]{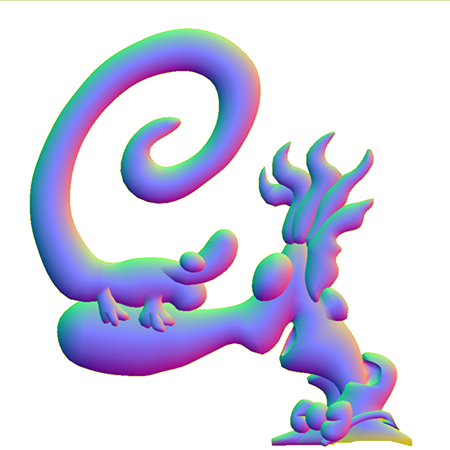}
  \caption{Mock3D shape: Normal Map.}
  \label{figBukalemunSM}
  \end{subfigure}
  \hfill
  \begin{subfigure}[t]{0.24\textwidth}
  \includegraphics[width=1.0\textwidth]{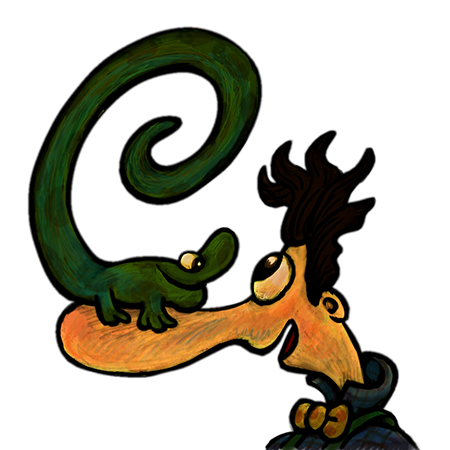}
  \caption{ Dark $I_0$ image.}
  \label{figBukalemunDI0}
  \end{subfigure}
  \hfill
  \begin{subfigure}[t]{0.24\textwidth}
  \includegraphics[width=1.0\textwidth]{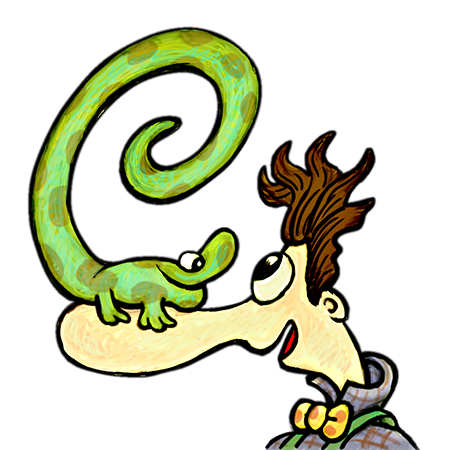}
  \caption{ Bright $I_1$ image.}
  \label{figBukalemunDI1}
  \end{subfigure}
  \hfill
\caption{An example of components of 2.5D Proxy Object that is used to make rendering shown in Figure~\ref{figtable}. (a) Original illustration of the artist of the ``object''; (b) normal map created from the original drawing \cite{Johnston2002, Shao2012}; (c) and (d) Two control images provided by the artist}
\label{figBukalemunChannels}
\end{figure}

Each of these rectangular layers consists of several components, each of which is a texture projected onto the rectangular surface. We call these components ``channels'' consistent, with the standard terminology used in image manipulation. On the other hand, the term channel in this case will refer to entities that are more general than simple color channels. One channel provides proxy shape information, which we call Mock3D shapes, that are either height fields or vector fields. The height fields are represented by images called depth maps, and the vector fields are represented by images called normal maps \cite{Akleman2014ir1}. The additional channels are control images that provide material properties of the objects. Two examples of channels are shown in Figures~\ref{figBukalemunChannels} and ~\ref{figEscherChannels}. 

\subsection{Depth Maps as Mock3D Shapes}
\label{section1.2}

Depth maps are given as mono-color images in the form $C(u,v)=(R, G, B)$ where $(u,v)$ are texture coordinates and $c$ is a single normalized color channel that corresponds to depth information, that is, $R=G=B$, which usually corresponds to the distance from the eye.  $C(u,v)$ can be used directly to represent shapes as bas-relief-type shape information. In computer graphics, depth maps were originally used to represent wrinkled surfaces and are generally known as bump maps \cite{blinn1978simulation}. 
The gradient of the depth map is used to obtain 3D vectors as follows:
\begin{equation} 
\vec{N}(u,v)=\nabla C(u,v). \label{eqgradient}
\end{equation}
These vector fields are called gradient fields, and they are conservative in the sense that any close path adds to zero. 
These gradient fields obtained by depth maps are used to manipulate the normal surface of the original to obtain a wrinkled look, which is the reason why they are commonly called bump maps in computer graphics practice \cite{blinn1978simulation}. In our case, the rectangle is already in the form of $\mathbf{P}(u,v)=(u,v,c)$. If we add the height field, we obtain $\mathbf{P'}(u,v)= (u,v,c+C(u,v))$. This is a function in the form of $z=H(x,y)$ If we replace $H(x,y) \rightarrow C(u,v)$, the gradient of this function gives us a normal vector as follows: 
\begin{equation} 
\vec{N}(u,v)= \left(\frac{\partial C(u,v)}{\partial u},\frac{\partial C(u,v)}{\partial v}, 1 \right). \label{eqgradient2}
\end{equation}
For unit normals, we need to normalize it, so:
\begin{equation}
\vec{N}(u,v) \longleftarrow \frac{\vec{N}(u,v)}{|\vec{N}(u,v)|}. \label{eqnormalize}
\end{equation}

\subsection{Normal Maps as Mock3D Shapes}
\label{section1.3}

Normal maps became popular as an alternative to bump maps \cite{Cohen1998}. The advantage of normal maps is that the computation of normal does not require any gradient computation. Normal maps can be any three-color image on a rectangle denoted by $C(u,v) = (r(u,v),g(u,v),b(u,v))$ where $(u,v)$ are texture coordinates. The unit normal for any given position $(u,v)$, $\vec{N}(u,v) = (x(u,v),y(u,v),z(u,v))$, is calculated from $(r(u,v),g(u,v),b(u,v))$ as follows:  $$\vec{N}(u,v)=(x=2r(u,v)-1,y=2g(u,v)-1,z=2b(u,v)-1).$$
Here, it is expected that $x^2(u,v)+y^2(u,v)+z^2(u,v)=1$ since $\vec{N}(u,v)$ is a unit normal. If normal maps are produced in real 3D shapes, this requirement is always satisfied. Even if $|\vec{N}(u,v)| \neq 1$, it can be easily corrected by normalizing using Equation~\ref{eqnormalize}. Normal maps were originally used to include details such as wrinkles, like bump maps. On the other hand, both can be used directly to define shapes. 

\section{Designing of Mock3D Shapes}
\label{section1.4}

The straightforward method of creating normal maps to represent shapes is to convert virtual 3D shapes into normal maps. The procedure for obtaining a normal map is a straightforward shading process that does not involve any illumination. The components $x$, $y$, and $z$ of the 3D normal vector of the visible point $(u,v)$ are simply converted to red, green, and blue colors of the image as follows: 
$$C(u,v)=0.5\vec{N}(u,v)+0.5=(R=0.5x+0.5,G=0.5y+0.5,B=0.5y+0.5).$$ 
Johnston's sketch-based system, Lumo, models normal maps by diffusing 2D normals from a line drawing \cite{Johnston2002}. Since then, only a few groups have investigated the potential use of normal maps as a shape representation, such as \cite{Okabe2006, Bezerra2005, Winnemoeller2009, Shao2012}. Sun et al. \cite{Sun07} introduced a gradient mesh to semi-automatically and quickly interpolate normals from edges, and Orzan et al. \cite{Orzan09} calculated a diffusion from edges by solving the Poisson equation. Sykora et al. \cite{Daniel09}, proposed Lazy-Brush, which can propagate scribbles to accelerate the definition of regions of constant color. Finch et al. \cite{Finch11} build thin-plate splines that provide smoothness everywhere except in user-specified tears and creases. The underlying splines are used to interpolate the normals. 

Wu et al. \cite{Wu07} proposed a shape palette, where the user can draw a simple 2D primitive in the 2D view and then specify its 3D orientation by drawing a corresponding primitive. This method also performs diffusion using a thin-plate spline. Recently, Shao et al. \cite{Shao2012} developed CrossShade, another sketch-based interface for designing complicated shapes such as normal maps. They used an explicit mathematical formulation of the relationships between cross-sectional curves and geometry. The specified cross-section is used as an extra control point to control the normals. Vergne et al. \cite{vergne2012surface} introduce surface flow from smooth differential analysis, which can be used to measure smooth luminance variations. Therefore, it is also possible to obtain shadows and other shading effects. 
Sykora et al. \cite{sykora2014ink} developed a user-assisted method to convert normal maps into Bass-Reliefs that can provide correct shadows in a commercial renderer. 

\begin{figure}[ht]
\centering  
  \begin{subfigure}[t]{0.24\textwidth}
  \includegraphics[width=1.0\textwidth]{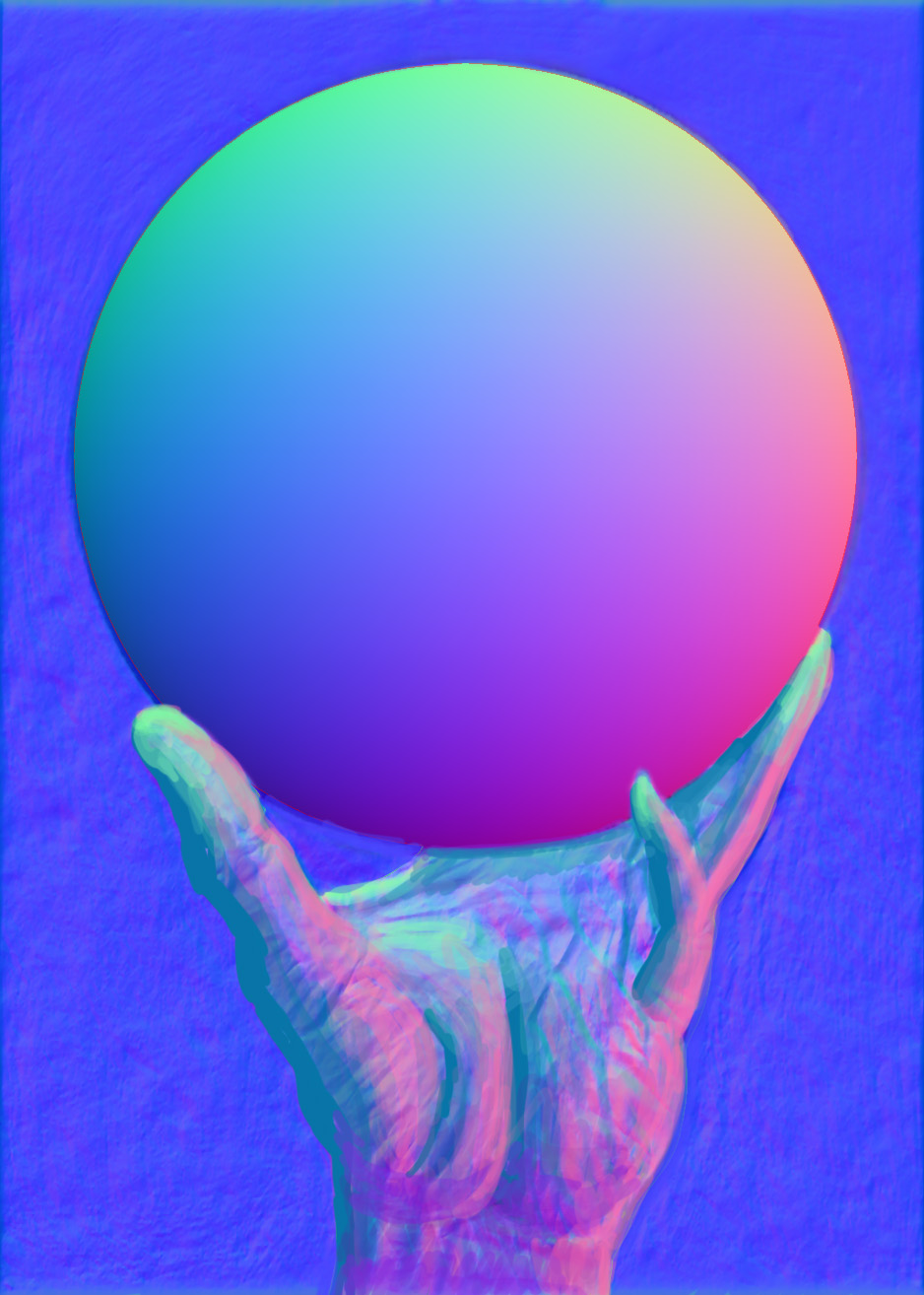}
  \caption{ Mock3D shape: Normal Map.}
  \label{figEscherSM}
  \end{subfigure}
  \hfill
  \begin{subfigure}[t]{0.24\textwidth}
  \includegraphics[width=1.0\textwidth]{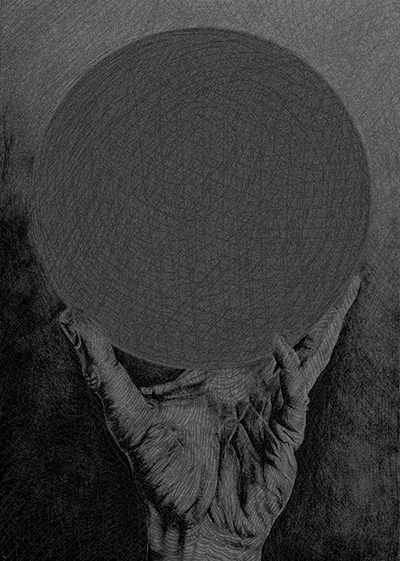}
  \caption{ Dark $I_0$ image.}
  \label{figEscherdark0}
  \end{subfigure}
  \hfill
  \begin{subfigure}[t]{0.24\textwidth}
  \includegraphics[width=1.0\textwidth]{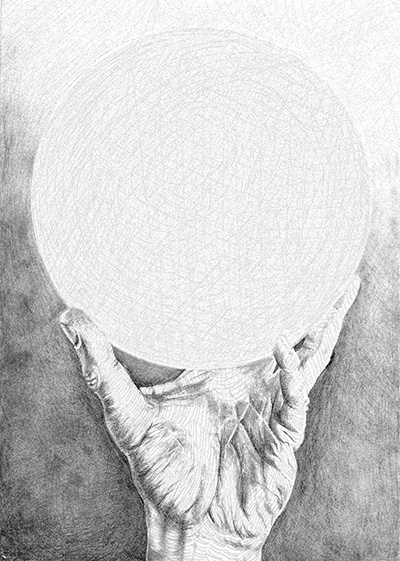}
  \caption{ Bright $I_1$ image.}
  \label{figEscherbright0}
  \end{subfigure}
  \hfill
  \begin{subfigure}[t]{0.24\textwidth}
  \includegraphics[width=1.0\textwidth]{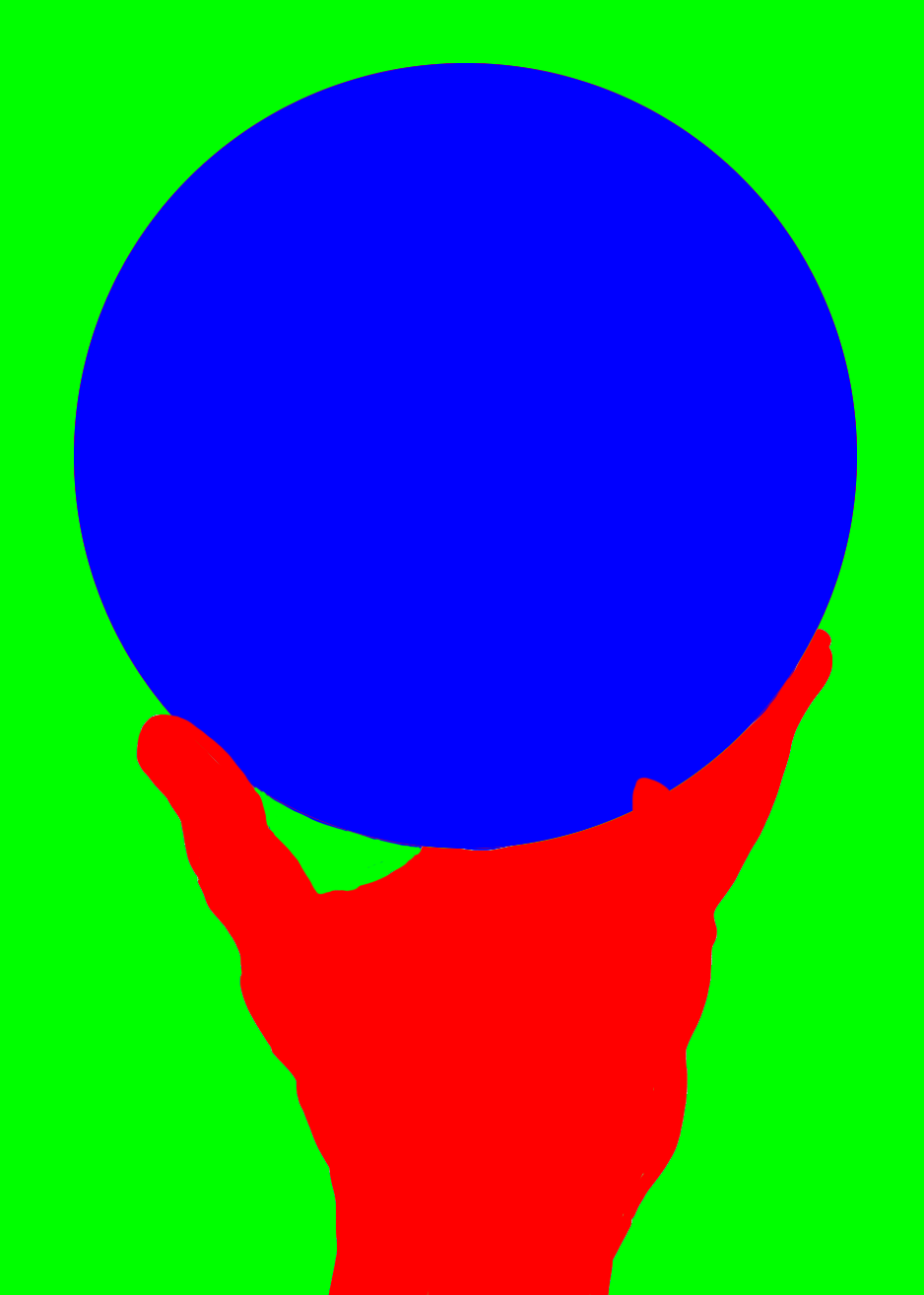}
  \caption{Reflection and Refraction Map.}
  \label{figEscheralpha0}
  \end{subfigure}
  \hfill
\caption{An example of components of 2.5D Proxy Object that represent M. C. Escher's lithograph, called ``Self-Portrait in Spherical Mirror'', in Figure~\ref{figEscher}. (a) a normal map created from the original drawing \cite{Johnston2002, Shao2012}; (b) and (c) Two control images provided by one artist. (d) Reflection and refraction map. The blue region is chosen to be reflective. }
\label{figEscherChannels}
\end{figure}

\begin{figure}[ht]
\centering  
  \begin{subfigure}[t]{0.24\textwidth}
  \includegraphics[width=1.0\textwidth]{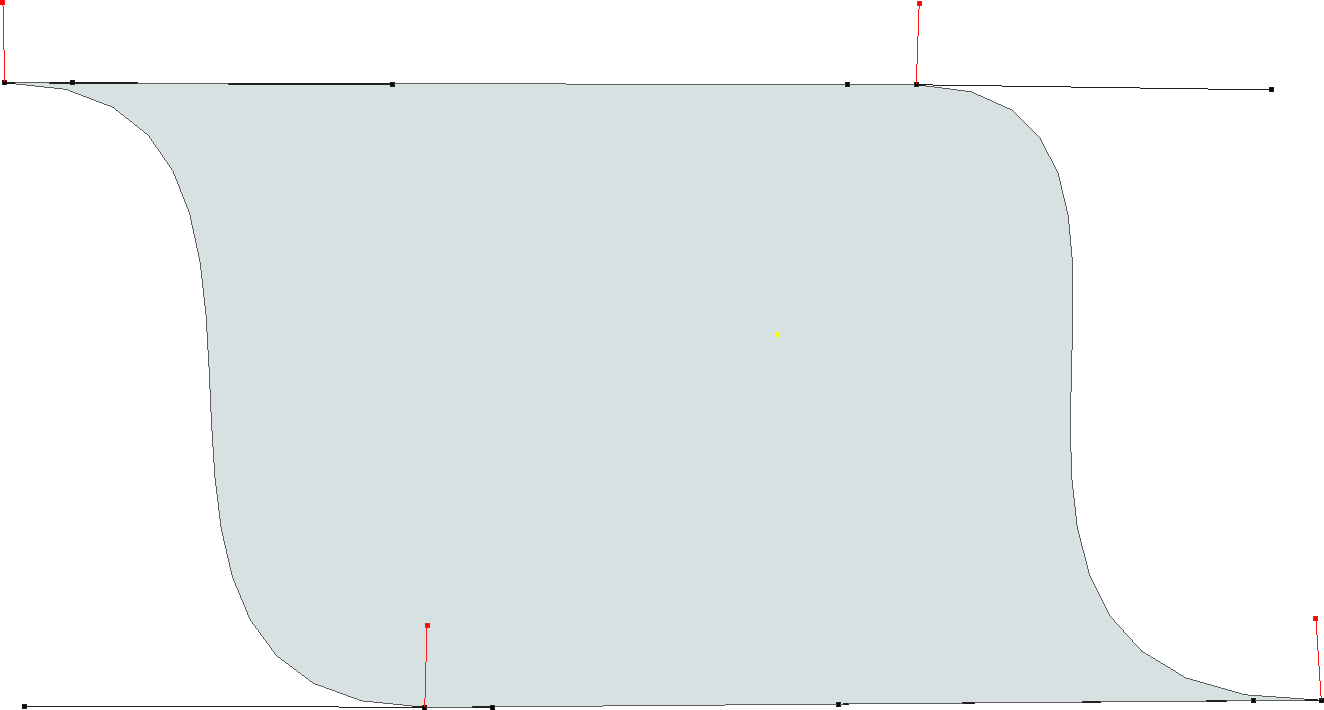}
  \caption{Control vectors.}
  \label{figvectorfield/1}
  \end{subfigure}
  \hfill
  \begin{subfigure}[t]{0.24\textwidth}
  \includegraphics[width=1.0\textwidth]{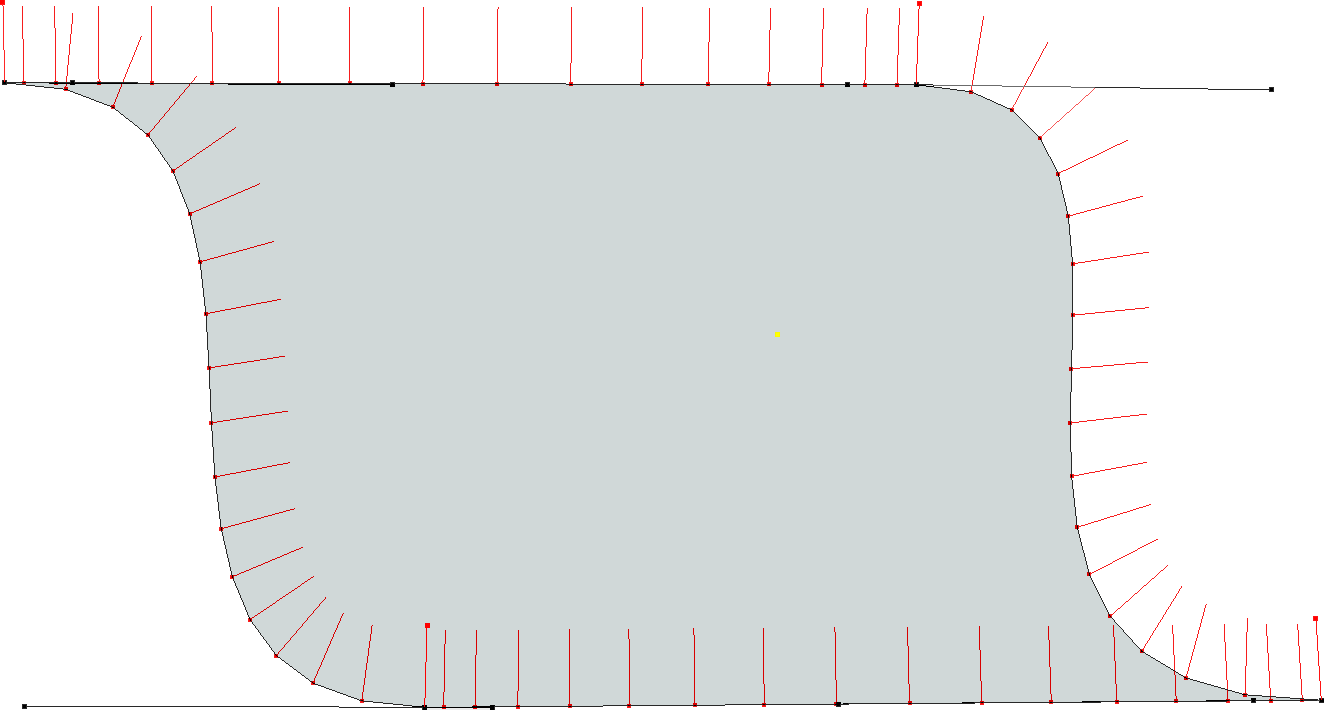}
  \caption{Boundary vectors.}
  \label{figvectorfield/2}
  \end{subfigure}
  \hfill  
  \begin{subfigure}[t]{0.24\textwidth}
  \includegraphics[width=1.0\textwidth]{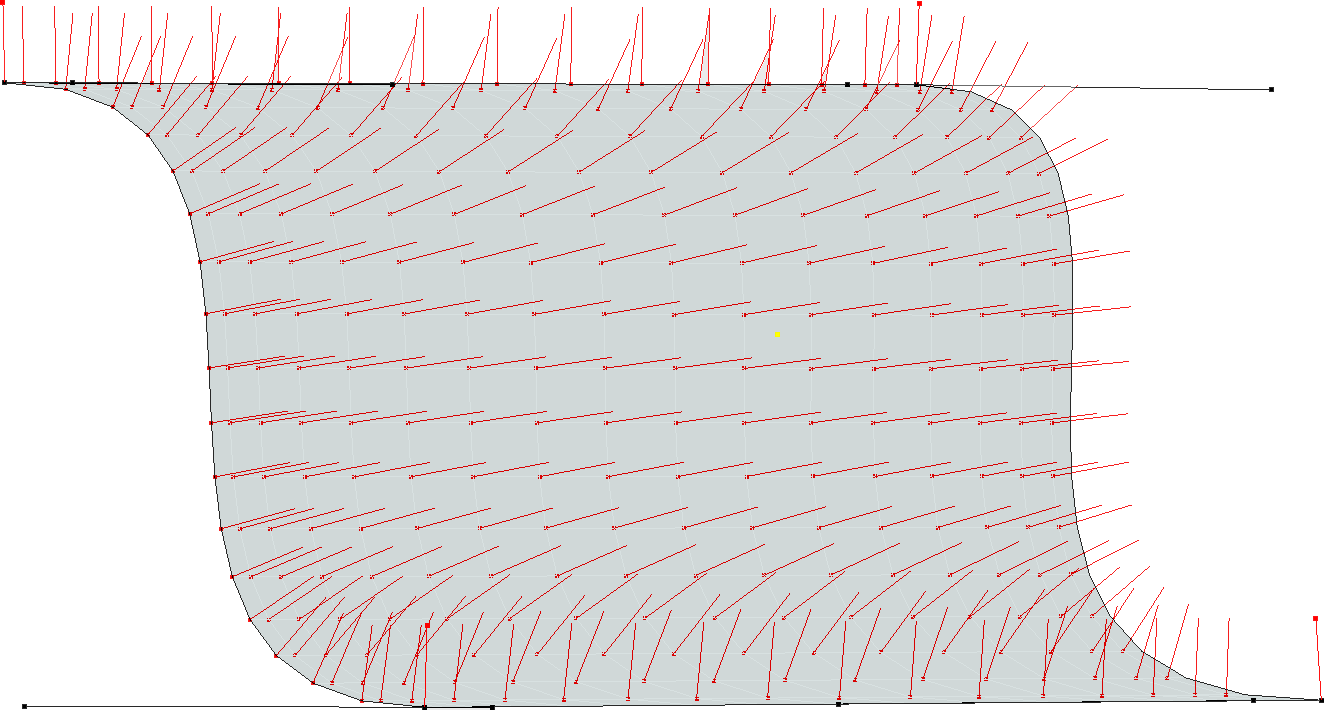}
  \caption{Coons Interpolation.}
  \label{figvectorfield/3}
  \end{subfigure}
  \hfill 
  \begin{subfigure}[t]{0.24\textwidth}
  \includegraphics[width=1.0\textwidth]{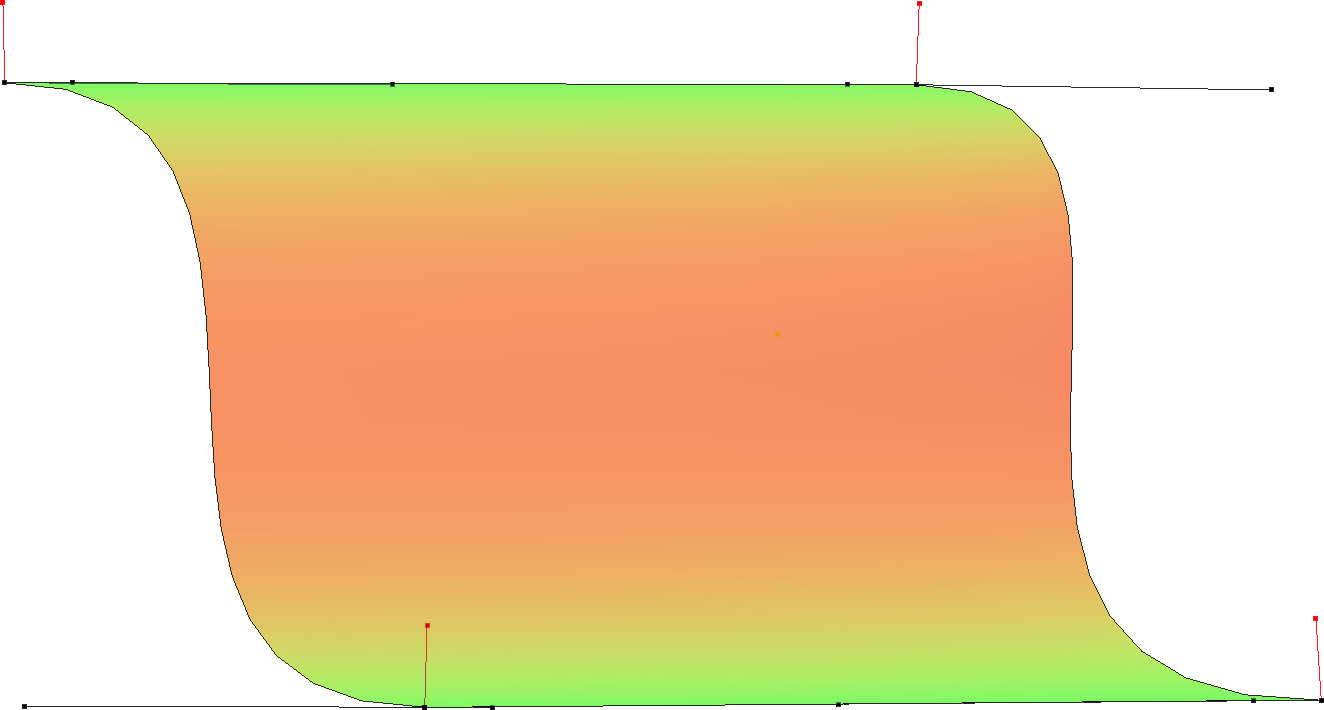}
  \caption{Normal Map.}
  \label{figvectorfield/4}
  \end{subfigure}
  \begin{subfigure}[t]{0.24\textwidth}
  \includegraphics[width=1.0\textwidth]{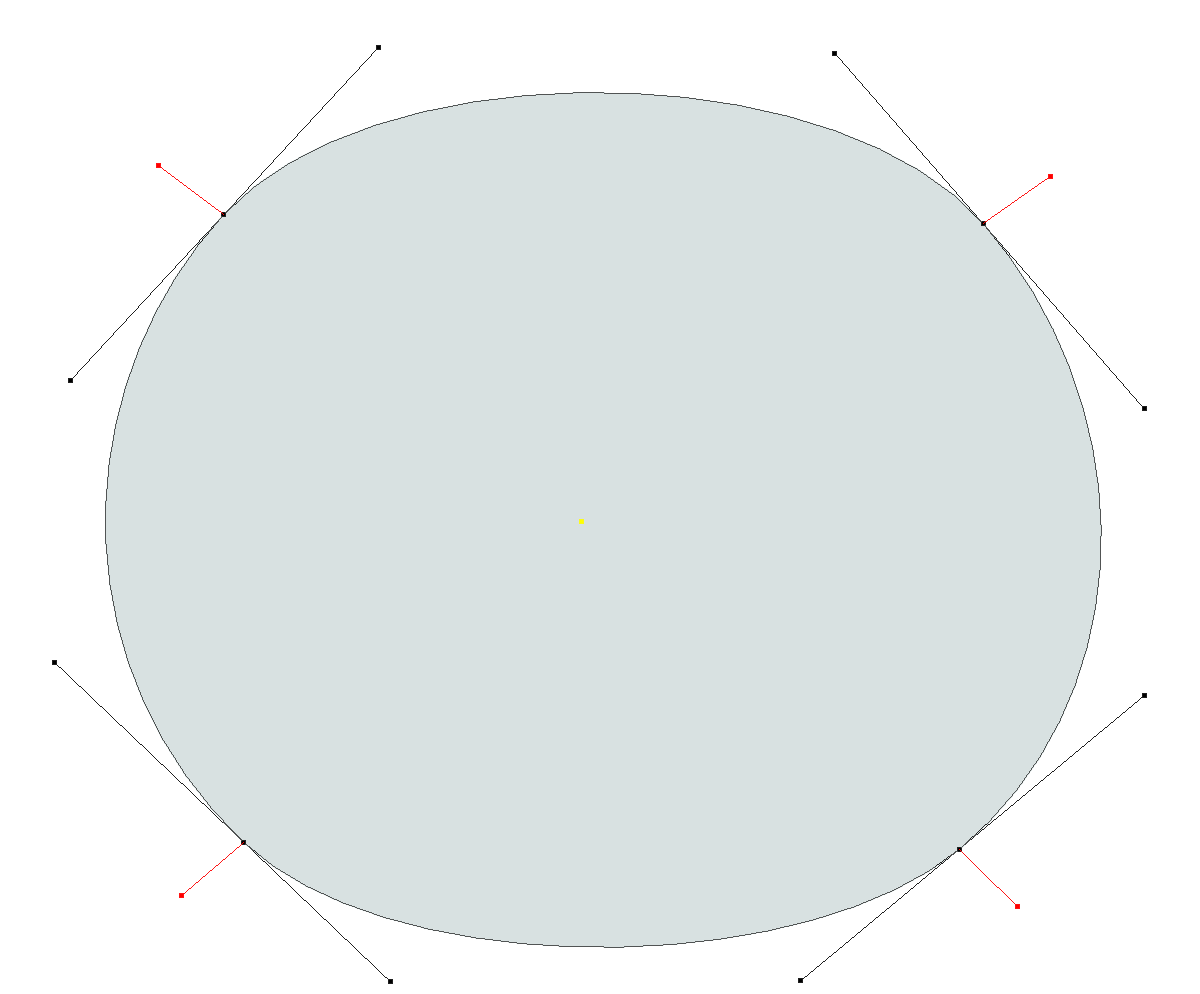}
  \caption{Control vectors.}
  \label{figvectorfield/a1}
  \end{subfigure}
  \hfill
  \begin{subfigure}[t]{0.24\textwidth}
  \includegraphics[width=1.0\textwidth]{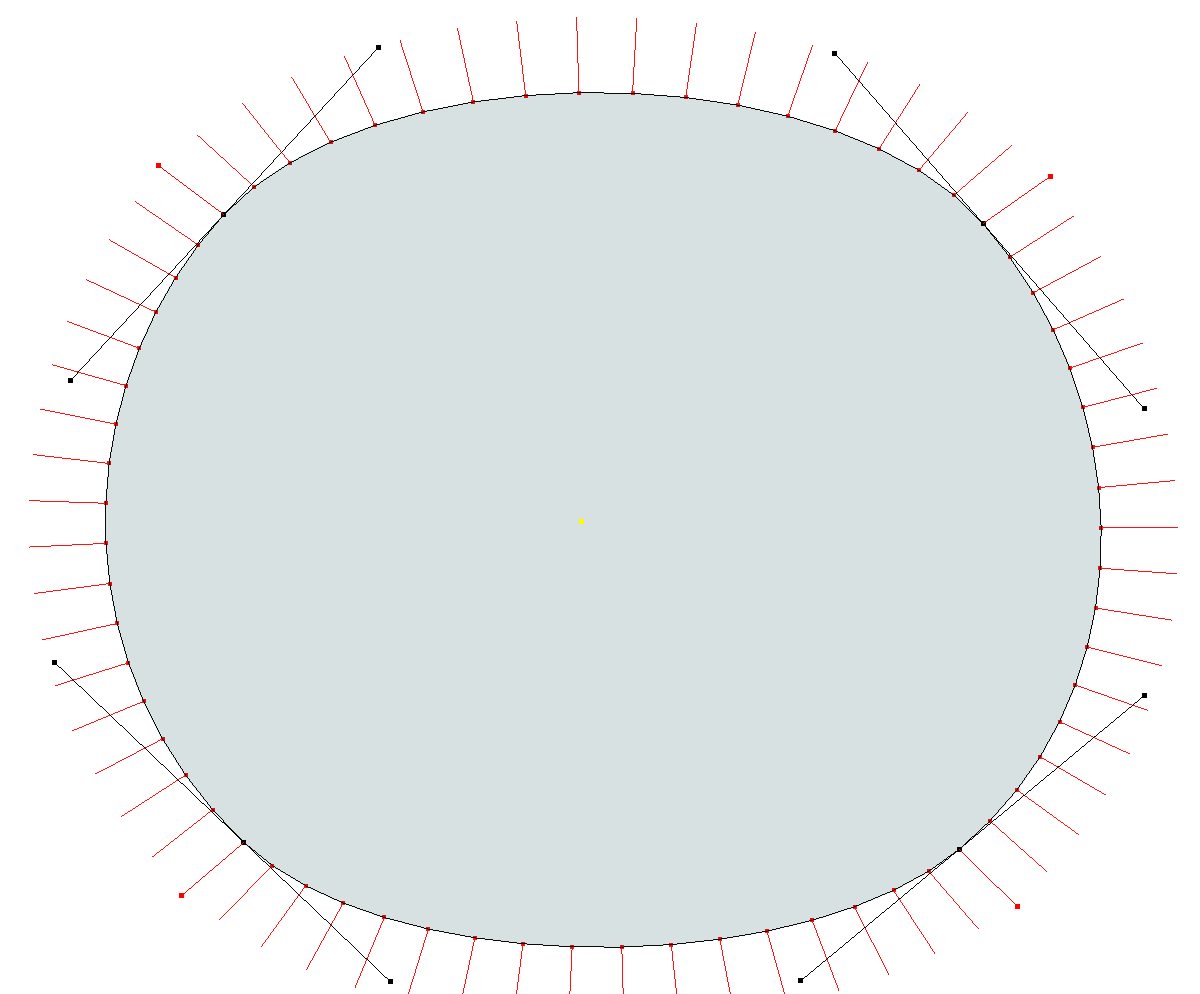}
  \caption{Boundary vectors.}
  \label{figvectorfield/a2}
  \end{subfigure}
  \hfill  
  \begin{subfigure}[t]{0.24\textwidth}
  \includegraphics[width=1.0\textwidth]{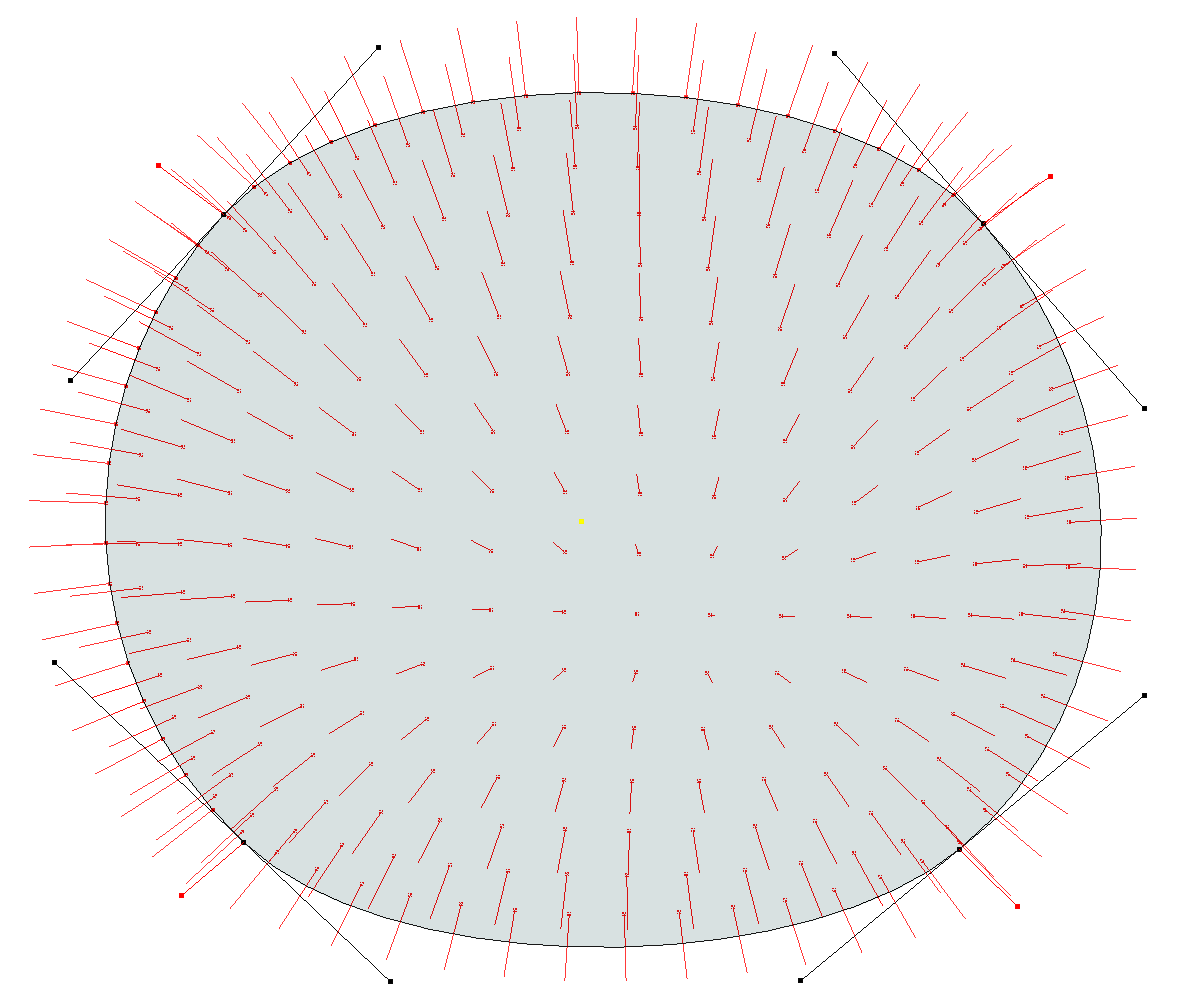}
  \caption{Coons Interpolation.}
  \label{figvectorfield/a3}
  \end{subfigure}
  \hfill 
  \begin{subfigure}[t]{0.24\textwidth}
  \includegraphics[width=1.0\textwidth]{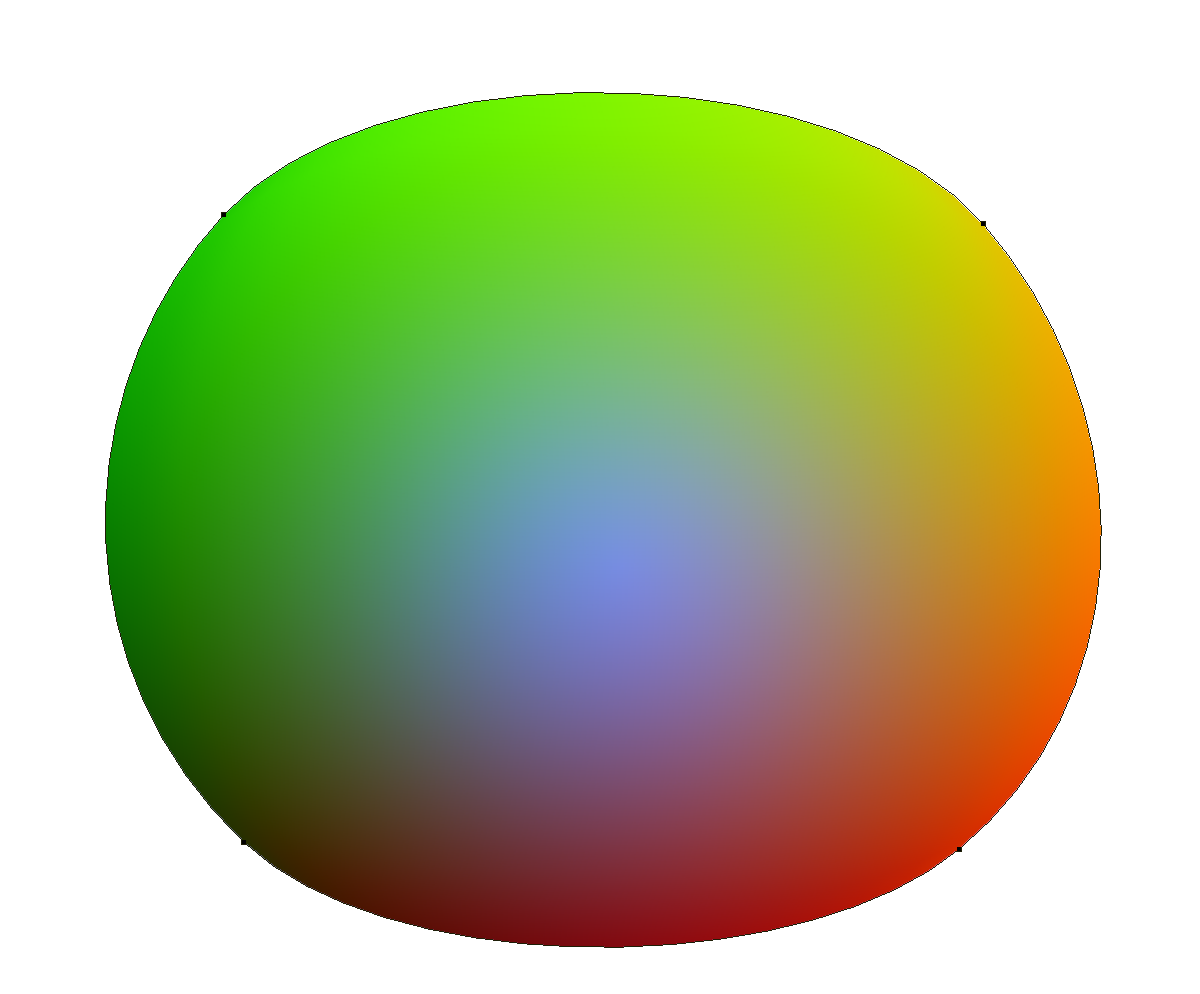}
  \caption{Normal Map.}
  \label{figvectorfield/a4}
  \end{subfigure}
   \begin{subfigure}[t]{0.24\textwidth}
  \includegraphics[width=1.0\textwidth]{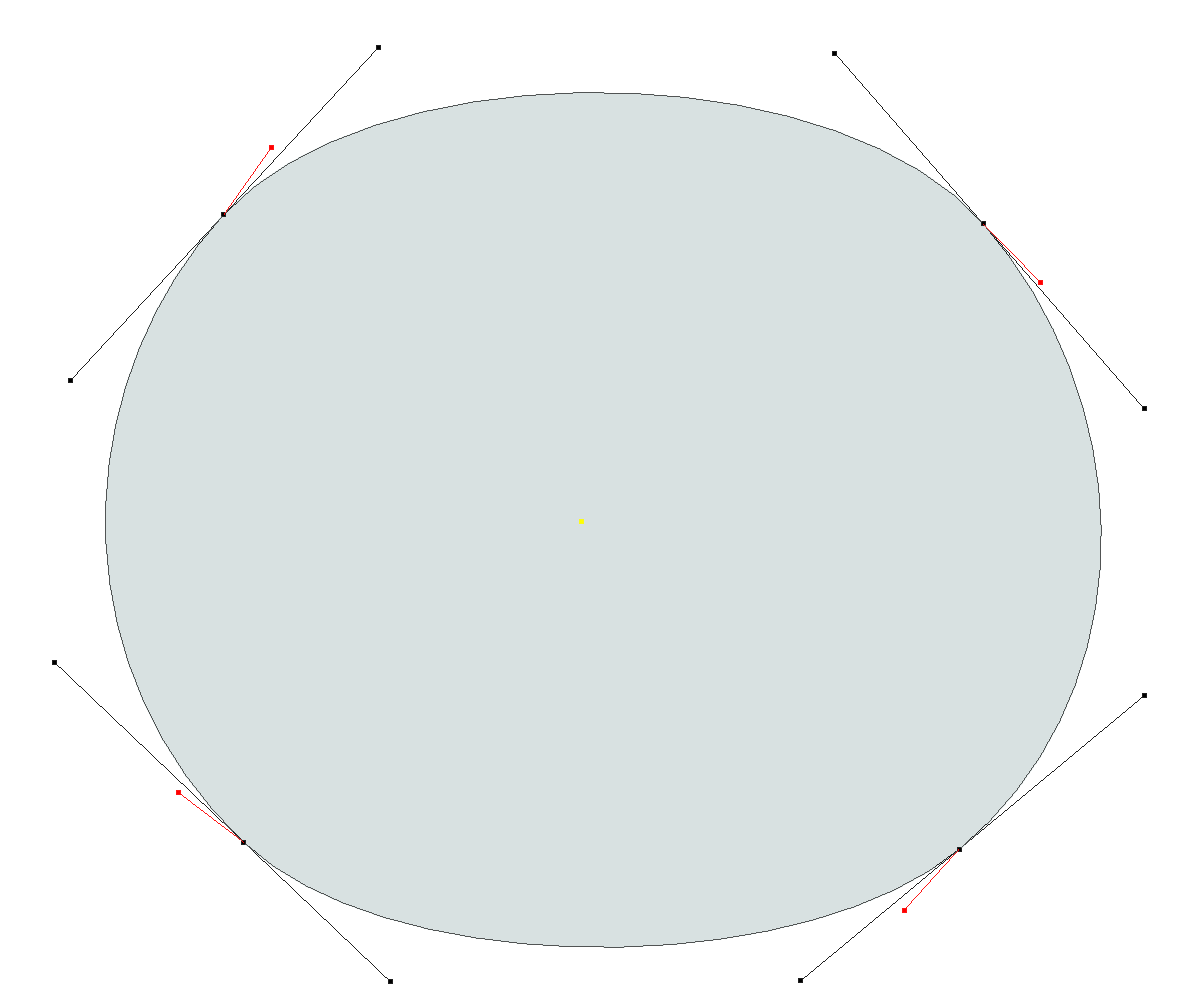}
  \caption{Control vectors.}
  \label{figvectorfield/c1}
  \end{subfigure}
  \hfill
  \begin{subfigure}[t]{0.24\textwidth}
  \includegraphics[width=1.0\textwidth]{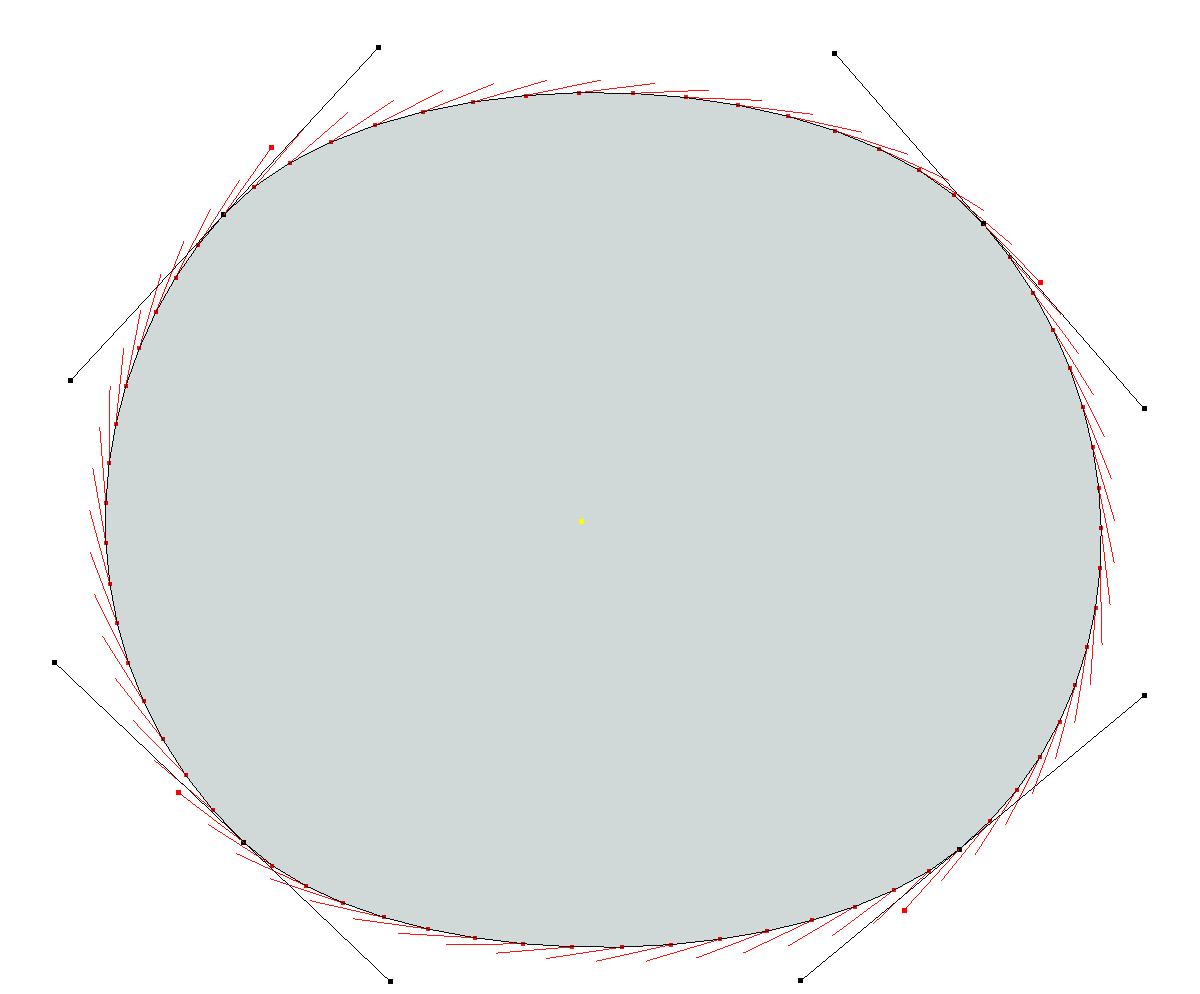}
  \caption{Boundary vectors.}
  \label{figvectorfield/c2}
  \end{subfigure}
  \hfill  
  \begin{subfigure}[t]{0.24\textwidth}
  \includegraphics[width=1.0\textwidth]{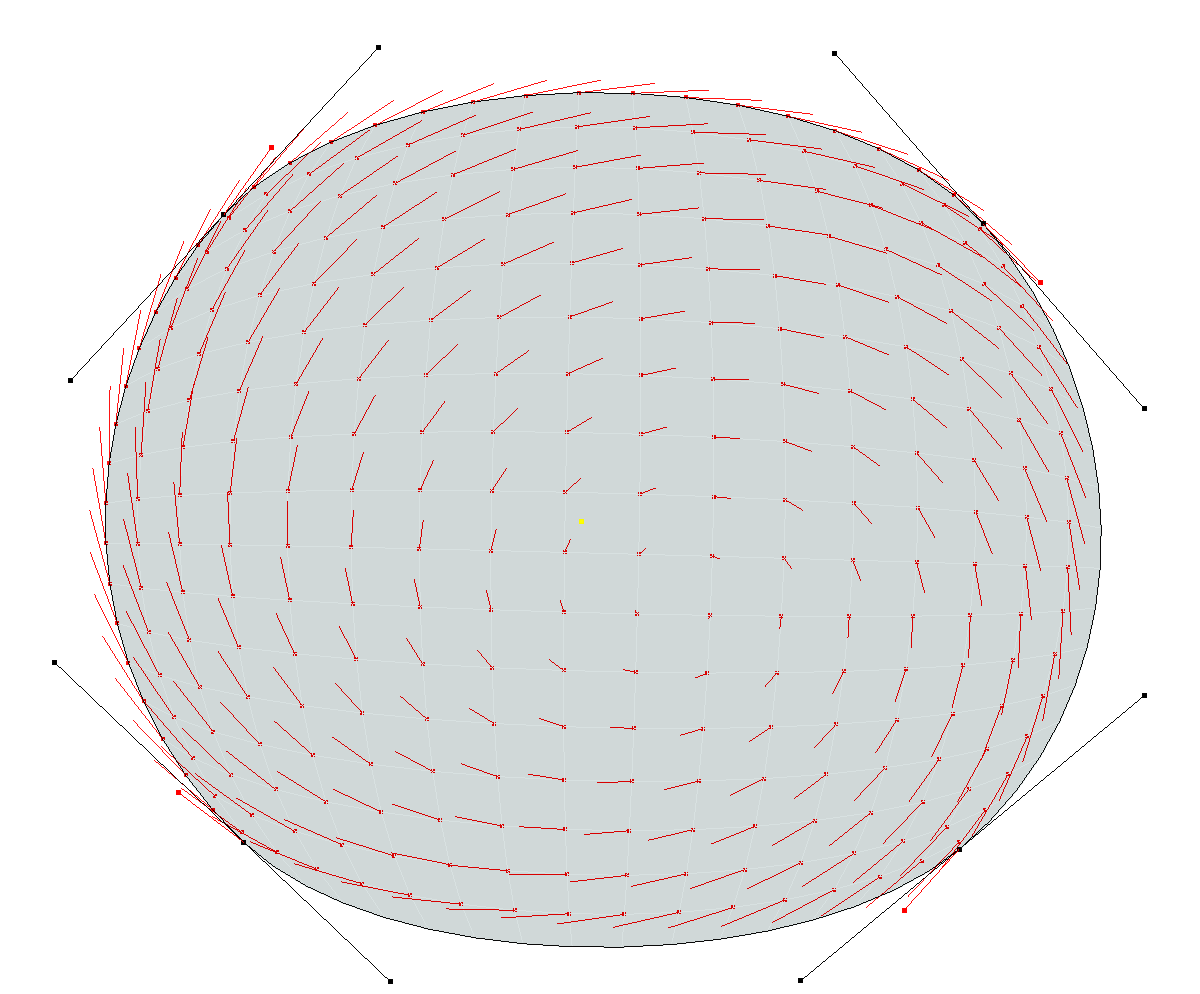}
  \caption{Coons Interpolation.}
  \label{figvectorfield/c3}
  \end{subfigure}
  \hfill 
  \begin{subfigure}[t]{0.24\textwidth}
  \includegraphics[width=1.0\textwidth]{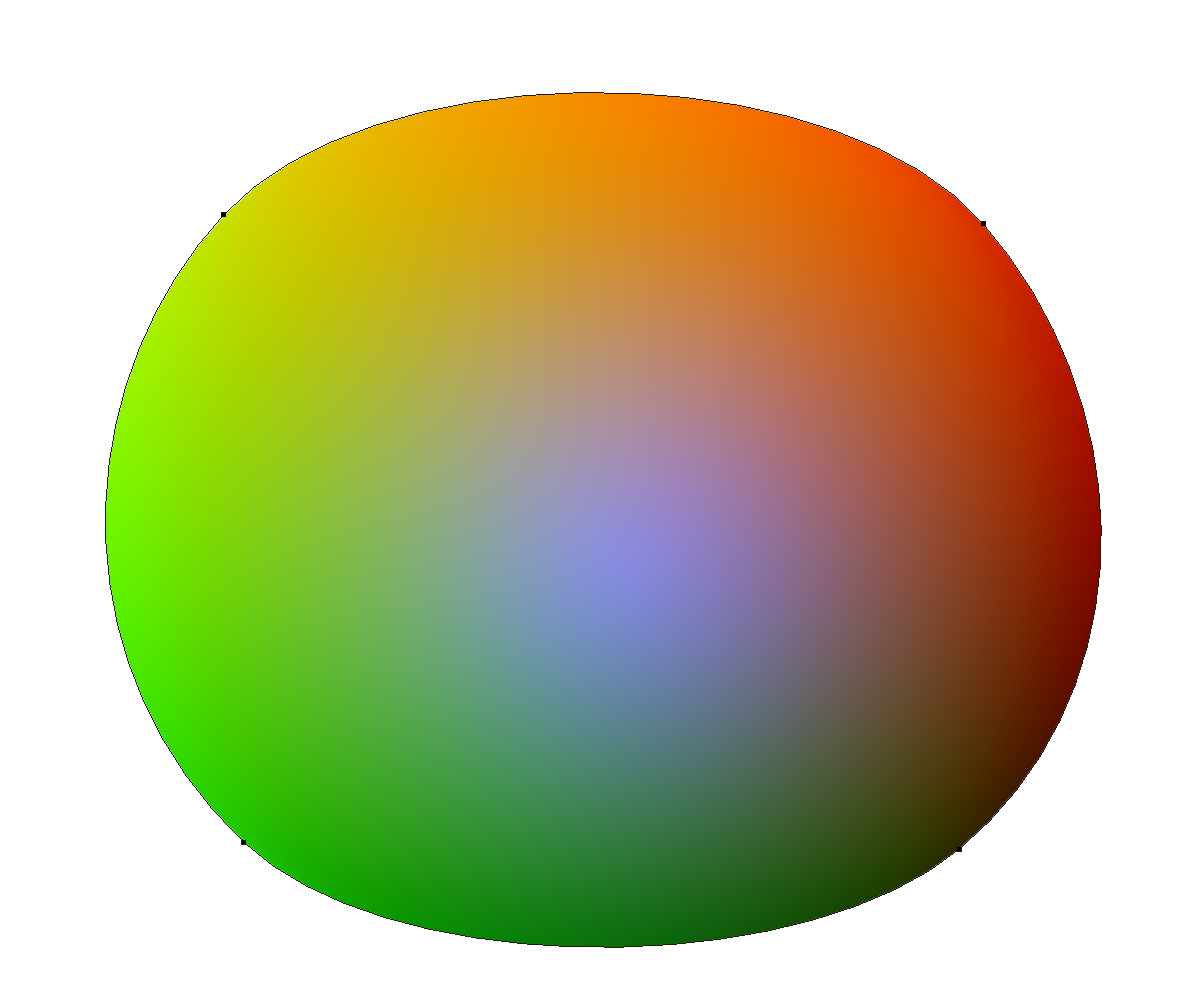}
  \caption{Impossible Shape.}
  \label{figvectorfield/c4}
  \end{subfigure}
\caption{
Examples of normal map creation on a B\'{e}zier patch by interpolating control vectors first along the edges, then inside of the patch \cite{gonen2016,wang2014global}.}
\label{figvectorfield}
\end{figure}

A big advantage of normal maps is that they can represent impossible shapes by constructing non-conservative vector fields \cite{wang2014,wang2014global}. Gonen et al. developed a sketch-based integrated mock-3D scene modeling system that can allow users to obtain any vector field by interpolating user-defined normals in quad dominant domains \cite{gonen2016,wang2014global}. The system looks and feels exactly like a 2D vector graphics system, in which users can only change the 2D positions of control points and provide normals in vertices. Figure~\ref{figvectorfield} three types of examples. Figure~\ref{figvectorfield/4} represents a normal map of a thin sheet or carpet, Figure~\ref{figvectorfield/a4} represents a realistic shape, and Figure~\ref{figvectorfield/c4} represents an impossible shape, which is a non-conservative vector field, which does not correspond to any height field. 

Normal maps can also be created using painting software such as Photoshop \cite{Knoll1988Photoshop}, GIMP \cite{Kimball998GIMP} or Procreate \cite{Pedchenko2021Procreate}. To paint normal maps, users can imagine an object that is illuminated with a directional red light from the left side and directional green light from the top. By ignoring shadows, they paint an image based on how much red and green light they want to see in every pixel. For example, a pixel color red=0.95 and green=0.75 means that the artist wants 95\% of the light from the left and 75\% of the light from the top to illuminate that particular pixel.  For our applications, it cannot be guaranteed that the sum of the squares is less than $1$ as in this example. This is not a problem for estimating surface normals or overall shape, as previously discussed. 

\begin{figure}[ht]
\centering  
  \begin{subfigure}[t]{0.24\textwidth}
  \includegraphics[width=1.0\textwidth]{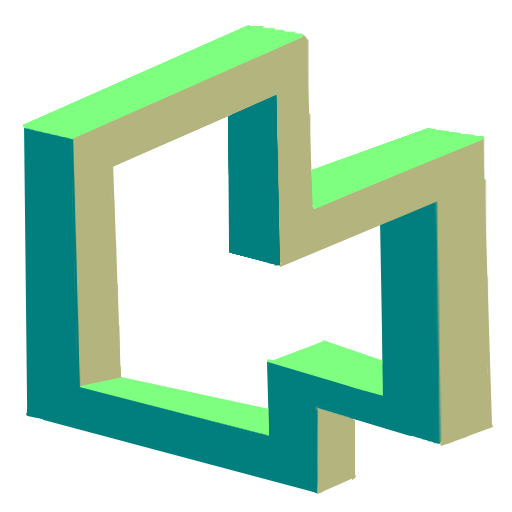}
  \caption{Normal map.}
  \label{figimpossible/0/sm}
  \end{subfigure}
  \hfill
  \begin{subfigure}[t]{0.24\textwidth}
  \includegraphics[width=1.0\textwidth]{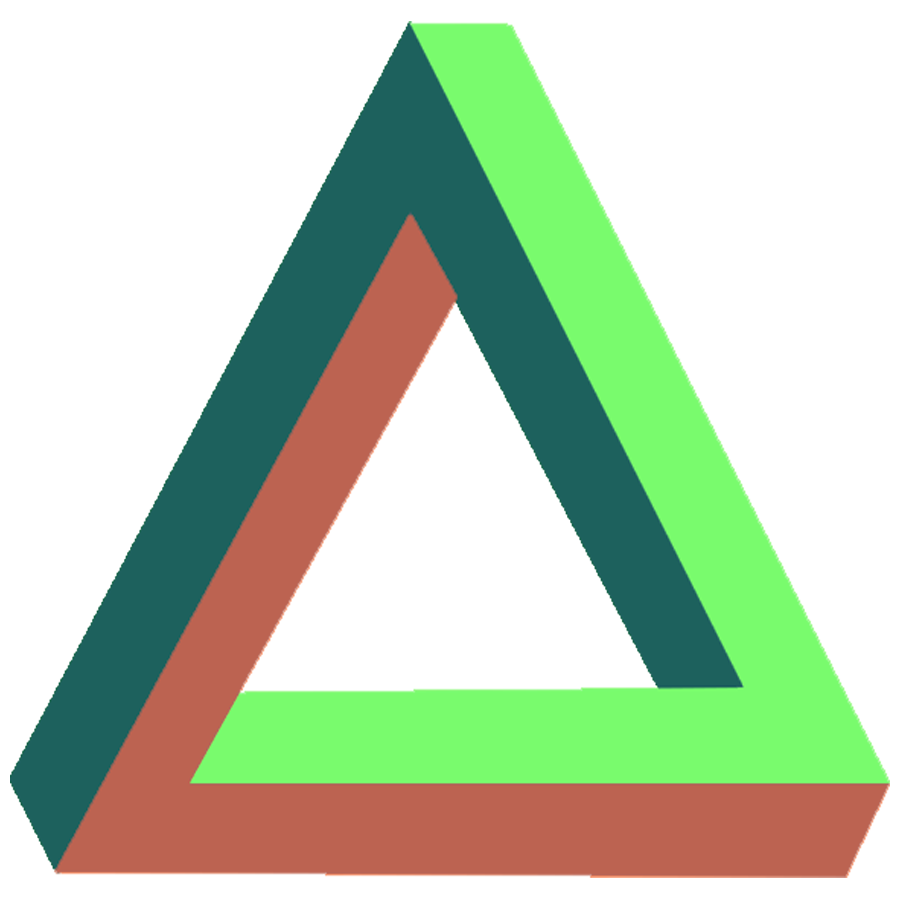}
  \caption{Normal map.}
  \label{figimpossible/1/sm}
  \end{subfigure}
  \hfill
  \begin{subfigure}[t]{0.24\textwidth}
  \includegraphics[width=1.0\textwidth]{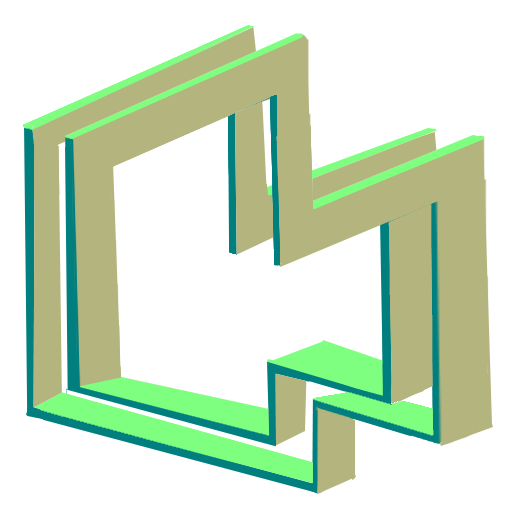}
  \caption{Normal map.}
  \label{figimpossible/2/sm}
  \end{subfigure}
  \hfill
  \begin{subfigure}[t]{0.24\textwidth}
  \includegraphics[width=1.0\textwidth]{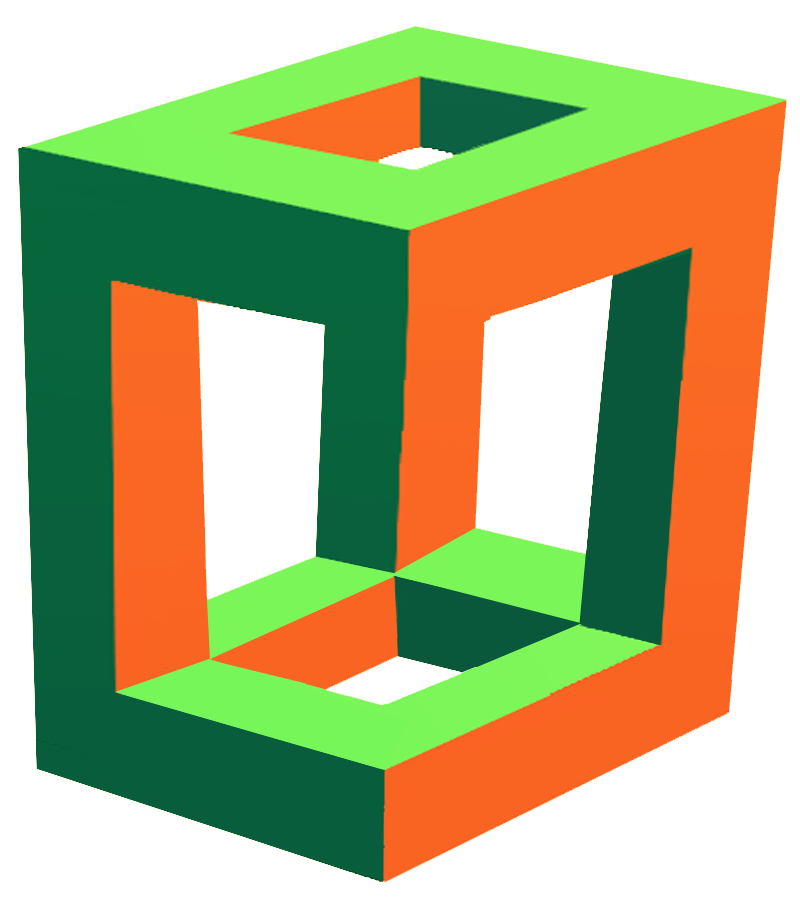}
  \caption{Normal map.}
  \label{figimpossible/3/sm}
  \end{subfigure}
  \hfill
\caption{
Examples of normal maps representing impossible shapes.}
\label{figimpossibleshapes}
\end{figure} 

\begin{figure}[ht]
\centering  
  \begin{subfigure}[t]{0.24\textwidth}
  \includegraphics[width=1.0\textwidth]{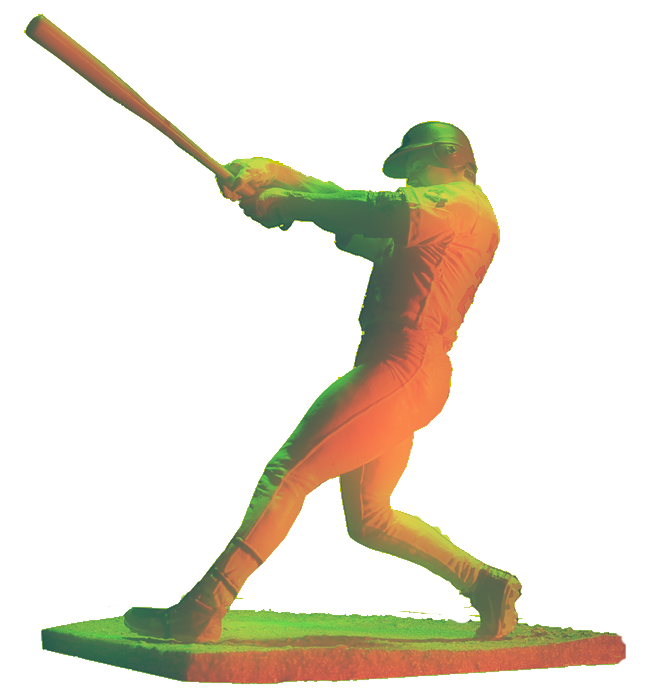}
  \caption{Normal map.}
  \label{figphotos/0/SM}
  \end{subfigure}
  \hfill
  \begin{subfigure}[t]{0.24\textwidth}
  \includegraphics[width=1.0\textwidth]{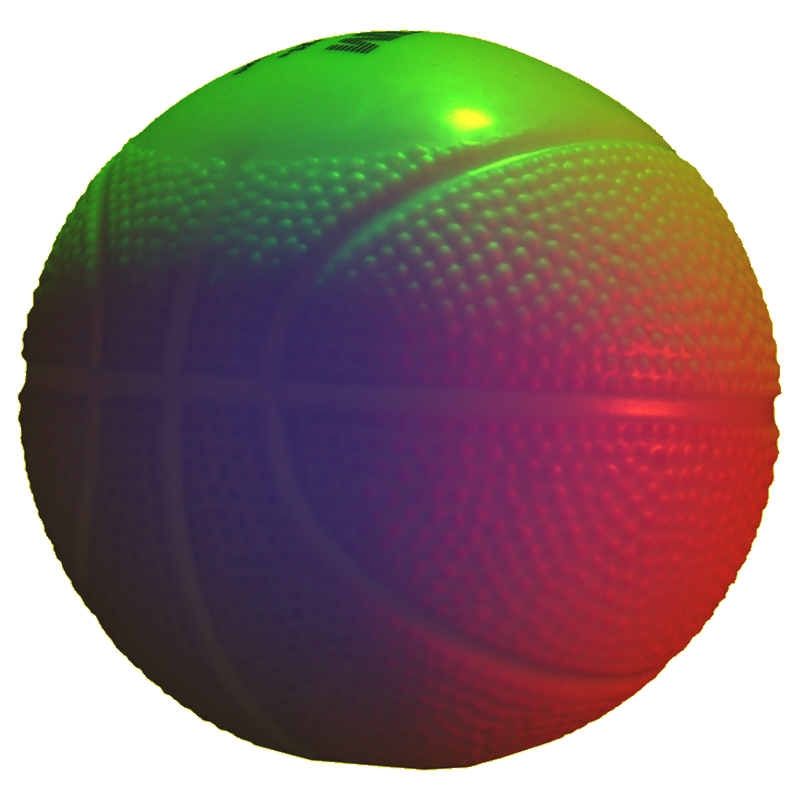}
  \caption{Normal map.}
  \label{figphotos/1/SM}
  \end{subfigure}
  \hfill
  \begin{subfigure}[t]{0.24\textwidth}
  \includegraphics[width=1.0\textwidth]{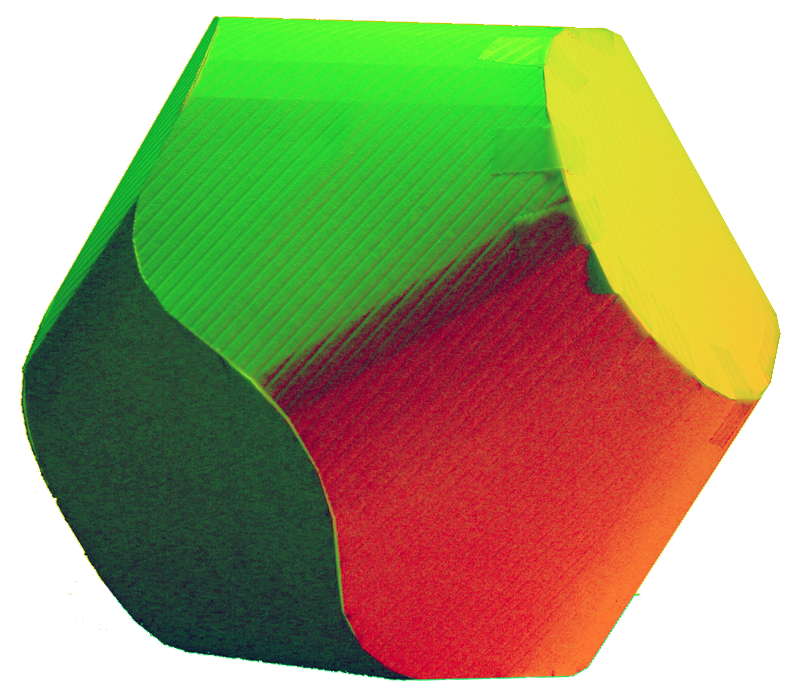}
  \caption{Normal map.}
  \label{figphotos/2/SM}
  \end{subfigure}
  \hfill
  \begin{subfigure}[t]{0.24\textwidth}
  \includegraphics[width=1.0\textwidth]{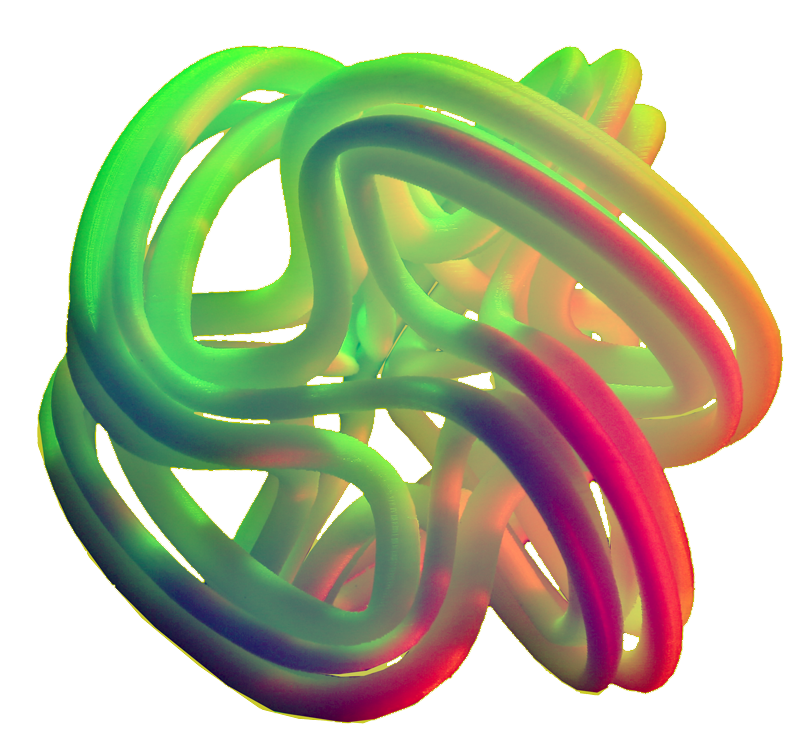}
  \caption{Normal map.}
  \label{figphotos/4/SM}
  \end{subfigure}
  \hfill
\caption{
Examples of normal maps are obtained by photographing real objects with a red light from the right side and green light from the top.}
\label{figshapemapphotos}
\end{figure}

Normal maps can also be obtained by photographing physical shapes using the fact that the red and green light vectors $(1,0)$ and $(0,1)$ are linearly independent. Therefore, any 2D light can be given a linear combination of the two as $(L.x, L.y)=L.x(1,0)+ L.y(0,1)$. Therefore, to compute illumination coming from an arbitrary parallel light, all we need to do is compute the contribution from two linearly independent components. This property allows for the creation of normal maps by photographing real objects using red and green lights, which can be used as a simple alternative to the environment mat \cite{Zongker1999}, as shown in some examples in Figure~\ref{figshapemapphotos}. 

\begin{figure}[ht]
\centering  
  \begin{subfigure}[t]{0.24\textwidth}
  \includegraphics[width=1.0\textwidth]{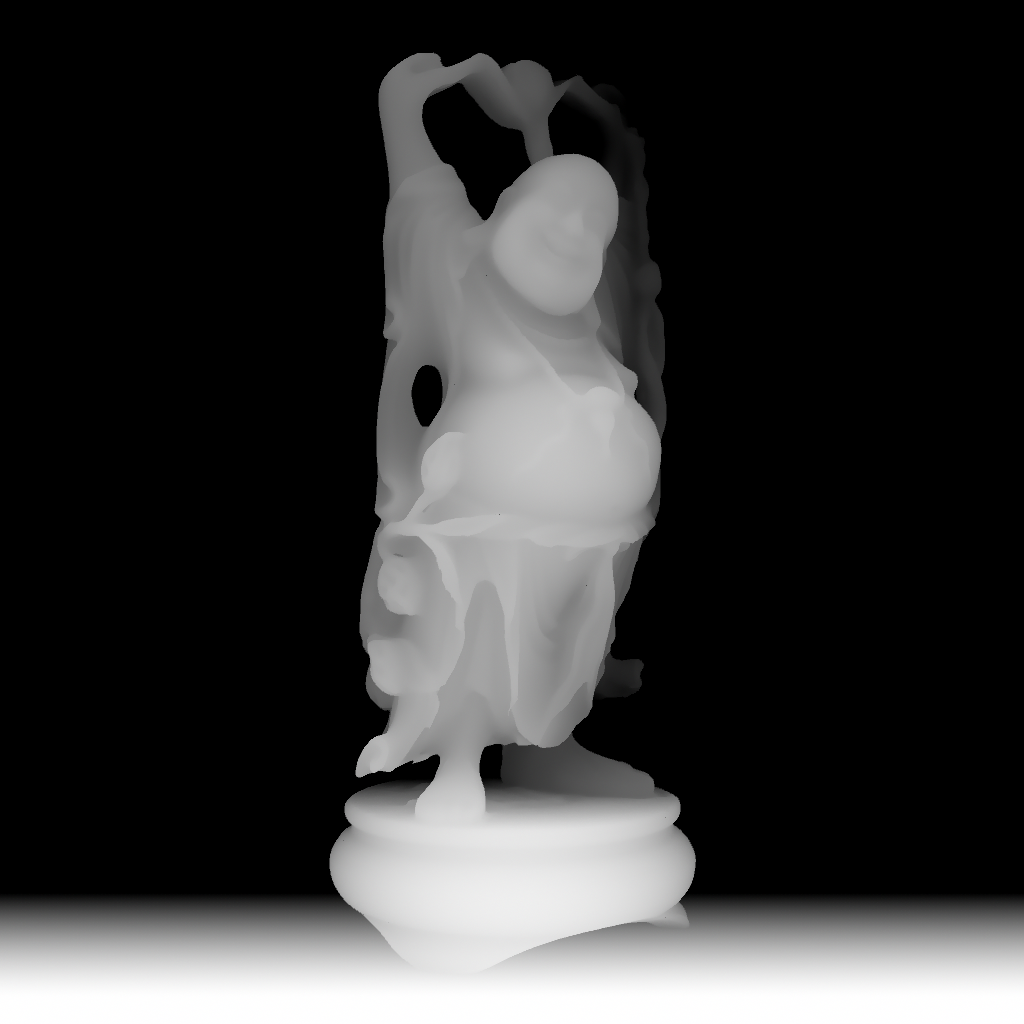}
  \caption{Buddha depth map.}
  \label{figHeightFields/buddha}
  \end{subfigure}
  \hfill
  \begin{subfigure}[t]{0.24\textwidth}
  \includegraphics[width=1.0\textwidth]{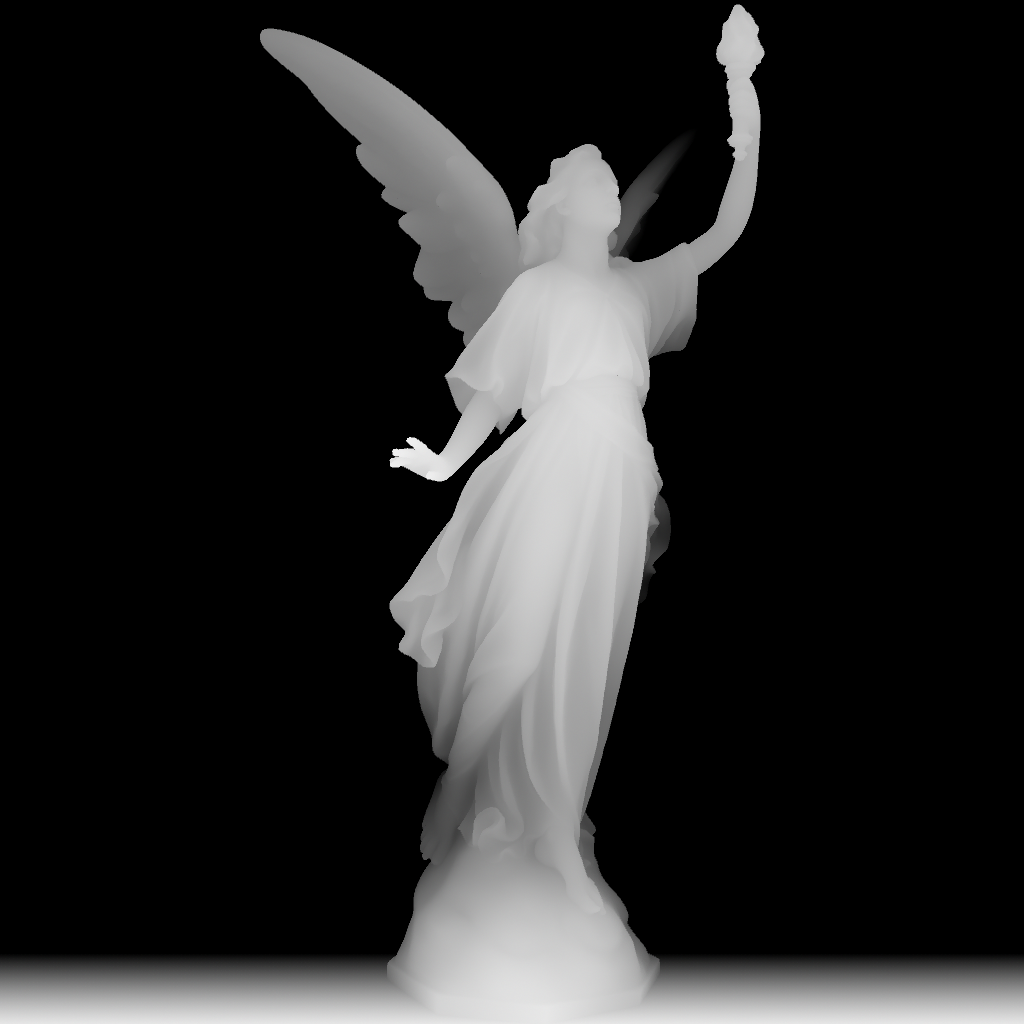}
  \caption{Lucy depth map.}
  \label{figHeightFields/lucy}
  \end{subfigure}
  \hfill
  \begin{subfigure}[t]{0.24\textwidth}
  \includegraphics[width=1.0\textwidth]{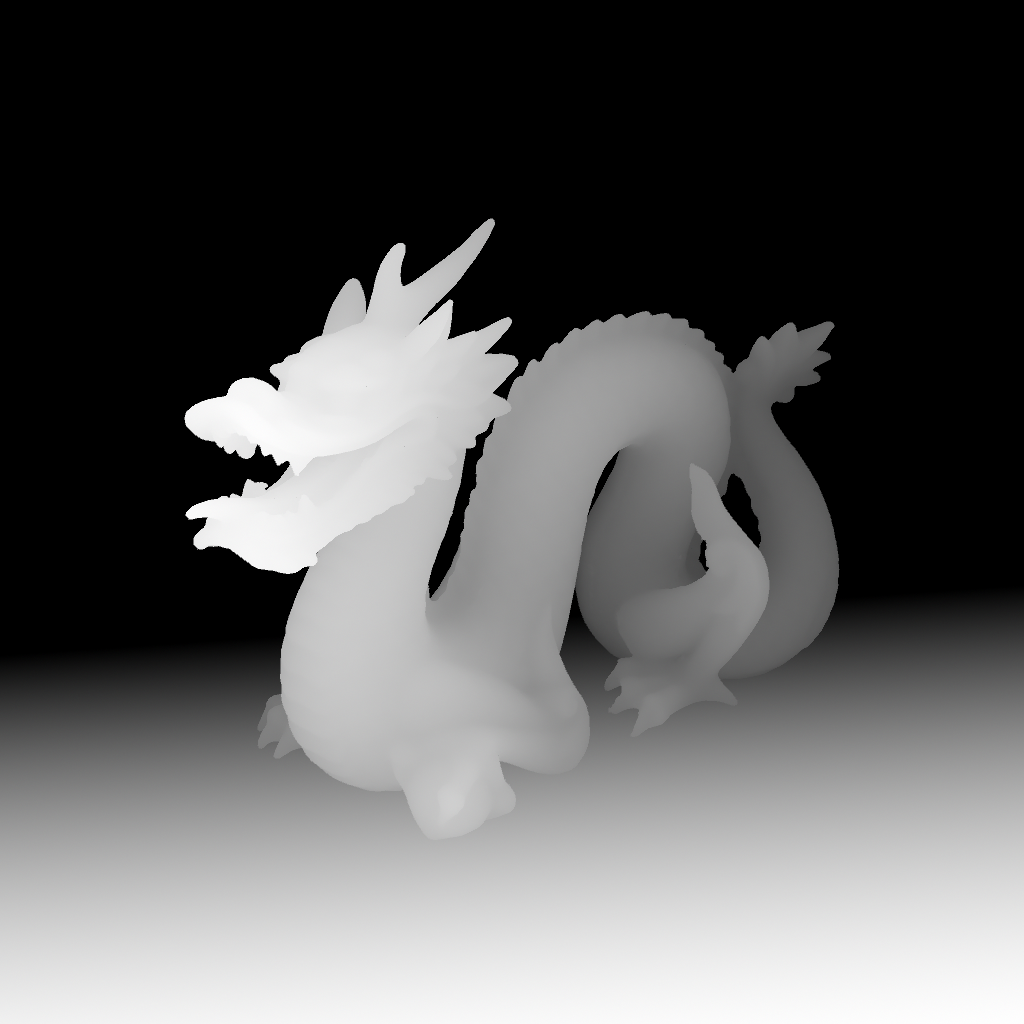}
  \caption{Dragon depth map.}
  \label{figHeightFields/dragon}
  \end{subfigure}
  \hfill
  \begin{subfigure}[t]{0.24\textwidth}
  \includegraphics[width=1.0\textwidth]{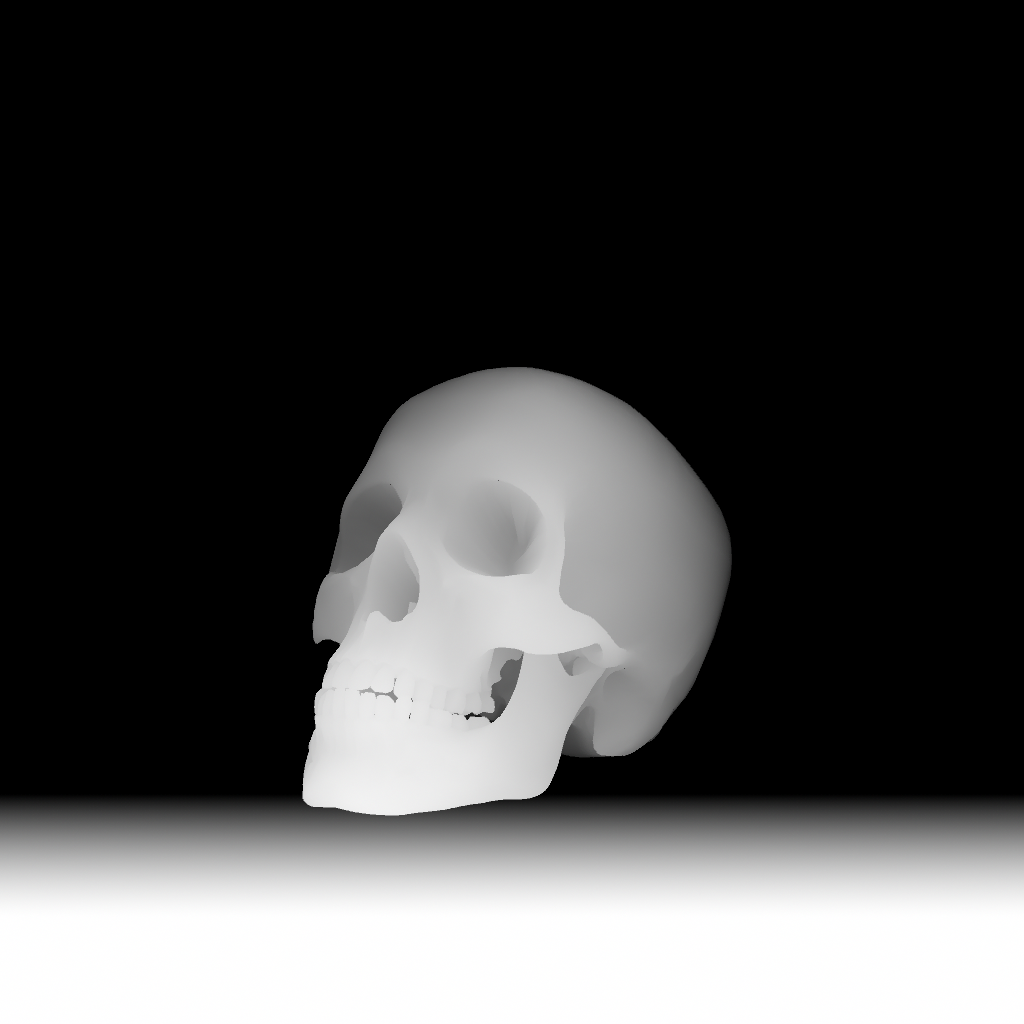}
  \caption{Skull depth map.}
  \label{figHeightFields/skull}
  \end{subfigure}
  \hfill
\caption{
Examples of depth maps obtained by rendering virtual 3D models.}
\label{figHeightFields}
\end{figure}

Many of the methods that are used to create normal maps can also be used to create height maps. They can be rendered from virtual 3D models as shown in Figure~\ref{figHeightFields}. They can be created by using vector-based drawing tools such as Lumo or Crossshade. There are even a few artists who paint depth maps, such as Kazuki Takamatsu \cite{takamatsu2021}. In conclusion, the creation of normal maps and depth maps is relatively easy. They can be created by using 2D methods such as painting or illustration. 

\section{Direct Illumination Effects with Mock3D Shapes}
\label{section2}

The major advantage of Mock3D shapes is that the illumination terms are simpler to compute and it can, therefore, be done in real time. In this section, we will discuss two basic illumination terms for each light: diffuse illumination and specular illumination. For barycentric shading, these must be scalars between $0$ and $1$. They are used as the parameters of barycentric shading equations. It is possible to use separate parts of illumination terms and control the color of each term separately. For example, it is possible to assign a specific color to the shadow region \cite{ross2022}. These are sometimes necessary to obtain a specific style of a painter \cite{ross2022}. 
We also need global illumination terms, which are not scalar but warping and transformations of images. They are used to create reflection and refraction. 

Diffuse and specular illumination terms are computed per light, where we view the light directions as interface parameters. Let the vector $\vec{L}(u,v)=(L.x, L.y, L.z)$ denote the direction of light for any given pixel, where $L.x^2+L.y^2+L.z^2=1$. For point lights, it is calculated using the vector between the light position $\mathbf{P}_{L}$ and the pixel position $\mathbf{P}(u,v)$  as follows: 
$$\vec{L}(u,v) = \frac{\mathbf{P}_{L}-\mathbf{P}(u,v) }{|\mathbf{P}_{L}-\mathbf{P}(u,v)|}$$
For directional lights and $\vec{L}(u,v)$ does not change for every pixel and can be simply written as $\vec{L}$. For a given light, there are two fundamentally different ways of computing the diffuse illumination. The first is a generalization of the classical diffuse illumination term and is provided in the next subsection, Section\ref{sec:CDI}. The second one is an integrated method that can provide diffuse illumination along with shadows. The second method is provided in Section \ref{sec:IDIwS}

\subsection{Classical Diffuse Illumination}
\label{sec:CDI}

For the generalization of the classical term $t=max(cos \theta,0)$ that needs a normal vector \cite{ebert2003texturing,foley1996computer}. For normal maps, the normal vector for every pixel position $(u,v)$ is given as $\vec{N}(u,v)$. For depth maps, $\vec{N}(u,v)$ needs to be pre-computed using Equation~\ref{eqgradient}. Now, the most important illumination term $\cos \theta$ can be computed as a dot product of $\vec{N}=(N.x, N.y, N.z)$ and $\vec{L}$ as follows, $\vec{N}\cdot \vec{L} = L.x N.x+L.y N.y+ L.z N.z,$, which is a very simple equation. The diffuse illumination term is given in Algorithm~\ref{AlgolDiffuse}. 

\begin{algorithm}
\caption{Calculation of diffuse illumination term}
\label{AlgolDiffuse}
\begin{algorithmic}
\REQUIRE Only $\vec{N}(u,v)=(N.x,N.y,N.z)$ and $\vec{L}(u,v)=(L.x,L.y,L.z)$ for $\cos \theta$ term. 
\ENSURE $|\vec{N}(u,v)|=N.x^2+N.y^2+N.z^2=1$ and $|\vec{L}(u,v)|=L.x^2+L.y^2+L.z^2=1$.
\STATE $t \leftarrow \vec{N}\cdot \vec{L} =  L.x N.x+L.y N.y+ L.z N.z$
\STATE $t =$\textbf{Clamp\&Step}$(t,t_0,t_1)$
\end{algorithmic}
\end{algorithm}

\begin{algorithm}
\caption{Clamp and Step functions with smooth-step, and smoother-step extensions}
\label{AlgolClampStep}
\begin{algorithmic}
\STATE  \textbf{Clamp\&Step}$(t,t_0,t_1,F,Type)$
\ENSURE Type is the Function type that can be either Smooth-Step or Smoother-Step. Any other name returns $t$ value without any modification. For these functions see \cite{perlin1985image,perlin2002improving,kesson2008pixar}. 
\STATE $ t \leftarrow (t-t_0)/(t_1-t_0)$
\IF{$t < 0$}
\STATE $t \leftarrow 0$
\ENDIF
\IF{$t > 1$}
\STATE $t \leftarrow 1$
\ENDIF
\IF{Type==Smooth-Step}
\STATE $t \leftarrow 3 t^2 - 2 t^3$
\ENDIF
\IF{Type==Smoother-Step}
\STATE $t \leftarrow 6 t^5 - 15 t^4 + 10 t^3$
\ENDIF
\RETURN $t$
\end{algorithmic}
\end{algorithm}

In this case, $t=0$ means that the shading point is not illuminated by a given light source, and $t=1$ means that the point is fully illuminated by the same light source. 
To have additional control when $t$ becomes zero or one, we use the clamp and step function \cite{apodaca2000advanced}. The function is given in Algorithm~\ref{AlgolClampStep}. 
Using step functions, we can allow the user to change the sizes of regions that are not illuminated and fully illuminated by changing the values of $t_0$ and $t_1$. This formulation also gives two essential forms used in computer graphics, namely (1) the classical diffuse term $t=max(cos \theta,0)$ with $t_0=0$ and $t_1=1$ \cite{foley1996computer}; and (2) Gooch \& Gooch term  $t=(cos \theta+1)/2$ with $t_0=-1$ and $t_1=1$ \cite{Gooch98,gooch2001non}. Moreover, we can also obtain cartoon shading by choosing $t_0=t_1$. In computer graphics, $t$ can be further manipulated by transformations such as the following to obtain continuity of the first and second degree in $t=0$ and $t=1$ that are called smooth step or smoother step functions \cite{perlin1985image,perlin2002improving,kesson2008pixar}.

\subsection{Barycentric Diffuse Shading}

To compute diffuse shading, we use barycentric formulas. The simplest of the Barycentric formulas is the linear interpolation, which is given in the form $x=x_0(1-t) + x_1 t$ where $x_0$ and $x_1$ can be any entity, but they need the same type. Using this basic concept, we can simply interpolate channels, i.e. images or textures that are mapped to Mock 3D shapes, as follows.
\begin{equation}C(u,v)= C_0(u,v) (1-t(u,v)) + C_1(u,v) t(u,v)
\label{eqDiffuseShade}
\end{equation}
where $C_0(u,v)=(R_0(u,v),G_0(u,v),B_0(u,v))$ and $C_1(u,v)=(R_1(u,v),G_1(u,v),B_1(u,v))$ are images such as $I_0$ and $I_1$ shown in Figure~\ref{figtable}. Therefore, they are considered to be material properties. Furthermore, $t(u,v)$ and $(1-t(u,v))$ are also images, which correspond to $W_1$ and $W_0$ in Figure~\ref{figtable}. The term $1$ in Equation~\ref{eqDiffuseShade} is actually a white image, and $t(u,v)$
is calculated using the algorithms in \ref{AlgolDiffuse} and \ref{AlgolClampStep} per pixel. Since light sources can be colorful, $W_0$ and $W_1$ can also be colorful, as shown in Figure~\ref{figtable}. Note that we are not limited by the linear interpolation equation. Any barycentric curve formula such as B-Spline curves can be used to obtain diffuse shading. The only requirement is that the equation must be a curve formula. 

\subsection{Integrated Diffuse Illumination with Shadows}
\label{sec:IDIwS}

The main problem with this particular diffuse term is that the shadows are not included. We still have some shadow regions that correspond to $\vec{N}\cdot \vec{L} \leq 0$. However, more interesting shadows are those that are caused by occlusions, and we need to obtain some styles. The problem is that these shadows can only be identified by considering the shape of the objects. Only normal information is not sufficient; we also need height information. 

\begin{figure}[htpb]
  \centering  
  \begin{subfigure}[t]{0.475\textwidth}
  \includegraphics[width=1.0\textwidth]{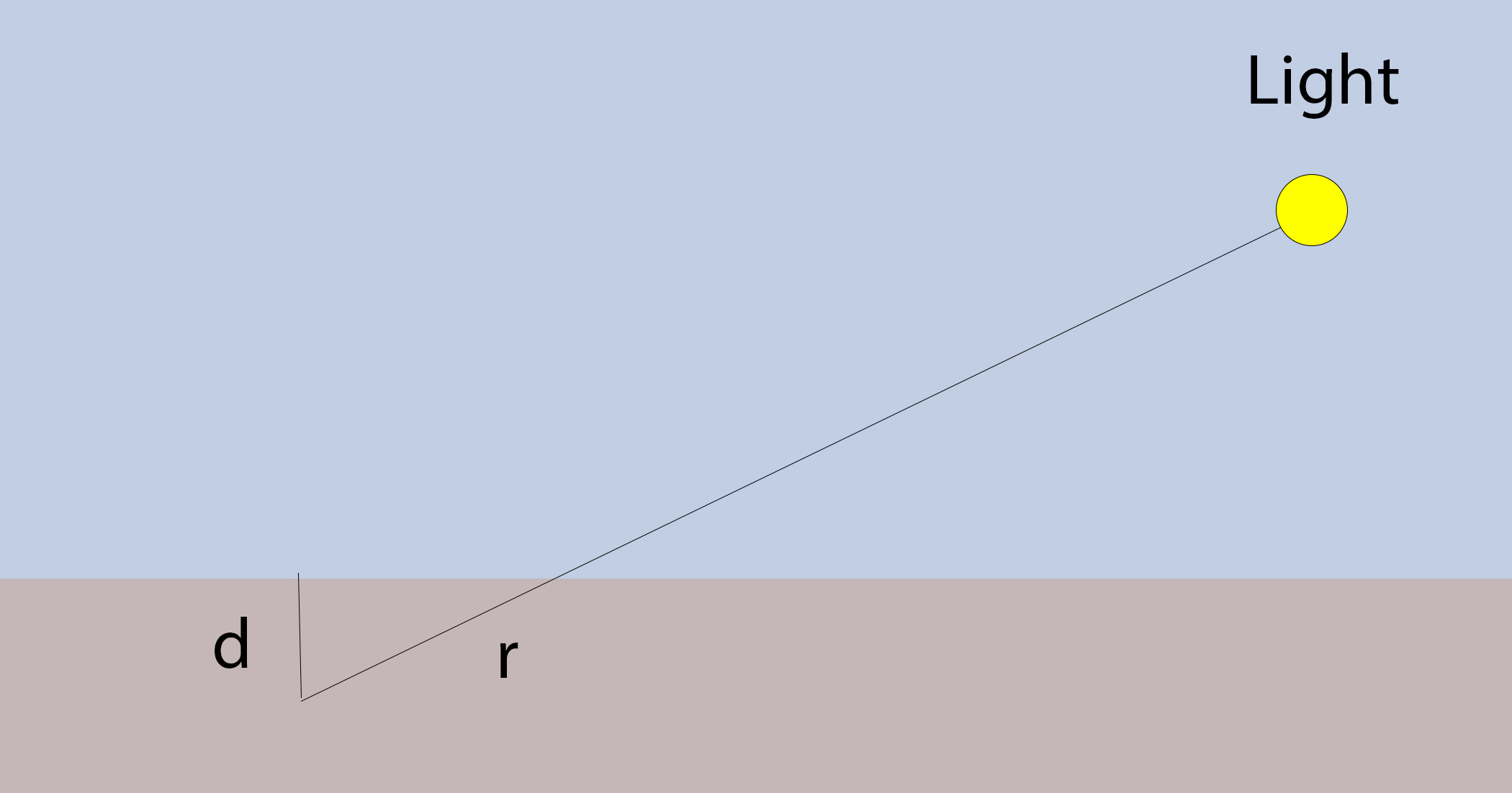}
  \caption{
  If the shading surface is a plane, $t=d/r$ gives $\cos \theta$.}
  \label{figcostetashadows/costheta}
  \end{subfigure}
  \hfill  
  \begin{subfigure}[t]{0.475\textwidth}
  \includegraphics[width=1.0\textwidth]{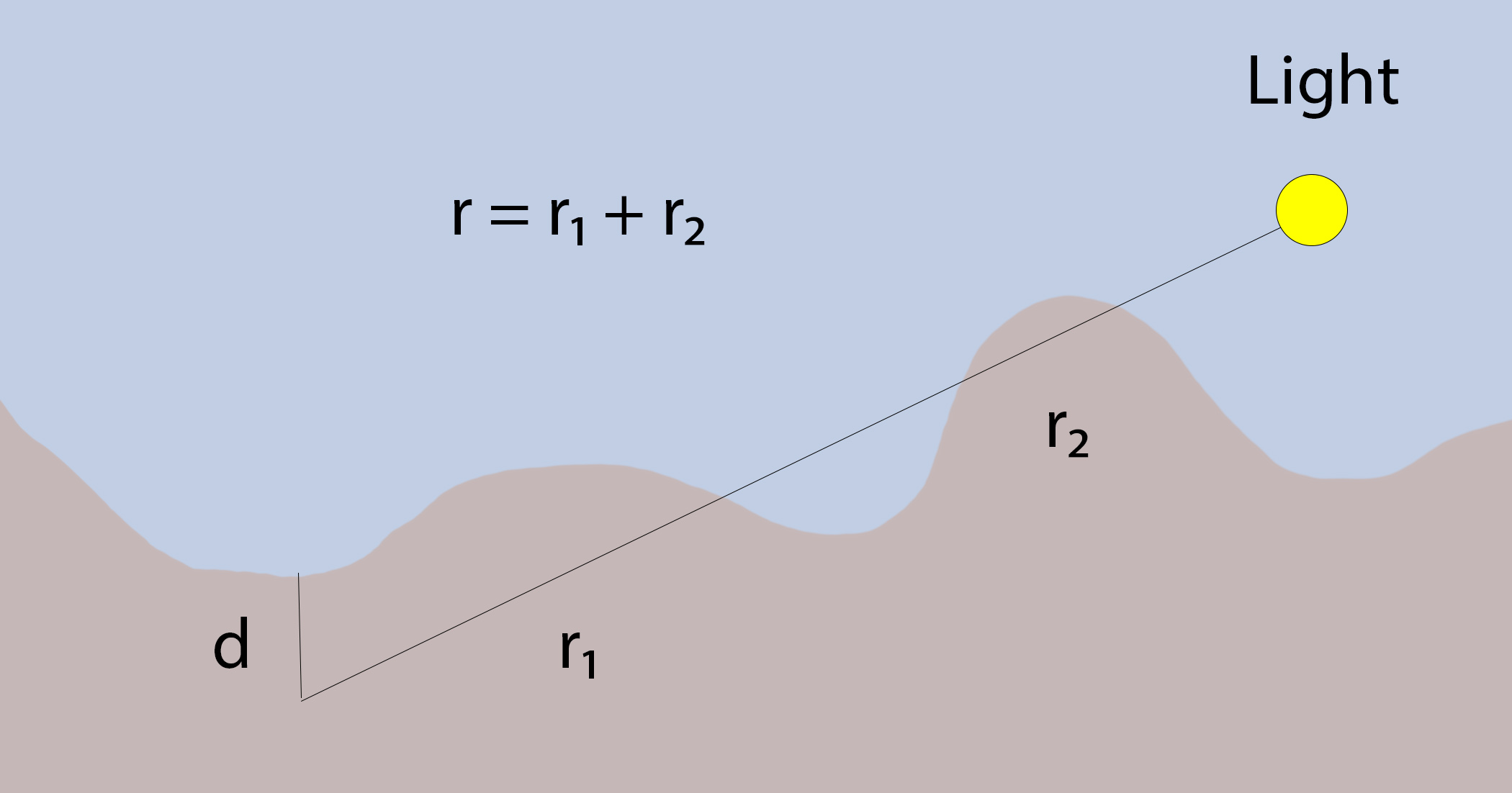}
  \caption{
  If the shading surface has bumps, $t=d/r$ provides shadows from local bumps.}
  \label{figcostetashadows/shadow}
  \end{subfigure}
  \hfill  
\caption{These two examples demonstrate how multiple illumination effects can be obtained with the $d/r$ formula.  }
\label{figcostetashadows/model}
\end{figure}

There are many methods for computing shadows using a variety of shadow maps \cite{kim2001opacity,aila2004alias,lokovic2000deep}. We use a 2D version of ray-tracing shadows, called $\cos \theta$ shadows, to render mock3D shapes \cite{akleman2017}. 
Let $\mathbf{P}_L$ denote the position of a point light and let $\mathbf{P}_S$ and $\vec{N}_S$ denote the position and the normal vector of a given shading point on a surface. We define a new shading position under the surface as $\mathbf{P'}_S = \mathbf{P}_S - d \vec{N}_S$. Let $L(\mathbf{P}_L,\mathbf{P'}_S)$ denote the line segment that starts from $\mathbf{P'}_S$ and ends at $\mathbf{P}_L$ and
let $r$ denote the length of $L(\mathbf{P}_L,\mathbf{P'}_S)$ that goes through the interior of the shapes. Then, the contribution of illumination is calculated simply as $t=d/r$ (see Algorithm~\ref{Algolcostetashadows2} and Figure~\ref{figcostetashadows/model} for an algorithmic and visual explanation). 

If the shading surface is a plane, it is easy to demonstrate that $t=d/r$ gives us $\cos \theta$ regardless of $d$ (see Figure~\ref{figcostetashadows/costheta}). The equation approaches $\cos \theta$ even for nonplanar regions when $d \rightarrow 0$ by assuming that the shape and line have only one intersection and the light cannot be inside of the object. This approach solves inconsistencies caused by shadows. As discussed earlier, we have two different sources of shadow: (1) self-shadows of convex shapes that are identified as $\cos \theta <0 $ regions; and (2) other shadows that are identified by occlusion of the light by other objects and other parts of the same but nonconvex object. This model combines both cases in one equation where $r_1$ qualitatively similar to the first type of shadows and $r_2$ qualitatively similar to the second type of shadows and $r=r_1+r_2$ logically combines both. Another advantage of this formulation is that it provides a qualitatively similar version of exponential attenuation. Therefore, this method also provides a visual appearance of subsurface scattering \cite{akleman2017,wang2014}. 

An important property of this approach is that the computation can be done in real-time in GPU for 2D height fields in the form of $z=H(x,y)$ with the Algorithm~\ref{Algolcostetashadows}. 

\begin{algorithm}
\caption{Calculation of $\cos \theta$ shadows as diffuse illumination term using depth maps (or height fields).}
\label{Algolcostetashadows}
\begin{algorithmic}
\REQUIRE A height function in the form of $z=H(x,y)$.
\REQUIRE A light direction $\textbf{N}_L=(L.x,L.y,L.z)$. It is computed if the light is point light. 
\REQUIRE The shading point position $\textbf{P}_S=(P.x,P.y,P.z)$ and normal $\vec{N}_S=(N.x,N.y,N.z)$.
\REQUIRE An offset amount: $d$
\REQUIRE The length of the step: $a$. This value must be smaller than $d$. 
\REQUIRE The maximum number of steps: $K$. It is given if directional light. It is computed using step size if it is point light. 
\ENSURE The first point $\textbf{P}_0=(x_0,y_0,z_0) \leftarrow \textbf{P}_S - d \vec{N}_S = (P.x-d N.x,P.y-d N.y, P.z-d N.z)$.
\STATE The first sample point: $\textbf{P}=(x,y,z) \leftarrow \textbf{P}_S+ \vec{\Delta \textbf{P}}/2=(P.x+dL.x,P.y+dL.y,P.z+dL.z)$
\STATE $r \leftarrow d$
\STATE The step vector: $\vec{\Delta \textbf{P}} = (\Delta x, \Delta y, \Delta z) \leftarrow a\textbf{N}_L = \left(aL.x, aL.y, aL.z \right)$
\WHILE{$K \geq 0$}
\STATE $K \leftarrow K-1$
\STATE $\textbf{P}=(x,y,z) \leftarrow \textbf{P}+ \vec{\Delta \textbf{P}}=(x+\Delta x,y+\Delta y,z+\Delta z)$
\IF{$H(x,y) > z$}
\STATE $r \leftarrow r+a$
\ENDIF
\ENDWHILE
\STATE $t \leftarrow d/r$
\STATE $t =$\textbf{Clamp\&Step}$(t,t_0,t_1)$
\end{algorithmic}
\end{algorithm}

Since we do the computation on the GPU, we can use an arbitrary step size. In GPU-level programming, the images are kept as dynamic textures, and we can access any step size in texture space. Here, for depth maps conversion from $(u,v)$ to $(x,y)$ is trivial; therefore, there is no need for additional explanation. On the other hand, the conversion of normal maps to $z=H(x,y)$ is not trivial, since normal maps may not be conservative and a corresponding height field may not necessarily exist. 

Our solution to this problem is to reconstruct a height field only in the line segment that connects $\textbf{P}_L$ and $\textbf{P}_S$. This does not give a globally consistent height field. However, this local height estimation is sufficient to compute reasonably acceptable shadows. 

 \begin{figure}
\begin{tabular}{cccccccccc}
  \hfill  \includegraphics[width=0.18\textwidth]{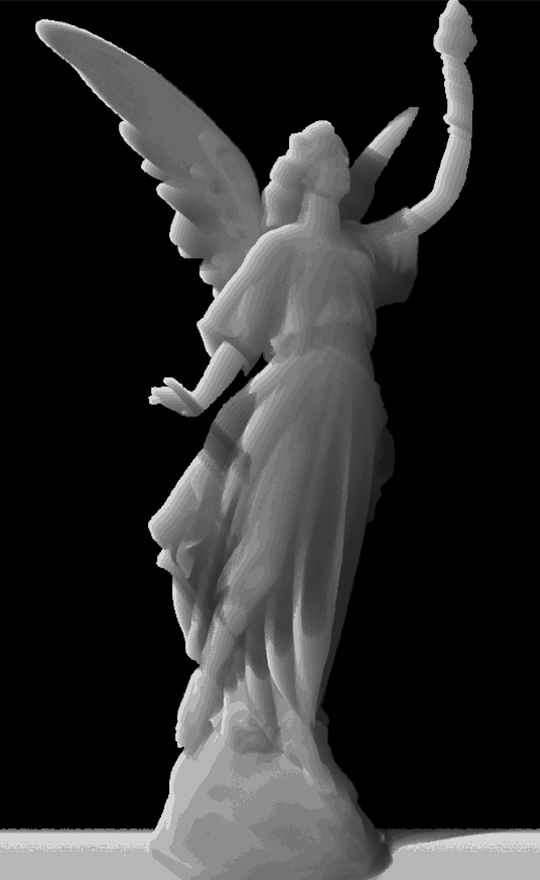}&
  \hfill  \includegraphics[width=0.18\textwidth]{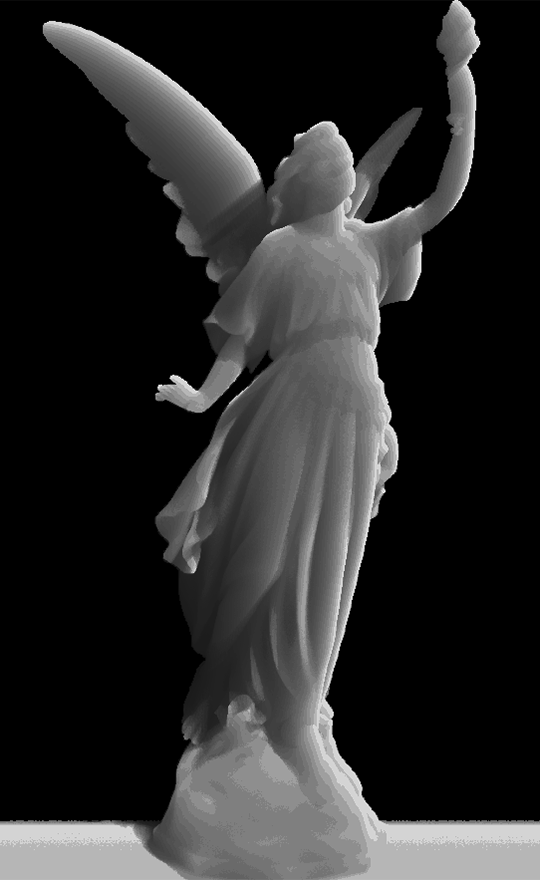}&
  \hfill  \includegraphics[width=0.18\textwidth]{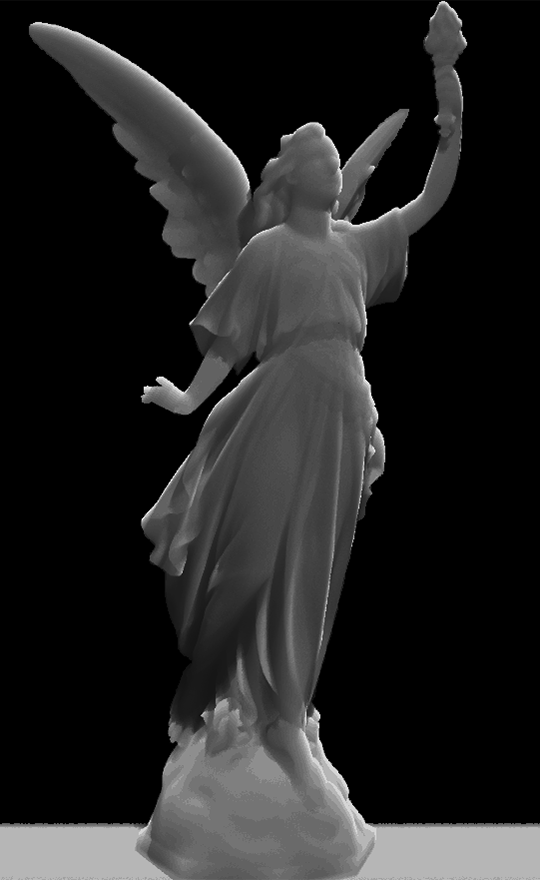}&
  \hfill  \includegraphics[width=0.18\textwidth]{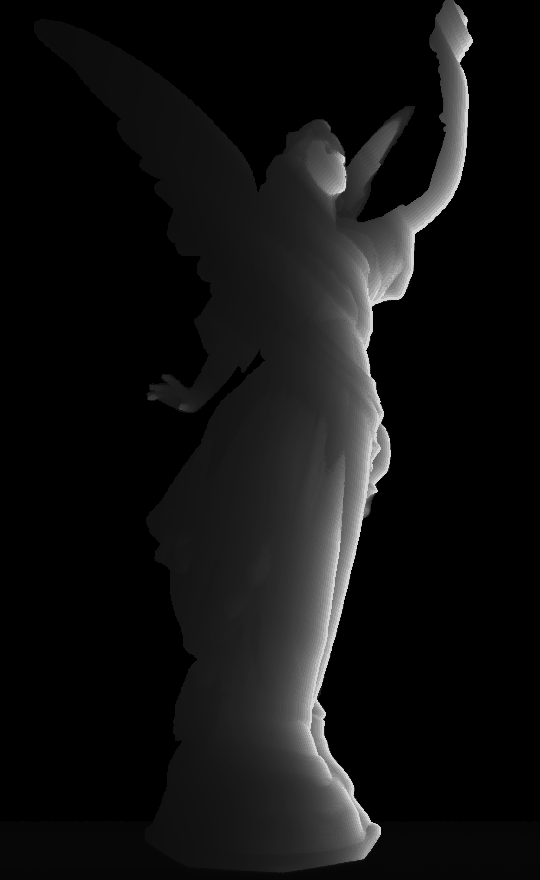}&
  \hfill  \includegraphics[width=0.18\textwidth]{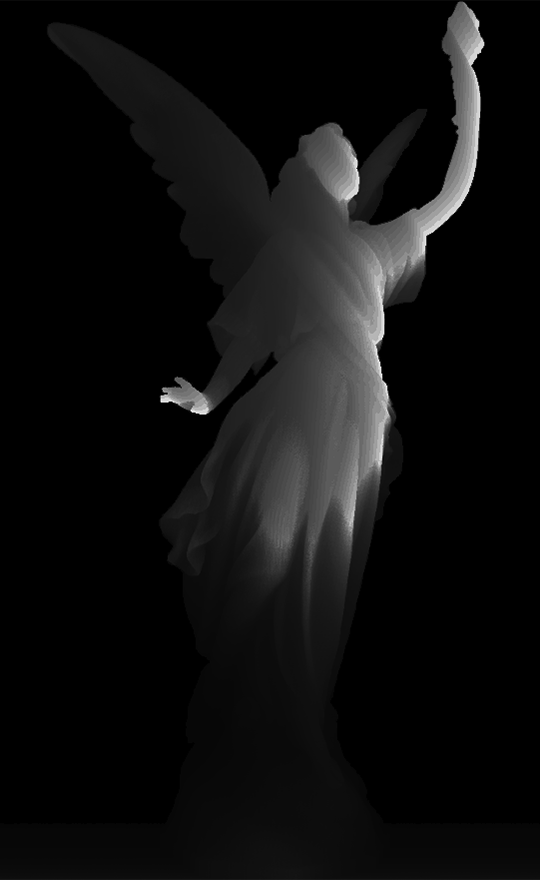}\\
\end{tabular}
\caption{An example of integrated real-time shading of shadow, direct illumination, and subsurface scattering using a depth map obtained from Lucy Sculpture shown in Figure~\ref{figHeightFields/lucy}. As shown in this sequence of images, a single parameter between $0$ and $1$ that is computed by using the $\cos \theta$ shadow method is sufficient to obtain all three effects consistently.  }
\label{figcostetashadows/lucy}
\end{figure}

\begin{algorithm}
\caption{Calculation of $\cos theta$ shadows as diffuse illumination term using normal maps (or vector fields).}
\label{Algolcostetashadows2}
\begin{algorithmic}
\REQUIRE A 2D function of normals $\vec{N}(u,v)=(N.x,N.y,N.z)$.
\REQUIRE A light direction $\vec{L}(u,v)=(L.x,L.y,L.z)$. For directed light, it is constant. It should be computed per pixel if the light is point light. 
\ENSURE An estimated shading point $\textbf{P}_S=(P.x,P.y,P.z)$. Here, $(P.x, P.y)$ are texture positions, but we do not know $P.z$. We assume that the $P.z$ is always the same constant for every shading point, usually $P.z=0$. 
\REQUIRE An offset amount: $d$
\REQUIRE The step length: $a$. This value has to be smaller than $d$. 
\REQUIRE The maximum number of steps: $K$. It is given if directional light. It is computed using step size if it is point light. 
\ENSURE The first point $\textbf{P}_0=(x_0,y_0,z_0) \leftarrow \textbf{P}_S - d \vec{N}_S = (P.x-d N.x,P.y-d N.y, P.z-d N.z)$.
\STATE The first sample point: $\textbf{P}=(x,y,z) \leftarrow \textbf{P}_S + \vec{\Delta \textbf{P}}/2=(P.x+d L.x,P.y+d L.y,P.z+d L.z)$
\STATE $r \leftarrow d$
\STATE The step vector: $\vec{\Delta \textbf{P}} = (\Delta x, \Delta y, \Delta z) \leftarrow a\textbf{N}_L = \left(a L.x, a L.y, a L.z \right)$
\STATE  $H \leftarrow 0$
\WHILE{$K \geq 0$}
\STATE $K \leftarrow K-1$
\STATE $\textbf{P}=(x,y,z) \leftarrow \textbf{P}+ \vec{\Delta \textbf{P}}=(x+\Delta x,y+\Delta y,z+\Delta z)$
\IF{$H > z$}
\STATE $r \leftarrow r+a$
\ENDIF
\STATE  $H \leftarrow \vec{\Delta \textbf{P}} - \left( \vec{N}_S(x,y) \dot \vec{\Delta \textbf{P}} \right) \vec{N}_S(x,y) $
\ENDWHILE
\STATE $t \leftarrow d/r$
\STATE $t =$\textbf{Clamp\&Step}$(t,t_0,t_1)$
\end{algorithmic}
\end{algorithm}

\begin{figure}
\includegraphics[width=0.24\linewidth]{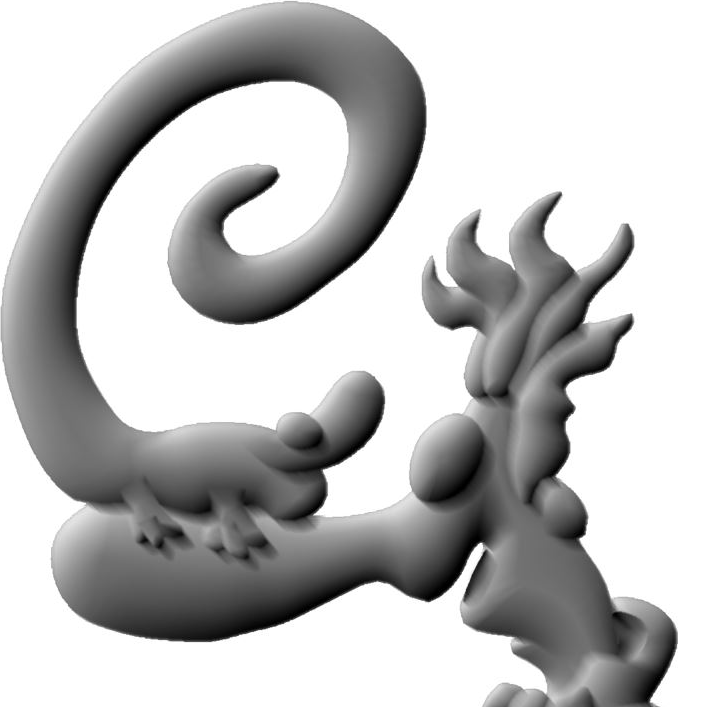}
\includegraphics[width=0.24\linewidth]{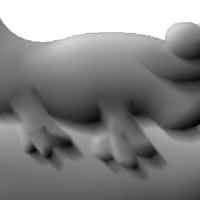}
\includegraphics[width=0.24\linewidth]{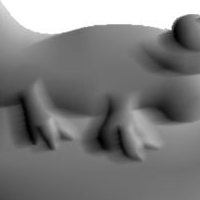}
\includegraphics[width=0.24\linewidth]{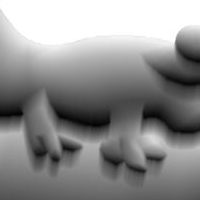}
\caption{$\cos \theta$ Shadows as diffuse illumination term  computed using normal maps. Three detailed images show how shadows change by changing the light position.}
\label{figBukalemun/shadow}
\end{figure}

\subsection{Barycentric Diffuse Shading with Shadows}

Shadow computation does not change how diffuse shading is computed. The only change is that the original $t(u,v)$ image will be replaced by an image that includes shadows. To obtain a consistent art style, it is critical to include shadows in the $t$ computation and compute shading using the same Barycentric formula. In this way, the image looks visually similar with or without shadow.  

\subsection{Specular Highlights}

The specular highlight term is another one that needs to be computed. In this case, the direction of the eye is fixed and is always $\vec{I}=(0,0,1)$. To compute the specular highlight term, we start with the basic Phong shading formula by computing $\cos \phi = \vec{I} \cdot \vec{R}_L$, where $\vec{R}_L$ is the reflection of a light vector according to the vector $\vec{N}$, which is computed as $\vec{R}_L = - \vec{L} + 2 \left(\vec{L} \cdot \vec{N}\right) \vec{N}$ (see Algorithm~\ref{Algolspecular}). 

\begin{algorithm}
\caption{Calculation of specular illumination term}
\label{Algolspecular}
\begin{algorithmic}
\REQUIRE $\vec{N}(u,v)=(N.x,N.y,N.z)$, $\vec{L}(u,v)=(L.x,L.y,L.z)$, and $\vec{L}(u,v)=(0,0,1)$ for $\cos \phi$ term. 
\ENSURE $|\vec{N}(u,v)|=L.x^2+L.y^2+L.z^2=1$ and $|\vec{L}(u,v)|=x^2+b^2+c^2=1$.
\STATE $\vec{R}_L \leftarrow - \vec{L} + 2 \left(\vec{L} \cdot \vec{N}\right) \vec{N} =  (2(L.xN.x+L.yN.y+L.zN.z)N.x-L.x, 2(L.xN.x+L.yN.y+L.zN.z)N.y-L.y, 2(L.xN.x+L.yN.y+L.zN.z)N.z-L.z)$
\STATE $s \leftarrow \vec{R}_L\cdot \vec{I} = 2(L.x N.x+L.y N.y+L.z N.z)N.z-L.zz$
\STATE $s =$\textbf{Clamp\&Step}$(s,s_0,s_1)$
\end{algorithmic}
\end{algorithm}

\subsection{Barycentric Shading with Specular Highlights}

Specular highlight term $s$ is linearly independent of diffuse term $t$. Therefore, the new equation must be a surface equation in the $(t,s) \in [0,1]^2$ domain rather than a curve equation in the $(t) \in [0,1]$ domain. The simplest of such equations is a cascading equation with a two bilinear term that requires only three data points as follows:  
\begin{eqnarray}
C(u,v) & \leftarrow & C_0(u,v) (1-t(u,v)) + C_1(u,v) t(u,v) \nonumber \\
C(u,v) & \leftarrow & C(u,v) (1-k_s(u,v) s(u,v)) + C_2(u,v) k_s(u,v) s(u,v)
\label{eqSpecShade}
\end{eqnarray}
where $C_2(u,v)=(R_2(u,v), G_2(u,v), B_2(u,v)) $ is the image that gives specular color in every position and it is generally white, i.e. $C_2(u,v)=(1,1,1) $. Furthermore, $k_s(u,v)$ is also an image that provides how the surface is reflected (see Figure~\ref{figEscheralpha0} as an example where the blue channel gives a value of $k_s$.). We can also use a true bilinear equation, but that form requires four data points, and the color of the specular highlight does not change with the value of $t$. Therefore, it is better to use a version of this cascading form. 

\section{Global Illumination Effects with Mock3D Shapes}

\begin{figure}[htpb]
  \centering  
  \begin{subfigure}[t]{0.535\textwidth}
  \fbox{\includegraphics[width=1.0\textwidth]{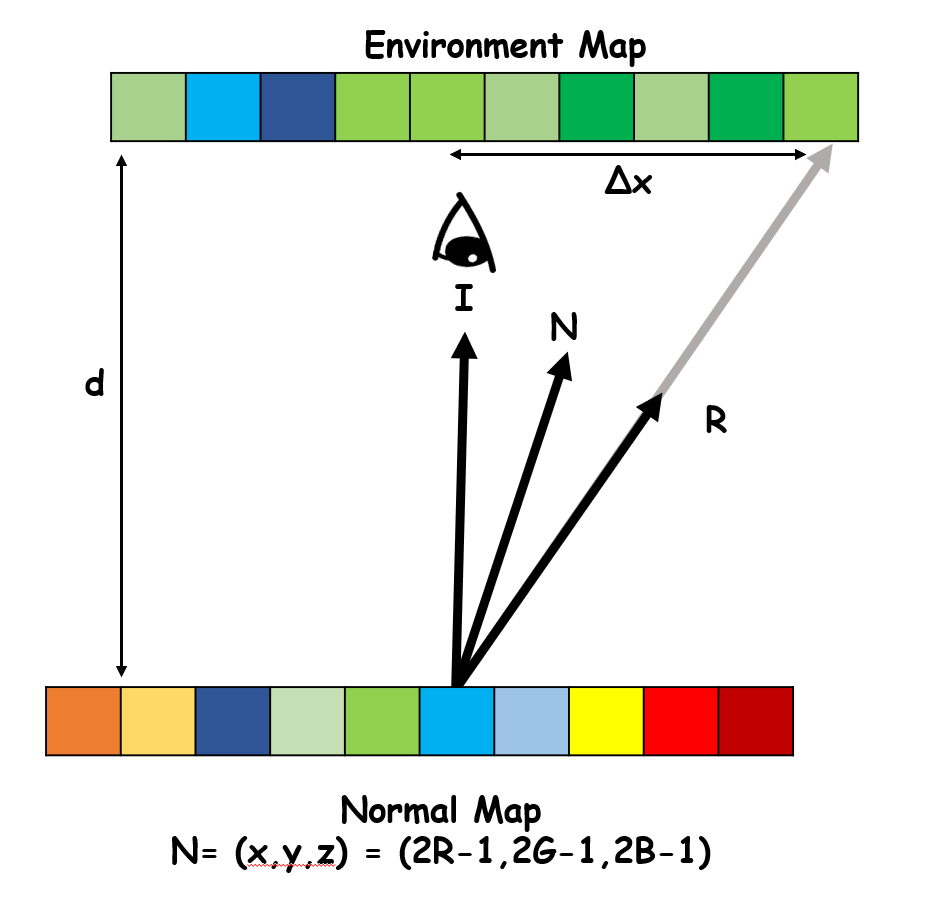}}
  \caption{
  2.5D reflections with Normal Maps. Side view.}
  \label{figimages/R0}
  \end{subfigure}
  \hfill  
  \begin{subfigure}[t]{0.432\textwidth}
  \fbox{\includegraphics[width=1.0\textwidth]{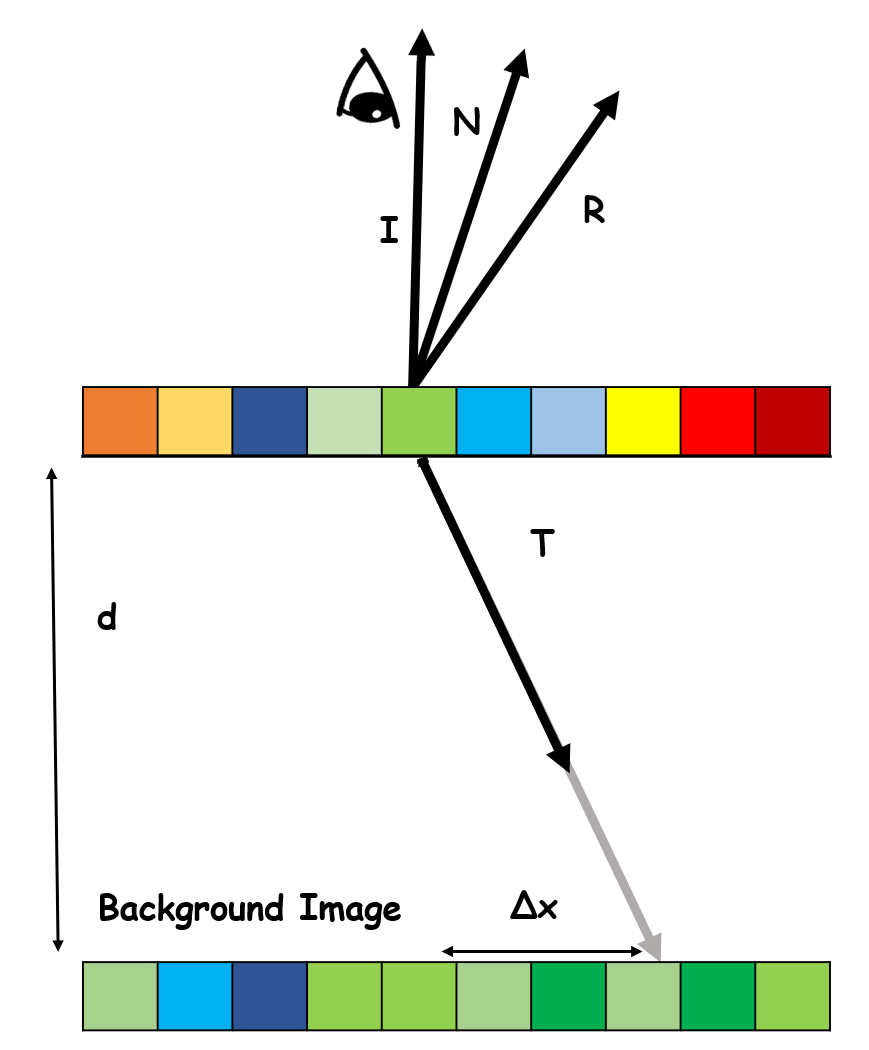}}
  \caption{
  2.5D Refraction with Normal Maps. Side view.}
  \label{figimages/R1}
  \end{subfigure}
  \hfill  
\caption{Computation of the reflection and refraction in 2.5D space using Normal Maps. Here $\vec{I}$ is the eye vector, $\vec{N}$ is the normal vector, $\vec{R}$ is the reflection vector, and $\vec{T}$ is the transmission vector. the term $d$ is the distance between the Normal Map and Environment Map or Background Image.  }
\label{figimages/R}
\end{figure}

Using Mock3D shapes, it is also possible to obtain physically plausible but stylistically expressive mirror, transparency, glossy, and translucency effects using the relatively simple framework shown in Figure~\ref{figimages/R}. All of these effects can be obtained by using relatively simple transformations that correspond to the reflection of a foreground image (or Environment Map) (see Algorithm~\ref{AlgolMirror}) and refraction of a background image (see Algorithm~\ref{AlgolRefraction}). Moreover, reflection and refraction terms can be combined using the Fresnel equation, which provides a weighted average of reflection and refraction based on the incident angle. The inclusion of such a physically plausible transparency operation significantly simplifies the pipeline. Instead of re-rendering the whole scene, artists can simply control the Fresnel function to obtain desired effects. This control can significantly improve the efficiency of the image synthesis process by moving the most important global illumination effects in 2.5D rendering. It can also provide 2D artists with the ability to design physically plausible painterly refraction and reflection effects during image manipulation.

\section{Shading with Global Illumination Effects}
\label{section3}

For shading, we need to have a weighted combination of mirror reflection $C_{M}(u,v)$ and refraction $C_{T}(u,v)$ terms that are computed using Algorithms~\ref{AlgolMirror} and~\ref{AlgolRefraction} using Fresnel. As a result, we obtain global effects by replacing $C_2(u,v)$ in Equation~\ref{eqSpecShade} as follows: 
\begin{eqnarray}
C(u,v) & \leftarrow & C_0(u,v) (1-t(u,v)) + C_1(u,v) t(u,v) \nonumber \\
C_2(u,v) & \leftarrow & Clamp\&Step(C_{M}(u,v)F + C_{T}(u,v) (1-F) + s(u,v) C_2(u,v)) \nonumber \\
C(u,v) & \leftarrow & C(u,v) (1-k_s(u,v) s(u,v)) + C_2(u,v) k_s(u,v) 
\label{eqGlobalShade}
\end{eqnarray}
where $Clamp\&Step()$ is applied $R$, $G$, and $B$ channels separately.  Equation~\ref{eqGlobalShade} can, therefore, combine specular highlights with mirror reflection and refractions. To obtain translucency and glossy effects, we applied blur filters to environment maps and background images. 

In Equation~\ref{eqGlobalShade}, the key term is the Fresnel term $F$. It can be computed as the weighted average of two types of Fresnel terms (s-polarized $F_s(\eta, \theta)$ and p-polarized $F_p(\eta, \theta)$), which are given as follows: 
\begin{eqnarray}
F_s(\eta, \theta) &=& \left(\frac{\cos \theta -  \sqrt{\eta (\eta - \sin^2 \theta } }{\cos \theta +  \sqrt{\eta (\eta - \sin^2 \theta}} \right)^2 \label{eqFresnelS} \\
F_p(\eta, \theta)  &=& \left(\frac{\sqrt{(1- \sin^2 \theta / \eta} - \eta  \cos \theta }{ \sqrt{(1- \sin^2 \theta / \eta} + \eta  \cos \theta} \right)^2 \label{eqFresnelP} 
\end{eqnarray}
where $\cos \theta= \vec{L} \cdot \vec{N}$ and $\eta = \frac{\eta_2}{\eta_1}$ is the index of refraction term that is given as the ratio of the two indexes of refractions for a ray going from the isotropic media with an index of refraction $\eta_1$, to another isotropic media with an index of refraction $\eta_2$. 

Another advantage of using $F$ to mix mirror $C_{M}(u,v)$ and refraction $C_{T}(u,v)$ terms is that we can obtain pure reflection or pure refraction by choosing $F=1$ and $F=0$, respectively. This is also useful since artworks can sometimes include only refraction by ignoring Fresnel effects. Further artistic effects can be obtained by using physically plausible approximations of the refraction vector and Fresnel. 

\begin{algorithm}
\caption{Calculation of Mirror Reflection Term $C_{M}(u,v)$}
\label{AlgolMirror}
\begin{algorithmic}
\REQUIRE $\vec{N}(u,v)=(N.x,N.y,N.z)$, $\vec{I}(u,v)=(0,0,1)$, $C_{E}(u,v)$ is the environment map and $d$ is the distance between environment and normal maps. 
\ENSURE $|\vec{N}(u,v)|=N.x^2+N.y^2+N.z^2=1$.
\STATE 
\begin{eqnarray}
\vec{R} &=&(R.x,R.y,R.z) \nonumber \\
& \leftarrow &- \vec{I} + 2 \left(\vec{I} \cdot \vec{N}\right) \vec{N} 
= (2N.z N.x, 2N.z N.y, 2 N.z^2-1) \nonumber\\
\Delta u & \leftarrow & d\frac{R.x}{R.z} 
= d \frac{2N.z N.x}{2 N.z^2-1} \nonumber\\
\Delta v & \leftarrow & d\frac{R.y}{R.z} 
= d \frac{2N.z N.x}{2 N.z^2-1}\nonumber\\ \nonumber\\ 
C_{M}(u,v)  &\leftarrow&  C_{E}((u+\Delta u)\%w,(v+\Delta u)\%h) \nonumber
\end{eqnarray}
\algorithmiccomment{Here, the terms $w$ and $h$ are width and height of the environment map and \% is the modulo operator.}
\end{algorithmic}
\end{algorithm}

\begin{algorithm}
\caption{Calculation of Refraction Term $C_{T}(u,v)$}
\label{AlgolRefraction}
\begin{algorithmic}
\REQUIRE Let normal vector $\vec{N}(u,v)=(N.x,N.y,N.z)$ and eye vector $\vec{I}(u,v)=(0,0,1)$ be given.
\REQUIRE Let $C_{B}(u,v)$ denote the background image and $d$ denote the distance between background image and normal maps.  
\ENSURE $|\vec{N}(u,v)|=N.x^2+N.y^2+N.z^2=1$.\newline
\algorithmiccomment{Let $\vec{T} =(T.x,T.y,T.z)$ denote transmittance vector}
\STATE 
\begin{eqnarray}
\vec{T} & \leftarrow & \frac{-1}{\eta} \vec{I} + \left( \frac{\vec{I} \cdot \vec{N}}{\eta} - \sqrt{\frac{ (\vec{I} \cdot \vec{N})^2 -1}{\eta^2} +1 }\right) \vec{N} \nonumber \\
\Delta u & \leftarrow & d\frac{T.x}{T.z} \nonumber\\
\Delta v & \leftarrow & d\frac{T.y}{T.z} \nonumber\\ \nonumber\\ 
C_{T}(u,v)  &\leftarrow&  C_{B}((u+\Delta u)\%w,(v+\Delta u)\%h) \nonumber 
\end{eqnarray}
\algorithmiccomment{The terms $w$ and $h$ are width and height of the background image and \% is the modulo operator.} \newline
\algorithmiccomment{The index of refraction term $\eta = \frac{\eta_2}{\eta_1}$ is the ratio of the two indexes of refractions for a ray going from the isotropic media with an index of refraction $\eta_1$, to another isotropic media with an index of refraction $\eta_2$.}
\end{algorithmic}
\end{algorithm}

\begin{figure}
\centering
\includegraphics[width=0.95\linewidth]{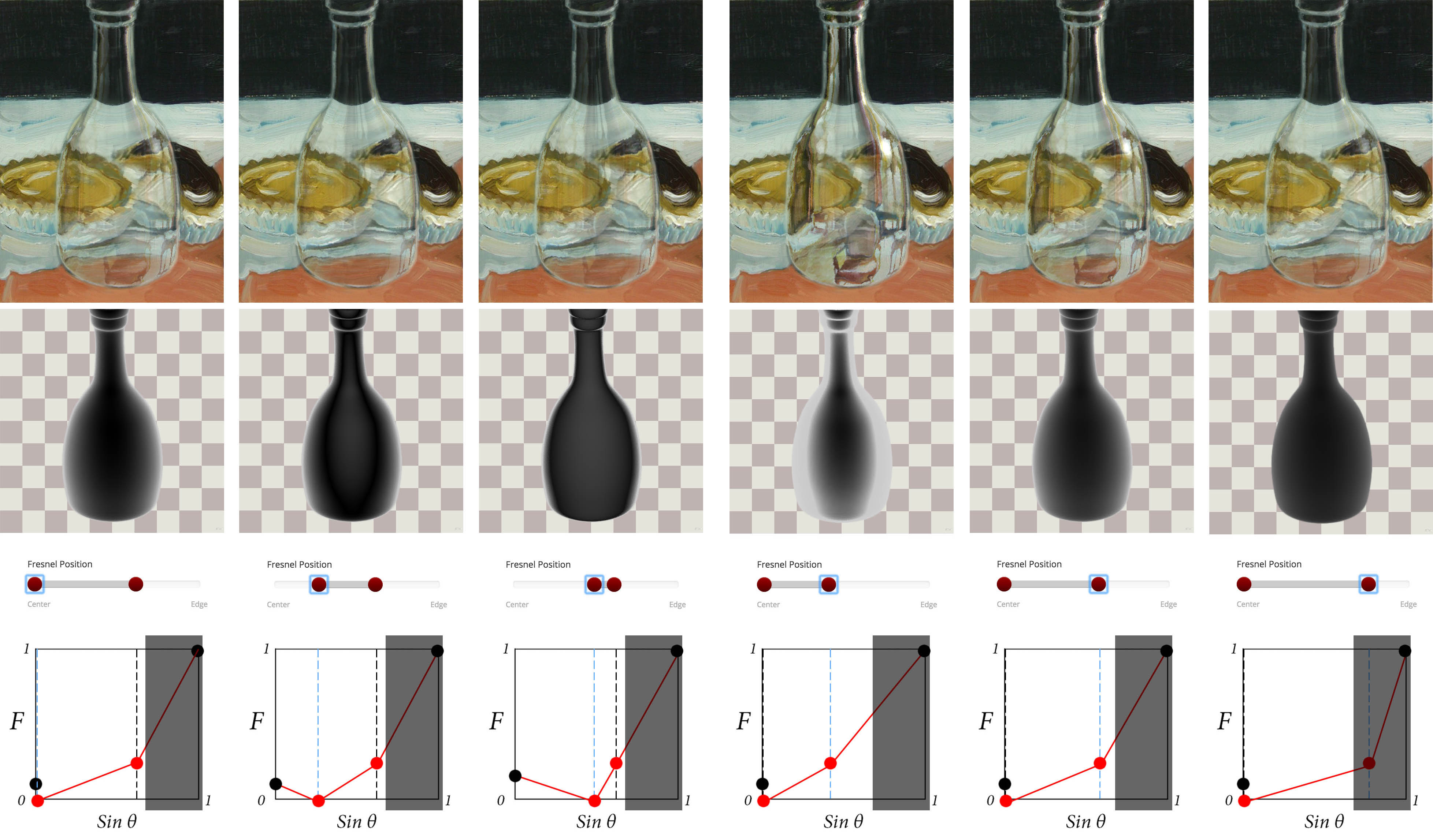}
\caption{Our Fresnel function allows us to combine reflection and refraction with an art-directed control.}
\label{figFresnel0} 
\end{figure}

\subsection{Art-Directed Refractions} 

For artistic applications, equation~\ref{Equ:TransVector} does not give intuitive control since the square root term in this equation can result in no reflection term. 
\begin{equation}\vec{T} =\frac{-1}{\eta} \vec{I} + \left( \frac{\vec{I} \cdot \vec{N}}{\eta} - \sqrt{\frac{ (\vec{I} \cdot \vec{N})^2 -1}{\eta^2} +1 }\right) \vec{N} \label{Equ:TransVector}
\end{equation}
We observe that it is possible to replace this standard refraction equation with linearized transformations for art-directed intuitive control. These linearized transformations provide warping effects that are visually similar to 3D realistic rendering, since they indirectly correspond to the underlying physical phenomena of refraction. To obtain this equation, we first linearize the term $\eta$. Note that this term is a division of two indices of refractions. Since air has an index of refraction $1$, water has an index of refraction $4/3$, and glass has an index of refraction $5/3$,  the term $\eta$ is usually between $3/5$ (glass to air) and $5/3$ (air to glass). These values never become smaller than $1/2$, and they cannot become larger than $2$. Note that these values are not symmetric around $1$. Now, if we take $\mu = \log_2 (\eta)$, the new value $\mu \in [-1,1]$ and the changes become equally spaced between $-1$ and $1$. Based on this new parameter and using the qualitative behavior of the transmission vector, we can write the following formula: 
\begin{eqnarray}
\mbox{if} \;\; \mu > 0: \; \; \; \; \; \; \vec{T} &=& \frac{\vec{V}(1-\mu)-\vec{N}\mu}{|\vec{V}(1-\mu)-\vec{N}\mu|} \nonumber \\
\mbox{if} \; \; \mu \leq 0: \; \; \; \; \; \; \vec{T} &=&  \frac{ ( \vec{V} - \vec{V} \cdot \vec{N}) \vec{N}) \mu  + \vec{V} (1+\mu) }{|( \vec{V} - \vec{V} \cdot \vec{N}) \vec{N}) \mu  + \vec{V} (1+\mu)| }
\label{Equ:LinTransVector}
\end{eqnarray}
This formula provides a linear change based on the values of $\mu$. It is also possible to provide Art-Directed Fresnel Functions. 

\begin{figure}
\centering
\includegraphics[width=0.95\linewidth]{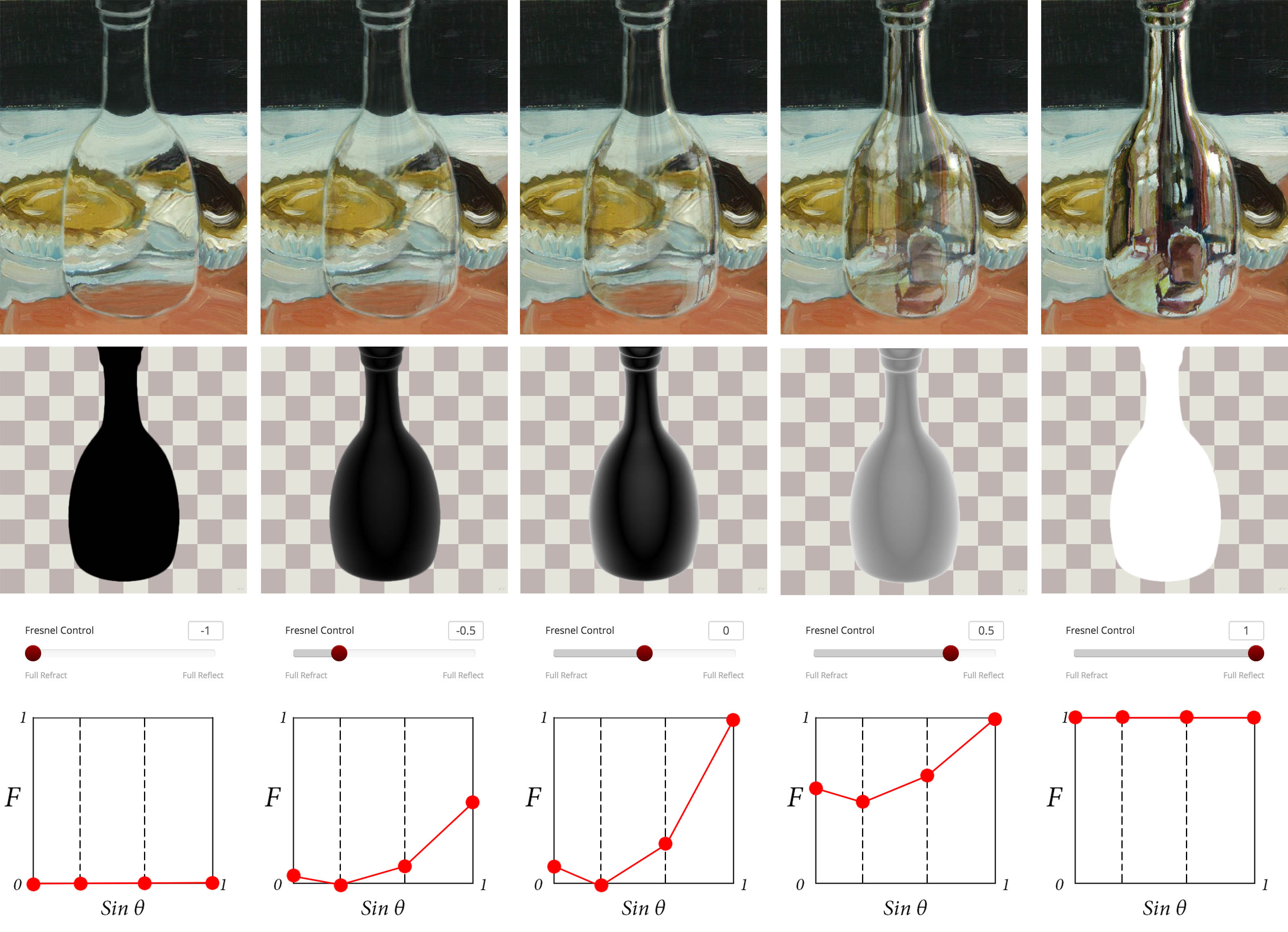}
\caption{Additional Fresnel control: from total refraction to total reflection. }
\label{figFresnel}
\end{figure}

\subsection{Art-Directed Fresnel Function} 

One of the key parts of the 2.5D pipeline is an art-directed Fresnel function that can allow for the physically plausible combination of reflections and refractions. It is possible to control the results using a single slider with two parameters. The first parameter controls the incident angle position where we want to obtain total refraction, and the second parameter controls the incident angle position where we want an equal mix. An additional control allows us to change Fresnel from total refraction to total reflection, as shown in Figure~\ref{figFresnel}.

\begin{figure*}[ht]
\begin{center}
\begin{tabular}{cccccccc}
\includegraphics[width=0.23\linewidth]{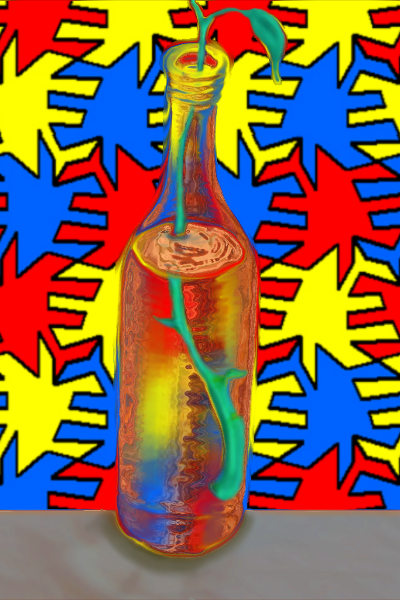}\hfill &
\includegraphics[width=0.23\linewidth]{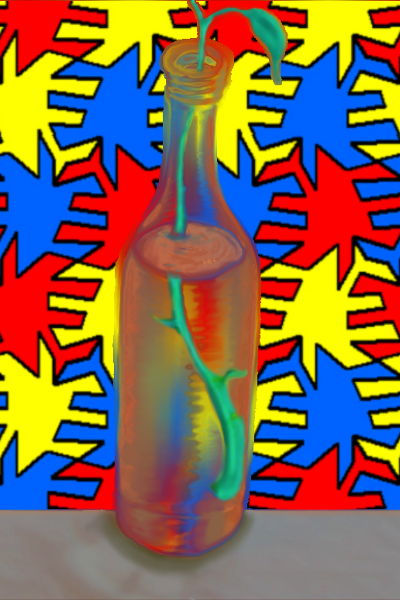}\hfill &
\includegraphics[width=0.23\linewidth]{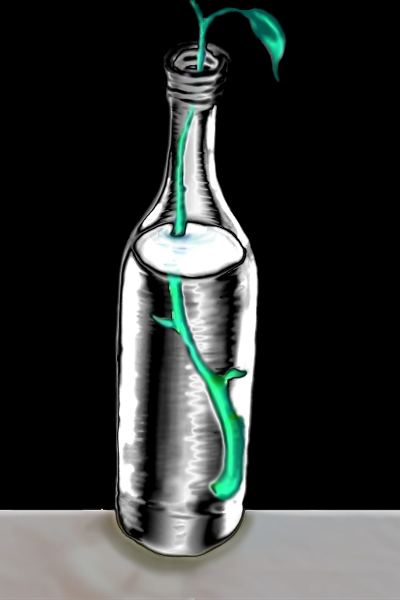}\hfill &
\includegraphics[width=0.23\linewidth]{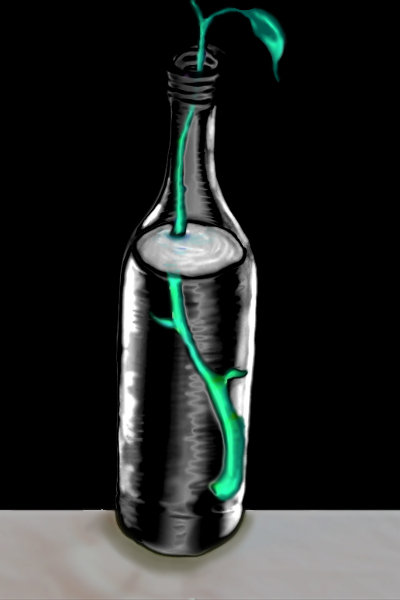}\\
\footnotesize (a) Refraction and  Reflection & 
\footnotesize (b) Translucent and Glossy & 
\multicolumn{2}{c}{\footnotesize (c) Fresnel control by Index of Refraction} \\
\end{tabular}
\end{center}
\caption{\it Another example that demonstrates the effect of thickness in addition to a normal map. (a) Reflection and refraction composited with our Fresnel function. (b) glossy reflection and translucent refraction combined with our Fresnel function. (c) Directly shows the Fresnel control with an index of refraction using a black background and a white environment map. }
\label{figBottle}
\end{figure*}

Our Fresnel function starts with a piecewise-linear approximation of the real Fresnel function. This particular sequence is provided for a single index of refraction. We allow users to change the incident angle positions (i.e. $sin \theta$) of two control points of this piecewise linear function, which turned out to be sufficient to obtain visually convincing Fresnel effects. The upper row of Figure~\ref{figFresnel0} shows the effects of the painterly composition. The middle row directly shows Fresnel control using a black background and a white environment map, in which full black means full refraction and full white means full reflection. In this case, the bottle is simply a normal map image. Gray is a weighted average of reflection and refraction. The last row shows actual Fresnel functions, in which gray areas are not reached to be used.

\section{Conclusion and Future Work} 

Our approach provides the ability for 2D artists to efficiently include refraction and refraction in their artworks and allow intuitive artistic control over visual results (see Figure~\ref{figBottle}). Using this approach, artists can create artificial but still believable versions of the original images, as well as original artwork that can be dynamically manipulated. Our system is available for any artist to create dynamic artwork. The link to our web-based software is mock3d.tamu.edu.  The system is developed using the Java script and WebGL \cite{xiong2016}. The interfaces of our system for normal and depth maps are shown in Figures~\ref{figMock3D0}, and~\ref{figMock3D1}. 

One limitation of this 2.5D pipeline is that it is based on LDR images and it is mainly suitable for relatively simple scenes with only a few objects that are confined in a small space, such as portraits. Using additional images, it is possible to obtain better global effects to create dynamic versions of complex scenes. Depth maps along with normal maps can provide local illumination, shadows, and subsurface scattering. Thickness maps help to accurately model thin objects such as sticks, trees, and humans. It is even useful to define if a certain part of the depth map can receive shadow using an additional alpha map. Our initial investigation suggests that it is possible to provide dynamic versions of complex scenes with a wider range of global effects in real-time. 

To include even more complicated scenes, there is a need to allow multiple 2.5D Proxy Objects. This approach can also allow one to animate scenes by appropriately moving each 2.5D proxy object. To be able to support more complicated scenes, there is a need to allow multiple 2.5D Proxy Objects in the same scene. They can still be included as quadrilateral planes and bilinear surfaces. However, additional layering information is also needed, such as local layering \cite{McCann2009}. 

The main limitation of the image-based approach is that it does not provide full-blown 3D scenes that can be viewed from any point of view. This limitation can be overcome if images are replaced by lumigraphs. Another limitation is to use only LDR images. By replacing LDR images with HDR images, the results can be significantly improved for some specific applications. Using HDR images can even improve the quality of depth maps. Another potential improvement can be to provide a native shading language that can allow better control to obtain the desired look and feel.

\bibliographystyle{unsrtnat}

\bibliography{references}

\begin{thebibliography}{70}
\providecommand{\natexlab}[1]{#1}
\providecommand{\url}[1]{\texttt{#1}}
\expandafter\ifx\csname urlstyle\endcsname\relax
  \providecommand{\doi}[1]{doi: #1}\else
  \providecommand{\doi}{doi: \begingroup \urlstyle{rm}\Url}\fi

\bibitem[Akleman et~al.(2022)Akleman, Shanker, Xiong, Barseghyan, and Fard]{akleman2022dynamic}
Ergun Akleman, Anusha Shanker, Yinan Xiong, Ani Barseghyan, and Motahareh Fard.
\newblock Dynamic paintings: Real-time interactive artworks in web.
\newblock In \emph{Proceedings of International Society of Electronic Arts 2022 (ISEA'2022)}, pages 1--12, June 2022.

\bibitem[Strothotte and Schlechtweg(2002)]{strothotte2002non}
Thomas Strothotte and Stefan Schlechtweg.
\newblock \emph{Non-Photorealistic Computer Graphics: Modeling, Rendering, and Animation}.
\newblock Morgan Kaufmann, San Francisco, CA, 2002.

\bibitem[Gooch and Gooch(2001)]{gooch2001non}
Bruce Gooch and Amy Gooch.
\newblock \emph{Non-photorealistic rendering}.
\newblock AK Peters, Natick, Massachusetts, 2001.

\bibitem[Hertzmann(1999)]{hertzmann1999silhouettes}
Aaron Hertzmann.
\newblock Introduction to {3D} non-photorealistic rendering: Silhouettes and outlines.
\newblock In \emph{Non-Photorealistic Rendering. SIGGRAPH 99 Course Notes}, pages 45--52, New York City, NW, 1999. ACM SIGGRAPH.

\bibitem[Gooch et~al.(1998{\natexlab{a}})Gooch, Gooch, Shirley, and Cohen]{gooch1998non}
Amy Gooch, Bruce Gooch, Peter Shirley, and Elaine Cohen.
\newblock A non-photorealistic lighting model for automatic technical illustration.
\newblock In \emph{Proceedings of the 25th annual conference on Computer graphics and interactive techniques}, pages 447--452, New York City, NW, 1998{\natexlab{a}}. ACM SIGGRAPH.

\bibitem[Deussen and Strothotte(2000)]{deussen2000computer}
Oliver Deussen and Thomas Strothotte.
\newblock Computer-generated pen-and-ink illustration of trees.
\newblock In \emph{Proceedings of the 27th Annual Conference on Computer Graphics and Interactive Techniques}, pages 13--18, New York City, NW, 2000. ACM SIGGRAPH.

\bibitem[Markosian et~al.(1997)Markosian, Kowalski, Goldstein, Trychin, Hughes, and Bourdev]{markosian1997real}
Lee Markosian, Michael~A Kowalski, Daniel Goldstein, Samuel~J Trychin, John~F Hughes, and Lubomir~D Bourdev.
\newblock Real-time nonphotorealistic rendering.
\newblock In \emph{Proceedings of the 24th annual conference on Computer graphics and interactive techniques}, pages 415--420, New York City, NW, 1997. ACM SIGGRAPH.

\bibitem[Du and Akleman(2017)]{du2017designing}
Yuxiao Du and Ergun Akleman.
\newblock Designing look-and-feel using generalized crosshatching.
\newblock In \emph{ACM SIGGRAPH 2017 Talks}, pages 1--2. ACM, 2017.

\bibitem[Litwinowicz(1997)]{litwinowicz1997processing}
Peter Litwinowicz.
\newblock Processing images and video for an impressionist effect.
\newblock In \emph{Proceedings of the 24th annual conference on Computer graphics and interactive techniques}, pages 407--414, New York City, NW, 1997. ACM SIGGRAPH.

\bibitem[Akleman(1998{\natexlab{a}})]{akleman1998}
Ergun Akleman.
\newblock Implicit painting of {CSG} solids.
\newblock In \emph{Proc. of CSG}, volume~98, pages 99--113, 1998{\natexlab{a}}.

\bibitem[Akleman(1998{\natexlab{b}})]{akleman1998a}
Ergun Akleman.
\newblock Implicit surface painting.
\newblock In \emph{Proc. of Implicit Surfaces’ 98}, pages 63--68, 1998{\natexlab{b}}.

\bibitem[Meadows and Akleman(2000)]{meadows2000a}
S.~Meadows and E.~Akleman.
\newblock Abstract digital paintings created with painting camera technique.
\newblock In \emph{Procedings D’ART 2000 / Information Visualization 2000}, pages 11--19, London, GB, 2000. D'ART.

\bibitem[Smith et~al.(2004)Smith, Akleman, Davison, Keyser, et~al.]{smith2004}
Jeffrey Smith, Ergun Akleman, Richard Davison, John Keyser, et~al.
\newblock Multicam: A system for interactive rendering of abstract digital images.
\newblock In \emph{Bridges: Mathematical Connections in Art, Music, and Science}, pages 265--272, Baltimore, MD, 2004. Bridges Organization.

\bibitem[Morrison and Akleman(2020)]{morrison2020remote}
Chris Morrison and Ergun Akleman.
\newblock Remote empathetic viewpoint: A novel approach to extending cubism.
\newblock In \emph{ACM SIGGRAPH 2020 Posters}, pages 1--2. ACM SIGGRAPH, New York City, NW, 2020.

\bibitem[Chan et~al.(2002)Chan, Akleman, and Chen]{chan2002two}
Ching Chan, Ergun Akleman, and Jianer Chen.
\newblock Two methods for creating chinese painting.
\newblock In \emph{10th Pacific Conference on Computer Graphics and Applications, 2002. Proceedings.}, pages 403--412, New York City, NW, 2002. IEEE Computer Society.

\bibitem[Liu and Akleman(2015)]{liu2015chinese}
Siran Liu and Ergun Akleman.
\newblock Chinese ink and brush painting with reflections.
\newblock In \emph{SIGGRAPH 2015: Studio}, SIGGRAPH '15, pages 8:1--8:1, New York City, NW, 2015. ACM SIGGRAPH.
\newblock ISBN 978-1-4503-3637-6.

\bibitem[Majumder and Gopi(2002)]{majumder2002real}
Aditi Majumder and M~Gopi.
\newblock Real time charcoal rendering using contrast enhancement operators.
\newblock In \emph{Proceedings of Symposium of Non Photorealistic Animation and Rendering}, pages 107--116, Goslar Germany, Germany, 2002. Eurographics Association.

\bibitem[Du and Akleman(2016)]{du2016charcoal}
Yuxiao Du and Ergun Akleman.
\newblock Charcoal rendering and shading with reflections.
\newblock In \emph{ACM SIGGRAPH 2016 Posters}, page~32, New York City, NW, 2016. ACM SIGGRAPH.

\bibitem[Lu et~al.(2002)Lu, Morris, Ebert, Rheingans, and Hansen]{lu2002non}
Aidong Lu, Christopher~J Morris, David~S Ebert, Penny Rheingans, and Charles Hansen.
\newblock Non-photorealistic volume rendering using stippling techniques.
\newblock In \emph{Proceedings of the Conference on Visualization '02}, pages 211--218, New York City, NW, 2002. IEEE Computer Society.

\bibitem[Vanderhaeghe and Collomosse(2013)]{vanderhaeghe2013stroke}
David Vanderhaeghe and John Collomosse.
\newblock Stroke based painterly rendering.
\newblock In \emph{Image and Video-Based Artistic Stylisation}, pages 3--21. Springer, New York City, NW, 2013.

\bibitem[Yeh and Ouhyoung(2002)]{yeh2002animals}
Jun-Wei Yeh and Ming Ouhyoung.
\newblock Non-photorealistic rendering in {Chinese} painting of animals.
\newblock \emph{Journal of System Simulation}, 14\penalty0 (6):\penalty0 1220--1224, 2002.

\bibitem[Hertzmann(1998)]{hertzmann1998painterly}
Aaron Hertzmann.
\newblock Painterly rendering with curved brush strokes of multiple sizes.
\newblock In \emph{Proceedings of the 25th Annual Conference on Computer Graphics and Interactive Techniques}, pages 453--460, New York City, NW, 1998. ACM SIGGRAPH.

\bibitem[Lin et~al.(2012)Lin, Zeng, Wang, Xu, and Zhu]{lin2012video}
Liang Lin, Kun Zeng, Yizhou Wang, Ying-Qing Xu, and Song-Chun Zhu.
\newblock Video stylization: Painterly rendering and optimization with content extraction.
\newblock \emph{IEEE Transactions on Circuits and Systems for Video Technology}, 23\penalty0 (4):\penalty0 577--590, 2012.

\bibitem[Curtis et~al.(1997)Curtis, Anderson, Seims, Fleischer, and Salesin]{curtis1997computer}
Cassidy~J Curtis, Sean~E Anderson, Joshua~E Seims, Kurt~W Fleischer, and David~H Salesin.
\newblock Computer-generated watercolor.
\newblock In \emph{Proceedings of the 24th annual conference on Computer graphics and interactive techniques}, pages 421--430, New York City, NW, 1997. ACM SIGGRAPH.

\bibitem[Gatys et~al.(2016)Gatys, Ecker, and Bethge]{gatys2016image}
Leon~A Gatys, Alexander~S Ecker, and Matthias Bethge.
\newblock Image style transfer using convolutional neural networks.
\newblock In \emph{Proceedings of the IEEE conference on computer vision and pattern recognition}, pages 2414--2423, New York City, NW, 2016. IEEE Computer Society.

\bibitem[Selim et~al.(2016)Selim, Elgharib, and Doyle]{selim2016painting}
Ahmed Selim, Mohamed Elgharib, and Linda Doyle.
\newblock Painting style transfer for head portraits using convolutional neural networks.
\newblock \emph{ACM Transactions on Graphics (ToG)}, 35\penalty0 (4):\penalty0 1--18, 2016.

\bibitem[Murphy and Galanter(2015)]{murphy2015developing}
Laura~K Murphy and Philip Galanter.
\newblock Developing stylized trees and landscapes inspired by eyvind earle.
\newblock In \emph{SIGGRAPH 2015: Studio}, pages 1--1. ACM SIGGRAPH, New York City, NW, 2015.

\bibitem[des Arts(2019)]{connaissance2019}
Connaissance des Arts.
\newblock \emph{Van Gogh, La Nuit Etoilee: Atelier des Lumieres}.
\newblock Societe Francaise de Promotion Artistique, Paris, 2019.

\bibitem[Justice and Akleman(2018)]{justice2018}
Matthew Justice and Ergun Akleman.
\newblock A process to create dynamic landscape paintings using barycentric shading with control paintings.
\newblock In \emph{ACM SIGGRAPH 2018 Posters}, page~30, New York City, NW, 2018. ACM SIGGRAPH.

\bibitem[Yan(2015)]{yan2015}
Zhao Yan.
\newblock Painterly shading ocean surface.
\newblock Master's thesis, Texas A\&M University, College Station, TX, 2015.
\newblock Retrieved from http://oaktrust.library.tamu.edu/handle/1969.1/15638.

\bibitem[Subramanian and Akleman(2020)]{subramanian2020painterly}
Meena Subramanian and Ergun Akleman.
\newblock A painterly rendering approach to create still-life paintings with dynamic lighting.
\newblock In \emph{ACM SIGGRAPH 2020 Posters}, pages 1--2. ACM SIGGRAPH, New York City, NW, 2020.

\bibitem[Ross and Akleman(2021{\natexlab{a}})]{ross2021Georgia}
Jessica Ross and Ergun Akleman.
\newblock Georgia o’keeffe shader: An approach to obtain shaders to simulate a given painterly style.
\newblock In \emph{Proceedings of Eurasia Graphics 2021 Seventh Computer Graphics Conference on Computer Graphics, Animation and Gaming Technologies}, pages 27--34. Eurasia Graphics, 2021{\natexlab{a}}.

\bibitem[Youyou(2014)]{wang2014}
Wang Youyou.
\newblock \emph{Qualitative Global Illumination of Mock-{3D} Scenes}.
\newblock PhD thesis, Texas A\&M University, College Station, TX, 2014.
\newblock Retrieved from http://oaktrust.library.tamu.edu/handle/1969.1/157921.

\bibitem[Wang et~al.(2014{\natexlab{a}})Wang, Gonen, and Akleman]{wang2014global}
Youyou Wang, Ozgur Gonen, and Ergun Akleman.
\newblock Global illumination for 2d artworks with vector field rendering.
\newblock In \emph{ACM SIGGRAPH 2014 Posters}, page~95, New York City, NW, 2014{\natexlab{a}}. ACM SIGGRAPH.

\bibitem[Akleman et~al.(2016)Akleman, Liu, and House]{akleman2016}
Ergun Akleman, S~Liu, and Donald House.
\newblock Barycentric shaders: Art directed shading using control images.
\newblock In \emph{Proceedings of the Joint Symposium on Computational Aesthetics and Sketch Based Interfaces and Modeling and Non-Photorealistic Animation and Rendering}, pages 39--49, Geneva, Switzerland, 2016. Eurographics Association.

\bibitem[Akleman et~al.(2017)Akleman, Perumal, and Wang]{akleman2017}
Ergun Akleman, Fermi Perumal, and Youyou Wang.
\newblock Cos $\theta$ shadows: an integrated model for direct illumination, subsurface scattering and shadow computation.
\newblock In \emph{ACM SIGGRAPH 2017 Posters}, page~38, New York City, NW, 2017. ACM SIGGRAPH.

\bibitem[Wang et~al.(2014{\natexlab{b}})Wang, Gonen, and Akleman]{Akleman2014ir1}
Youyou Wang, Ozgur Gonen, and Ergun Akleman.
\newblock Global illumination for 2d artworks with vector field rendering.
\newblock In \emph{ACM SIGGRAPH 2014 Posters}, SIGGRAPH '14, pages 95:1--95:1, New York, NY, USA, 2014{\natexlab{b}}. ACM.

\bibitem[Blinn(1978)]{blinn1978simulation}
James~F Blinn.
\newblock Simulation of wrinkled surfaces.
\newblock \emph{ACM SIGGRAPH computer graphics}, 12\penalty0 (3):\penalty0 286--292, 1978.

\bibitem[Cohen et~al.(1998)Cohen, Olano, and Manocha]{Cohen1998}
Jonathan Cohen, Marc Olano, and Dinesh Manocha.
\newblock Appearance-preserving simplification.
\newblock In \emph{Proceedings of the 25th annual conference on Computer graphics and interactive techniques}, SIGGRAPH '98, pages 115--122, New York, NY, 1998. ACM SIGGRAPH.

\bibitem[Johnston(2002)]{Johnston2002}
Scott~F. Johnston.
\newblock Lumo: illumination for cel animation.
\newblock In \emph{Proceedings of the 2nd international symposium on Non-photorealistic animation and rendering}, NPAR '02, pages 45--52, Geneva, Switzerland, 2002. Eurographics Association.

\bibitem[Okabe et~al.(2006)Okabe, Zeng, Matsushita, Igarashi, Quan, and Shum]{Okabe2006}
Makoto Okabe, Gang Zeng, Yasuyuki Matsushita, Takeo Igarashi, Long Quan, and Heung-Yeung Shum.
\newblock Single-view relighting with normal map painting.
\newblock In \emph{Proceedings of Pacific Graphics}, pages 27--34, New York, NY, 2006. IEEE Computer Society.

\bibitem[Bezerra et~al.(2005)Bezerra, Feijo, and Velho]{Bezerra2005}
H.~Bezerra, B.~Feijo, and L.~Velho.
\newblock An image-based shading pipeline for 2d animation.
\newblock In \emph{18th Brazilian Symposium on Computer Graphics and Image Processing, 2005. SIBGRAPI 2005}, pages 1--9, Natal, Brazil, 2005. IEEE \& SIBGRAPI.

\bibitem[Winnemoeller et~al.(2009)Winnemoeller, Orzan, Boissieux, and Thollot]{Winnemoeller2009}
H.~Winnemoeller, A.~Orzan, L.~Boissieux, and J.~Thollot.
\newblock Texture design and draping in 2d images.
\newblock \emph{Computer Graphics Forum}, 28\penalty0 (4):\penalty0 1091--1099, 2009.

\bibitem[Shao et~al.(2012)Shao, Bousseau, Sheffer, and Singh]{Shao2012}
Cloud Shao, Adrien Bousseau, Alla Sheffer, and Karan Singh.
\newblock Crossshade: shading concept sketches using cross-section curves.
\newblock \emph{ACM Transactions on Graphics (TOG)}, 31\penalty0 (4):\penalty0 45:1--45:11, 2012.

\bibitem[Sun et~al.(2007)Sun, Liang, Wen, and Shum]{Sun07}
J.~Sun, L.~Liang, F.~Wen, and H.~Shum.
\newblock Image vectorization using optimized gradient meshes.
\newblock \emph{ACM Transactions on Graphics (TOG)}, 26\penalty0 (11):\penalty0 11:1--11:7, 2007.

\bibitem[Orzan et~al.(2008)Orzan, Bousseau, Winnemoller, Barla, Thollot, and Salesin]{Orzan09}
Alexandrina Orzan, Adrien Bousseau, Holger Winnemoller, Pascal Barla, Joelle Thollot, and David Salesin.
\newblock Diffusion curves: A vector representation for smooth-shaded images.
\newblock \emph{ACM Transactions on Graphics (TOG)}, 27\penalty0 (3):\penalty0 92:1--92:8, 2008.

\bibitem[S\'{y}kora et~al.(2009)S\'{y}kora, Dingliana, and Collins]{Daniel09}
Daniel S\'{y}kora, John Dingliana, and Steven Collins.
\newblock Lazy- brush: Flexible painting tool for hand-drawn cartoons.
\newblock \emph{Computer Graphics Forum}, 28\penalty0 (2):\penalty0 599--608, 2009.

\bibitem[Finch et~al.(2011)Finch, Snyder, and Hoppe]{Finch11}
M.~Finch, J.~Snyder, and H.~Hoppe.
\newblock Freeform vector graphics with controlled thin-plate splines.
\newblock \emph{ACM Transactions on Graphics (TOG)}, 30:\penalty0 166:1--166:10, 2011.

\bibitem[Wu et~al.(2007)Wu, Tang, Brown, and Shum]{Wu07}
T.~Wu, C.~Tang, M.~Brown, and H.~Shum.
\newblock Shapepalettes: Interactive normal transfer via sketching.
\newblock \emph{ACM Transactions on Graphics (TOG)}, 26\penalty0 (3):\penalty0 44:1--44:5, 2007.

\bibitem[Vergne et~al.(2012)Vergne, Barla, Fleming, and Granier]{vergne2012surface}
Romain Vergne, Pascal Barla, Roland~W Fleming, and Xavier Granier.
\newblock Surface flows for image-based shading design.
\newblock \emph{ACM Transactions on Graphics (TOG)}, 31\penalty0 (4):\penalty0 1--9, 2012.

\bibitem[S{\`y}kora et~al.(2014)S{\`y}kora, Kavan, {\v{C}}ad{\'\i}k, Jamri{\v{s}}ka, Jacobson, Whited, Simmons, and Sorkine-Hornung]{sykora2014ink}
Daniel S{\`y}kora, Ladislav Kavan, Martin {\v{C}}ad{\'\i}k, Ond{\v{r}}ej Jamri{\v{s}}ka, Alec Jacobson, Brian Whited, Maryann Simmons, and Olga Sorkine-Hornung.
\newblock Ink-and-ray: Bas-relief meshes for adding global illumination effects to hand-drawn characters.
\newblock \emph{ACM Transactions on Graphics (TOG)}, 33\penalty0 (2):\penalty0 1--15, 2014.

\bibitem[Gonen(2016)]{gonen2016}
Ozgur Gonen.
\newblock \emph{Quad Dominant 2-Manifold Mesh Modeling}.
\newblock PhD thesis, Texas A\&M University, College Station, TX, 2016.
\newblock Retrieved from http://http://oaktrust.library.tamu.edu/handle/1969.1/161442.

\bibitem[Knoll and Knoll(1988)]{Knoll1988Photoshop}
Thomas Knoll and John Knoll.
\newblock Photoshop, early history.
\newblock https://en.wikipedia.org/wiki/Adobe\_Photoshop\#Early\_history, 1988.

\bibitem[Kimball and Mattis(1988)]{Kimball998GIMP}
Spencer Kimball and Peter Mattis.
\newblock Gimp, early history.
\newblock https://www.gimp.org/about/, 1988.

\bibitem[Pedchenko(2019)]{Pedchenko2021Procreate}
Ksenia Pedchenko.
\newblock Interview with james cuda and lloyd bottomley, procreate, savage interactive.
\newblock https://thedesignest.net/interview-james-cuda-procreate/, 2019.

\bibitem[Zongker et~al.(1999)Zongker, Werner, Curless, and Salesin]{Zongker1999}
Douglas~E Zongker, Dawn~M Werner, Brian Curless, and David~H Salesin.
\newblock Environment matting and compositing.
\newblock In \emph{Proceedings of the 26th annual conference on Computer graphics and interactive techniques}, pages 205--214, New York City, NW, 1999. ACM SIGGRAPH.

\bibitem[Takamatsu(2021)]{takamatsu2021}
Kazuki Takamatsu.
\newblock The depth maps of kazuki takamatsu.
\newblock https://www.artsy.net/artist/kazuki-takamatsu, 2021.

\bibitem[Ross and Akleman(2021{\natexlab{b}})]{ross2022}
Jessica Ross and Ergun Akleman.
\newblock Georgia o'keeffe shader: An approach to obtain shaders to simulate a given painterly style.
\newblock In \emph{Proceedings of Eurasia Graphics 2022}, pages 27--34, Ankara, Turkey, 2021{\natexlab{b}}. Eurasia Graphics.

\bibitem[Ebert et~al.(2003)Ebert, Musgrave, Peachey, Perlin, and Worley]{ebert2003texturing}
David~S Ebert, F~Kenton Musgrave, Darwyn Peachey, Ken Perlin, and Steven Worley.
\newblock \emph{Texturing \& modeling: a procedural approach}.
\newblock Morgan Kaufmann, San Francisco, CA, 2003.

\bibitem[Foley et~al.(1996)Foley, Van, Van~Dam, Feiner, Hughes, and Hughes]{foley1996computer}
James~D Foley, Foley~Dan Van, Andries Van~Dam, Steven~K Feiner, John~F Hughes, and J~Hughes.
\newblock \emph{Computer graphics: principles and practice}, volume 12110.
\newblock Addison-Wesley, Boston, MA, 1996.

\bibitem[Perlin(1985)]{perlin1985image}
Ken Perlin.
\newblock An image synthesizer.
\newblock \emph{ACM Siggraph Computer Graphics}, 19\penalty0 (3):\penalty0 287--296, 1985.

\bibitem[Perlin(2002)]{perlin2002improving}
Ken Perlin.
\newblock Improving noise.
\newblock In \emph{Proceedings of the 29th annual conference on Computer graphics and interactive techniques}, pages 681--682, New York City, NW, 2002. ACM SIGGRAPH.

\bibitem[Kesson(2008)]{kesson2008pixar}
Malcolm Kesson.
\newblock Pixar's renderman.
\newblock In \emph{ACM SIGGRAPH ASIA 2008 courses}, pages 1--138. ACM SIGGRAPH, New York City, NW, 2008.

\bibitem[Apodaca et~al.(2000)Apodaca, Gritz, and Barzel]{apodaca2000advanced}
Anthony~A Apodaca, Larry Gritz, and Ronen Barzel.
\newblock \emph{Advanced RenderMan: Creating CGI for motion pictures}.
\newblock Morgan Kaufmann, San Francisco, CA, 2000.

\bibitem[Gooch et~al.(1998{\natexlab{b}})Gooch, Gooch, Shirley, and Cohen]{Gooch98}
Amy Gooch, Bruce Gooch, Peter Shirley, and Elaine Cohen.
\newblock A non-photorealistic lighting model for automatic technical illustration.
\newblock \emph{ACM Transactions on Graphics (TOG)}, 31\penalty0 (94):\penalty0 447--452, 1998{\natexlab{b}}.

\bibitem[Kim and Neumann(2001)]{kim2001opacity}
Tae-Yong Kim and Ulrich Neumann.
\newblock Opacity shadow maps.
\newblock In \emph{Eurographics Workshop on Rendering Techniques}, pages 177--182, New York City, NW, 2001. Springer.

\bibitem[Aila and Laine(2004)]{aila2004alias}
Timo Aila and Samuli Laine.
\newblock Alias-free shadow maps.
\newblock \emph{Rendering techniques}, 2004:\penalty0 15th, 2004.

\bibitem[Lokovic and Veach(2000)]{lokovic2000deep}
Tom Lokovic and Eric Veach.
\newblock Deep shadow maps.
\newblock In \emph{Proceedings of the 27th annual conference on Computer graphics and interactive techniques}, pages 385--392, New York City, NW, 2000. ACM SIGGRAPH.

\bibitem[Xiong(2016)]{xiong2016}
Yinan Xiong.
\newblock Mock-{3D} web application: Interactive lighting, rendering and shading for {2D} artwork.
\newblock Master's thesis, Texas A\&M University, College Station, TX, 2016.
\newblock Retrieved from http://oaktrust.library.tamu.edu/handle/1969.1/15794.

\bibitem[McCann and Pollard(2009)]{McCann2009}
James McCann and Nancy Pollard.
\newblock Local layering.
\newblock \emph{ACM Transactions on Graphics (TOG)}, 28\penalty0 (3):\penalty0 841--847, 2009.

\end{thebibliography}

\end{document}